\documentclass[epj]{svjour}
\usepackage{graphics}
\usepackage[dvips]{graphicx}
\usepackage[dvips]{color}
\usepackage{dcolumn}
\usepackage{bm}
\usepackage{url}
\usepackage{amsmath}
\usepackage{amssymb}
\usepackage{booktabs}

\begin{document}
\title{Color confinement, chiral symmetry breaking, and catalytic effect induced by monopole and instanton creations}%Chiral symmetry breaking and catalytic effect induced by monopole and instanton creations in QCD}

\author{Masayasu Hasegawa%\inst{1}% etc
\thanks{\emph{Present address: hasegawa@theor.jinr.ru}}%
}                     % Do not remove
\institute{Bogoliubov Laboratory of Theoretical Physics, Joint Institute for Nuclear Research, Dubna, Moscow 141980, Russia}
\date{Received: date / Revised version: date}
\abstract{Our research reveals the relations among monopoles, color confinement, instantons, and chiral symmetry breaking which experiments can detect, by numerical calculations of lattice gauge theory. We first add a monopole and an anti-monopole varying their magnetic charges to the gauge field configurations in the quenched approximation of quantum chromodynamics (QCD), by applying the monopole creation operator and investigate the effects of the added monopoles and anti-monopoles on color confinement. Second, we reveal the quantitative relations among instantons, anti-instantons, and observables using the eigenvalues and eigenvectors of the overlap Dirac operator, which are calculated using the normal configurations and the configurations with the additional monopoles and anti-monopoles. Finally, we ascertain the outcomes by comparing them with the predictions. We have already discovered the catalytic effect: the decay width of the charged pion becomes wider and its lifetime becomes shorter than the experimental outcomes by increasing the number density of instantons and anti-instantons. However, the outcomes in the previous study were obtained using one lattice volume and lattice spacing. In this research, we improve the previous study using a variety of configurations of different lattice volumes and values of the lattice spacing from low to finite temperatures. The main purposes of this study are to inspect the influences of the finite lattice volume and discretization on the observables and quantitative relations that we have obtained in our previous research and to acquire the interpolated results at the continuum limit.
}

\titlerunning{Color confinement, chiral symmetry breaking, and catalytic effect}
\authorrunning{Masayasu Hasegawa}
\maketitle

\section{Introduction}

There are two interesting phenomena of the low-energy region of the QCD. These are color confinement and chiral symmetry breaking. Proving color confinement is one of the most important assignments in elementary particle physics and nuclear physics~\cite{Clay1}. To illuminate the mechanism of color confinement, a great number of studies based on the ideas of the dual superconductor picture of Mandelstam~\cite{Mandelstam1} and 't Hooft~\cite{tHooft2} has already been carried out. The achieved results support the theoretical explanation that magnetic monopoles condensing in the vacuum are the key ingredient for causing the dual Meissner effect and confining color-charged particles~\cite{Kronfel1,Maedan1,Brandstaeter1,Hioki1,DiGiacomo1,Kitahara1,DiGiacomo2,DiGiacomo3,Bornyakov4}.

Currently, experiments to discover magnetic monopoles have been attempted. In condensed matter physics, a research group makes the Dirac monopole in a Bose-Einstein condensate and observes it~\cite{Ray1,Ray2}. P. A. M. Dirac first referred to an isolated magnetic pole or Dirac monopole in 1931. This outcome shows the possibility that a Dirac monopole exists in the real world.

In high-energy physics, the ``Monopole and Exotics Detector at the LHC (MoEDAL)'' experiment, which aims to search for magnetic monopoles and other highly-ionizing particles in proton-proton collisions at the Large Hadron Collider (LHC), has begun. The magnetic monopoles that are produced in high-energy collisions were explored~\cite{Moedal1,Moedal2}.

These experiments are significant challenges to advance the frontiers of science.

Spontaneous breaking of chiral symmetry is another interesting phenomenon of the low-energy QCD~\cite{Nambu1,Nambu2,Goldstone1,Goldstone2,Gross1,Kugo1}. When chiral symmetry is spontaneously broken, the NG (Nambu-Goldstone) boson that is the pion appears through the axial-vector current. The chiral condensate is an order parameter of chiral symmetry breaking and has a nonzero value, and the quarks obtain small masses. Once the quarks obtain the masses, the pion obtains the mass from the assumption of PCAC~\cite{Weinberg1}.

The instanton~\cite{Belavi1} is a configuration of QCD that breaks chiral symmetry~\cite{Dyakonov6,Shuryak2}. The instanton model demonstrates that in an instanton vacuum, the effective mass of a quark is nonzero, and the massless pole of the NG boson appears as a pion; moreover, the chiral condensate and the pion decay constant are reasonably estimated from the number density of the instantons and anti-instantons~\cite{Dyakonov1,Dyakonov2,Dyakonov3,Dyakonov4}.

These magnetic monopoles and instantons are strongly tied to each other and are closely related to quarks and gluons. We suppose that the mechanism of color confinement and the spontaneous breaking of chiral symmetry are connected through magnetic monopoles and instantons. On the one hand, investigating the relations among the magnetic monopoles, instantons, quarks, and gluons by phenomenological calculations is challenging because of the strong interaction in the low-energy region of the QCD. On the other hand, demonstrating the effects of magnetic monopoles and instantons on physical quantities that experiments can detect is fascinating. Therefore, we first carry out lattice gauge theory simulations and investigate the effects.
\begin{table*}[htbp]
  \begin{center}
  \caption{The simulation parameters. The results of the lattice spacing $a$ are calculated with the analytic formula~\cite{Necco1}. The Sommer scale is $r_{0} = 0.5$ [fm]. The monopole and anti-monopole locations are indicated as the four-dimensional coordinates $(\frac{t}{a}, \frac{x}{a}, \frac{y}{a}, \frac{z}{a})$ in the lattice unit.}\label{tb:lattice}
  \begin{tabular}{|c|c|c|c|c|c|c|c|}\hline
    $\beta$ & $a$ [fm] & $V$ $(V_{s}\times T)$ &$V_{\text{phys}}$ [fm$^{4}$]& $m_{c}$ & Monopole & Anti-monopole & $D$ [fm]\\\hline
    5.8457 & 0.1242               & $12^{3}\times$24 & 9.8682 & 0-5 & (12, 9.5, 9.5, 6.5) & (12, 3.5, 3.5, 5.5) & 1.06 \\\hline
    5.9256 & 0.1065               & $14^{3}\times$28 & 9.8682 & 0-5 & (14, 11.5, 11.5, 7.5) & (14, 4.5, 4.5, 6.5) & 1.06 \\\hline
    6.0000 & 9.315$\times10^{-2}$ & $14^{3}\times$28 & 5.7845 & 0-5 & (14, 12.5, 12.5, 8.5) & (14, 4.5, 4.5, 7.5) & 1.06 \\\hline
    6.0000 & 9.315$\times10^{-2}$ & $16^{3}\times$32 & 9.8682 & 0-5 & (16, 12.5, 12.5, 8.5) & (16, 4.5, 4.5, 7.5) & 1.06 \\\hline
    6.0522 & 8.527$\times10^{-2}$ & $18^{3}\times$32 & 9.8682 & 0-5 & (18, 14.5, 14.5, 9.5) & (18, 5.5, 5.5, 8.5) & 1.09 \\\hline
    6.1366 & 7.452$\times10^{-2}$ & $20^{3}\times$40 & 9.8682 & 4-5 & (20, 15.5, 15.5, 10.5) & (20, 5.5, 5.5, 9.5) & 1.06 \\\hline
  \end{tabular}
  \end{center}
\end{table*}

In lattice gauge theory, various kinds of research on monopoles and instantons have already been attempted~\cite{Bornyakov1,Sasaki1,Sasaki2,Kitahara2}. However, fermions that do not preserve chiral symmetry have been mainly used as quarks.

Accordingly, we carry out numerical calculations using the eigenvalues and eigenvectors of the overlap Dirac operator that preserves the exact chiral symmetry in lattice gauge theory~\cite{Ginsparg1,Neuberger1,Neuberger2,Lusher1,Chandrasekharan1} and quantitatively demonstrate the monopole and instanton effects in QCD on chiral symmetry breaking and hadrons~\cite{DiGH3,Hasegawa2,Hasegawa3}. The primary purposes of this study are to inspect the influences of the finite lattice volume and discretization on the observables and quantitative relations that we have obtained in our previous research and to derive the interpolated results at the continuum limit.

First, we add one pair of monopole and anti-monopole with the magnetic charges to the SU(3) gauge field configurations in the quenched approximation of QCD, by applying the monopole creation operator to the vacuum~\cite{DiGH3}. We increase the number of monopoles and anti-monopoles in the configurations by varying the monopole and anti-monopole magnetic charges. We generate the normal configurations and the configurations to which the monopoles and anti-monopoles are added and investigate the effects of the additional monopoles and anti-monopoles on the long monopole loops and color confinement.

Second, we calculate the eigenvalues and eigenvectors of the overlap Dirac operator using the normal configurations and the configurations with the added monopoles and anti-monopoles. We compute observables using eigenvalues and eigenvectors and demonstrate the quantitative relations among the number density of the instantons and anti-instantons, chiral condensate, light quark masses, decay constants, and light meson masses. Finally, we ascertain the quantitative relations by comparing them with the predictions. 

In the previous research, we found the following quantitative relations~\cite{DiGH3,Hasegawa2}.
\begin{enumerate}
\item The added monopole with a magnetic charge of +1 and anti-monopole with a magnetic charge of -1 make one instanton or anti-instanton.
  
\item The value of the chiral condensate, which is an order parameter of symmetry breaking, is defined as a negative number and decreases in direct proportion to the square root of the number density of the instantons and anti-instantons.

\item The light quark masses increase in direct proportion to the square root of the number density of the instantons and anti-instantons.
  
\item The masses and decay constants of the pion and kaon increase in direct proportion to the one-fourth root of the number density of the instantons and anti-instantons.
  
\item We analytically estimated the catalytic effect of the monopoles and instantons on the charged pion using these outcomes as input values. The lifetime of the charged pion becomes shorter than the experimental result because the decay width of the charged pion becomes wider with increasing number density of the instantons and anti-instantons.
\end{enumerate} 
We have confirmed that these results are consistent with the predictions. However, we performed simulations using one lattice of volume $V = 18^{3}\times32$ and parameter $\beta = 6.0522$ for the lattice spacing and obtained these outcomes.

In this research, we generate and use various types of configurations from low to finite temperatures and improve the previous study. We first analyze the number density of the long monopole loops.

At finite temperatures, we calculate the average and absolute values of the Polyakov loop operator as an order parameter of the color deconfinement phase transition in quenched QCD. We demonstrate the relation between the number density of the long monopole loops and the order parameter of the color deconfinement. We then determine the transition temperature from the color confinement phase to the color deconfinement phase and demonstrate that the transition temperature rises by increasing the magnetic charges of the additional monopoles and anti-monopoles.

Next, to demonstrate that the previous results are not caused by lattice artifacts, we inspect the influences of the finite lattice volume and discretization on the outcomes and obtain more precise results by interpolating the outcomes to the continuum limit. 

To inspect the influence of the finite lattice volume, we add the monopole and anti-monopole to the configurations of two different lattice volumes $V$ = $14^{3}\times 28$ ($V_{\text{phys}}$ =  5.7845 [fm$^{4}$] ) and $V$ = $16^{3}\times 32$ ($V_{\text{phys}}$ = 9.8682 [fm$^{4}$]) of the parameter $\beta$ = 6.0000 for the lattice spacing\footnote{The results of $V = 14^{4}$, $\beta$ = 6.0000 have already been reported in our previous research~\cite{DiGH3}.}.

To inspect the discretization influence and interpolate the outcomes to the continuum limit, we set a physical volume to $V_{\text{phys}}$ = 9.8682 [fm$^{4}$]. We add the monopole and anti-monopole to the configurations of five sets of lattice volumes $V$ and parameter values $\beta$ as follows: (i) $V = 12^{3}\times24$ of $\beta = 5.8457$. (ii) $V = 14^{3}\times28$ of $\beta = 5.9256$. (iii) $V$ = $16^{3}\times 32$ of $\beta$ = 6.0000\footnote{The preliminary results were reported in~\cite{DiGH5}.}. (iv) $V = 18^{3}\times32$ of $\beta = 6.0522$\footnote{The numerical results were provided in~\cite{Hasegawa2}.}. (v) $V = 20^{3}\times40$ of $\beta = 6.1356$.

The simulation parameters are presented in Table~\ref{tb:lattice}. In addition, we consider the renormalization constant $\hat{Z}_{S}$ for the scalar density, which is obtained by the nonperturbative calculations~\cite{Wennekers1}, for the computations of the observables. Moreover, to verify whether our outcomes are proper, we ascertain the outcomes by comparing them with the predictions. 

This article is composed of seven sections, and each content of the section is as follows. In section~\ref{sec:2}, we generate the normal configurations and the configurations with the additional monopoles and anti-monopoles from low to high temperatures. We confirm that the monopole creation operator produces monopoles and anti-monopoles by measuring the number density of the long monopole loops. We investigate the effects of the additional monopoles and anti-monopoles on color confinement.

In section~\ref{sec:3}, we compute the number of instantons and anti-instantons created by the added monopoles and anti-monopoles and compare the results with the predictions.

In section~\ref{sec:4}, we examine the effects of the additional monopoles and anti-monopoles on the spectrum of the overlap Dirac operator by comparing them with random matrix theory (RMT).

In section~\ref{sec:5}, we first calculate the correlations of the pseudoscalar density and scalar density and confirm the PCAC relation. We then compute the renormalization constant for the scalar density and evaluate the renormalized chiral condensate in the $\overline{\text{MS}}$-scheme at 2 [GeV] using the scale parameter obtained by comparison with chiral random matrix theory (chRMT). 

In section~\ref{sec:6}, to compare the numerical results with the experimental results, we determine the normalization factor for the pion mass and decay constant by matching the numerical results with the experimental outcomes. We re-evaluate the renormalized chiral condensate in the $\overline{\text{MS}}$-scheme at 2 [GeV] and estimate the renormalized average mass of the light quarks in the $\overline{\text{MS}}$-scheme at 2 [GeV], pion mass, and decay constants. Finally, we evaluate the catalytic effect on the charged pion. We investigate the influences of the finite lattice volume and discretization on the outcomes and evaluate the interpolated results at the continuum limit.

In section~\ref{sec:7}, we provide a summary and conclusions.

We have already presented the preliminary results at international conferences~\cite{Hasegawa3,Hasegawa6}. We put some tables relating to this article in a data file~\cite{Hasegawa5} and disclose it on ``figshare.''
%%%%%%%%%%%%%%%%%%%%%%%%%%%%
%%%% SEC 1 END %%%%%%%%%%%%%%%%%%
%%%%%%%%%%%%%%%%%%%%%%%%%%%%

\section{Monopoles}\label{sec:2}

In this section, we explain monopole and anti-monopole creations in configurations. We then detect the long monopole loops and calculate the number density of the long monopole loops. We investigate the effects of the additional monopoles and anti-monopoles on Abelian dominance, monopole dominance, and color confinement.
%%%%%%%%%%%%%%%%%%%%%%%%%%%%

\subsection{The creation of monopoles and anti-monopoles}\label{sec:monopole_ins}
\begin{figure*}[htbp]
  \begin{center}
    \includegraphics[width=160mm]{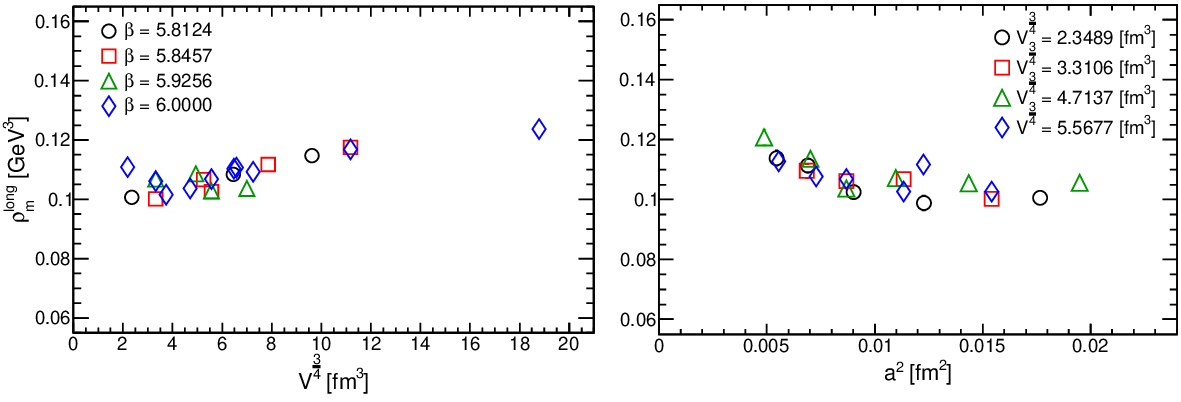}
  \end{center}
  \setlength\abovecaptionskip{-1pt}
  \caption{The number density of the monopoles and anti-monopoles of the standard configurations. The left panel shows the effects of the finite lattice volume on the numerical results of the configurations of $\beta$ = 5.8124, 5.8457, 5.9256, and 6.0000. The right panel shows the discretization effects on the numerical results of three-dimensional physical volumes $V_{\text{phys}}^{\frac{3}{4}}$ = 2.3489, 3.3106, 4.7137, and 5.5677 [fm$^{3}$].}\label{fig:mono_dens_or}
\end{figure*}

We generate the normal gauge field configurations and the configurations with an additional monopole and an anti-monopole in the quenched approximation of QCD using the standard techniques of the heat bath algorithm and the over-relaxation method. In this study, a pair of monopoles and anti-monopoles is added by applying the monopole creation operator to the vacuum of the quenched SU(3) by using the same technique as in our previous study~\cite{DiGH3}.

The additional monopole and anti-monopole are three-dimensional objects. The additional monopole has a positive integer charge from 0 to 5, and the additional anti-monopole has a negative integer charge from 0 to -5. The total magnetic charges of the additional monopoles and anti-monopoles are added to the vacuum to be zero. The magnetic charges are varied to increase the numbers of monopoles and anti-monopoles in the vacuum. Hereafter, the magnetic charge $m_{c}$ indicates that both the magnetic charge $m_{c}$ of the monopole and the magnetic charge $-m_{c}$ of the anti-monopole are added.

We put the monopole and anti-monopole on the dual-sites in the spatial volume $V_{s}$ at the time-slice $\frac{T}{2}$, maintaining the three-dimensional distance $D$ between them to be approximately 1.1 [fm]. $T$ is the time component of the lattice volume $V$. This proper distance is determined by checking the stability of the density of the monopoles and anti-monopoles, which is explained in our previous study~\cite{DiGH3}.

We confirm that the added monopole and anti-monopole do not affect the numerical results of the lattice spacing, which are calculated from the static potential. The numerical results are presented in Table 1 in~\cite{Hasegawa5}. Therefore, we use the outcomes of the lattice spacing that are computed with the analytic formula~\cite{Necco1} and the Sommer scale $r_{0}$ = 0.5 [fm] in this study.

The simulation parameters of $\beta$ for the lattice spacing $a$, lattice volumes $V$, magnetic charges $m_{c}$, locations of the monopole and ant-monopole, and distances $D$ are presented in Table~\ref{tb:lattice}.

To confirm whether we correctly add the monopole and anti-monopole to the configurations, we diagonalize the SU(3) matrices under the condition of the maximal Abelian gauge and perform the Abelian projection~\cite{tHooft3}. We restrain the Gribov copy from influencing the numerical results using the simulated annealing algorithm and analyze the Abelian monopoles.
\begin{figure*}[htbp]
  \begin{center}
    \includegraphics[width=160mm]{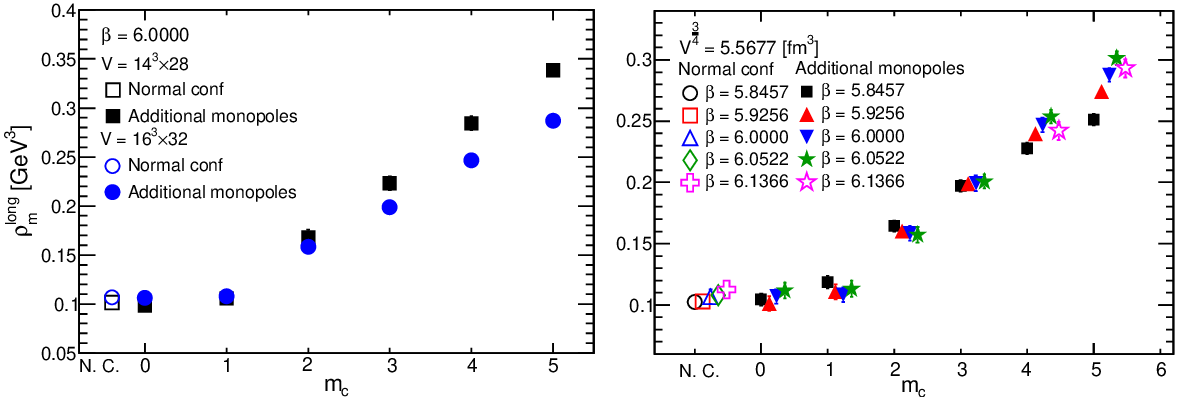}
  \end{center}
  \setlength\abovecaptionskip{-1pt}
  \caption{The number density of the additional monopoles and anti-monopoles. The left panel shows the results of the configurations of $V = 14^{3}\times28$ and $V = 16^{3}\times32$, and the right panel compares the numerical results of the additional monopoles and anti-monopoles of three-dimensional physical volume $V_{\text{phys}}^{\frac{3}{4}} = 5.5677$ [fm$^{3}$] with the results of the standard configurations. The symbols in the right panel are slightly shifted along the horizontal axis to show the difference clearly.}\label{fig:mono_dens_add}
\end{figure*}

Our previous study confirmed that the monopole creation operator makes only long monopole loops~\cite{DiGH3}, which condense in vacua and are closely related to the mechanism of color confinement~\cite{Kitahara1,Kronfel2}. We first count the number of monopoles and anti-monopoles that compose the closed long loops $C$ in the four-dimensional volumes $V$~\cite{Bode1}. The monopole is a three-dimensional object. Therefore, we define the three-dimensional number densities of the monopoles and anti-monopoles by counting the number of monopole currents composing the long monopole loops as follows:
\begin{equation}
  \rho_{m}^{long} \equiv \left(\sum_{i, \mu}\sum_{^{*}n \in C}\frac{|k_{\mu}^{i}(^{*}n)|}{12V}\right)^{\frac{3}{4}} \ \ [\text{GeV}^{3}]\label{eq:mono_dens}
\end{equation}
$k_{\mu}^{i}(^{*}n)$ is the monopole current defined on the dual-sites $^{*}n$~\cite{Kronfel1,Kitahara2,DeGrand1}. It satisfies the current conservation law and forms closed loops~\cite{Bode1}. The indices $i$ and $\mu$ indicate the color $i$ = 1, 2, and 3 and directions $\mu$ = 1, 2, 3, and 4, respectively.

First, we calculate the number density of the monopoles and anti-monopoles using the standard configurations. The numerical results of the standard configurations are listed in Table 2 in~\cite{Hasegawa5}.

To check the finite lattice volume effect on the numerical results, we set the values of the parameter $\beta$ to $\beta$ = 5.8124, 5.8457, 5.9256, and 6.0000, and vary the lattice volumes. The left panel of Fig.~\ref{fig:mono_dens_or} shows that the effects of the finite lattice volume appear when we use the fine lattice spacing ($\beta$ = 6.0000) and three-dimensional physical volumes $V_{\text{phys}}^{\frac{3}{4}}$ from 2 to 4 [fm$^{3}$]; moreover, the calculated results slightly become larger as the physical volumes become larger.

Then, to inspect the discretization effects, we set the values of the parameter $\beta$ and the lattice volumes $V$ so that three-dimensional physical volumes are $V_{\text{phys}}^{\frac{3}{4}}$ = 2.3489, 3.3106, 4.7137, and 5.5677 and vary the values of the lattice spacing and the lattice volumes. The right panel of Fig.~\ref{fig:mono_dens_or} shows that the numerical results of the different physical volumes are consistent; the results, however, slightly rise near the continuum limit. Therefore, in this study, we add the monopoles and anti-monopole to the vacua of the three-dimensional physical volume $V_{\text{phys}}^{\frac{3}{4}} = 5.5677$ [fm$^{3}$] ($V_{\text{phys}} = 9.8682$ [fm$^{4}$]) and vary the values of the parameter $\beta$ to confirm the discretization effects and obtain the results at the continuum limit by interpolation. 

The monopole creation operator, which has a certain number of magnetic charges, makes a certain number of monopoles and anti-monopoles in vacua; therefore, when the lattice volumes of $\beta = 6.0000$ increase from $V = 14^{3}\times28$ to $V = 16^{3}\times32$, the number densities of the monopoles and anti-monopoles of each magnetic charge reduce as shown in the left panel of Fig.~\ref{fig:mono_dens_add}. 
\begin{table*}[htb]
  \begin{center}
    \caption{The ratios of the string tensions of the non-Abelian $\sigma_{\text{non-Abe}}$, Abelian $\sigma_{\text{Abe}}$, monopole $\sigma_{\text{Mon}}$, and photon $\sigma_{\text{Pho}}$. The lattice is $V = 16^{3}\times32$ of $\beta = 6.0000$. The superscripts N. C. and $m_{c}$ represent the computed results of the normal configuration and magnetic charges $m_{c}$, respectively. The string tensions of the Abelian, monopole, and photon are determined when the size of the temporal direction of the Wilson loop is at $T/a$ = 6, the fitting range is $R_{I}/a$ = 1-8, and one-hundred diagonalized configurations for each type of configuration are used for these calculations.}
  \begin{tabular}{|c|c|c|c|c|c|c|} \hline
    $m_{c}$ & $\frac{\sigma_{\text{Abe}}^{m_{c}}}{\sigma_{\text{non-Abe}}^{\text{N.C.}}}$ & $\frac{\sigma_{\text{Mon}}^{m_{c}}}{\sigma_{\text{Abe}}^{\text{N.C.}}}$ & $\frac{\sigma_{\text{Pho}}^{m_{c}}}{\sigma_{\text{Abe}}^{\text{N.C.}}}$ & $\frac{\sigma_{\text{Abe}}^{m_{c}}}{\sigma_{\text{non-Abe}}^{m_{c}}}$ & $\frac{\sigma_{\text{Mon}}^{m_{c}}}{\sigma_{\text{Abe}}^{m_{c}}}$ & $\frac{\sigma_{\text{Pho}}^{m_{c}}}{\sigma_{\text{Abe}}^{m_{c}}}$ \\ \hline
    N.C. & 0.74(4) & 0.69(6) & 0.27(5) & - & - & - \\ \hline
    0  & 0.73(5) & 0.69(7) & 0.28(6) & 0.76(5) & 0.70(7) & 0.28(6) \\ \hline
    1  & 0.74(4) & 0.70(6) & 0.23(5) & 0.73(4) & 0.70(6) & 0.23(5) \\ \hline
    2  & 0.77(4) & 0.73(6) & 0.20(5) & 0.76(4) & 0.70(6) & 0.19(4) \\ \hline
    3  & 0.79(5) & 0.76(7) & 0.19(5) & 0.77(5) & 0.71(6) & 0.16(5) \\ \hline
    4  & 0.79(5) & 0.78(6) & 0.16(5) & 0.81(5) & 0.73(6) & 0.15(5) \\ \hline
    5  & 0.78(5) & 0.75(6) & 0.19(5) & 0.80(6) & 0.72(6) & 0.18(5) \\ \hline
  \end{tabular}
\label{tb:sigma_3_ov_abe_t6}
  \end{center}
\end{table*}

The number densities of the monopoles and anti-monopoles computed from the configurations of the same three-dimensional physical volume $V_{\text{phys}}^{\frac{3}{4}} = 5.5677$ [fm$^{3}$] are shown in the right panel of Fig.~\ref{fig:mono_dens_add}. The figure shows that the finite lattice volume effects appear when we add the magnetic charges $m_{c} =$ 4 and 5 to the lattice volume $V = 12^{3}\times24$. Therefore, we generate configurations of the larger lattice volume $V = 20^{3}\times40$ of the magnetic charges $m_{c} =$ 4 and 5 and obtain the interpolated results at the continuum limit by fitting the curves.

The numerical results of the number density of the additional monopoles and anti-monopoles are shown in Table 3 in~\cite{Hasegawa5}.
%%%%%%%%%%%%%%%%%%%%%%%%%%%%%%%%%%%%%%%%%%%%
%%%%%%%%%%%%%%%%%%%%%%%%%%%%%%%%%%%%%%%%%%%%

\subsection{Abelian dominance and monopole dominance}

The string tension of the linear quark potential computed from the non-Abelian gauge links is reproduced by the string tension of the static potential calculated from the Abelian contribution. This is Abelian dominance. Similarly, the string tension of the linear potential computed from the Abelian contribution is reproduced by the string tension of the static potential calculated from the monopole contribution. This is monopole dominance. The rest of the contribution is the photon. The string tension of the static potential calculated from the photon part becomes approximately zero. Abelian dominance and monopole dominance are important features supporting that monopoles are responsible for color confinement~\cite{Hioki1,Smit1,Ejiri1,Suzuki3}.

After smearing to the spatial components of the non-Abelian link variables, we diagonalize the configurations of $V = 16^{3}\times32$ of $\beta = 6.0000$ under the condition of the maximal Abelian gauge as discussed in subsection~\ref{sec:monopole_ins}. The same smearing parameters are used as presented in Table 1 in~\cite{Hasegawa5}. We use the same method in reference~\cite{Bornyakov4}~\footnote{We do not perform smearing on the spatial components of the Abelian link variables after performing the Abelian projection because we do not observe any improvements in the Abelian dominance or monopole dominance by the smearing.} and derive the Abelian, monopole, and photon parts from the non-Abelian gauge fields. To inspect the impact of the additional monopole and anti-monopoles on the Abelian dominance and monopole dominance, we compute the Wilson loops from the Abelian, monopole, and photon parts and analyze their static potentials. We determine their string tensions $\sigma_{\text{Abe}}$, $\sigma_{\text{Mon}}$, and $\sigma_{\text{Pho}}$, by fitting the same curve
\begin{equation}
  V(R) = V_{0} + \sigma R - \frac{\alpha}{R}\label{eq:st_poten}
\end{equation}
in section 1~\cite{Hasegawa5}, respectively, and compare the ratios of the string tensions. The numerical results in Table~\ref{tb:sigma_3_ov_abe_t6} show that the Abelian dominance $\frac{\sigma_{\text{Abe}}}{\sigma_{\text{non-Abe}}}$ and monopole dominance $\frac{\sigma_{\text{Mon}}}{\sigma_{\text{Abe}}}$ slightly increase with the increase in the magnetic charges of the additional monopoles and anti-monopoles. However, if considering the errors in the numerical results, we do not observe the effects of the additional monopoles and anti-monopoles on the Abelian dominance and monopole dominance.
%%%%%%%%%%%%%%%%%%%%%%%%%%%%%%%%%%%%%%%%%%%%
%%%%%%%%%%%%%%%%%%%%%%%%%%%%%%%%%%%%%%%%%%%%

\subsection{Monopole effects on color confinement and deconfinement}
 
In this subsection, we generate the standard configuration and the configurations with the additional monopoles and anti-monopoles at finite temperatures and inspect the effects of the additional monopoles and anti-monopoles on color confinement and deconfinement.

The critical temperature of the Wilson gauge action at the continuum limit is evaluated as follows~\cite{Necco2}:
\begin{equation}
 T_{c}r_{0} = 0.750(5)\label{eq:chi_temp}
\end{equation}
Therefore, the transition temperature of the standard configuration is estimated as follows:
\begin{equation}
T_{c}^{\text{sta}} = 296(2) \ \ [\text{MeV}]\label{eq:trans_temp}
\end{equation}
The temperature of the configurations is estimated from the following equation:
\begin{equation}
Tr_{0} = \frac{1}{N_{t}\left(a/r_{0}\right)}.
\end{equation}
The values of the lattice spacing $a/r_{0}$ are yielded using the analytic function (3.10) in reference~\cite{Necco2}\footnote{We suppose that the lattice spacing is not affected by the additional monopoles and anti-monopoles from the discussion in subsection~\ref{sec:monopole_ins}.}.
\begin{table}[htb]
  \begin{center}
      \caption{The simulation parameters for the configurations of the finite temperature.}\label{tb:conf_info_tc_18xx3x6}
    \begin{tabular}{|c|c|c|c|} \hline
      $T/T_{c}^{\text{sta}}$ & $V_{s}\times N_{t}$ & $\beta$ & $a/r_{0}$ \\ \hline
      0.8282 & $14^{3}\times6$ & 5.8074 & 0.2684 \\ \hline          
      0.9155 & $16^{3}\times6$ & 5.8563 & 0.2428 \\ \hline          
      1.0000 & $18^{3}\times6$ & 5.9019 & 0.2223 \\ \hline           
      1.0822 & $20^{3}\times6$ & 5.9445 & 0.2054 \\ \hline           
      1.1624 & $22^{3}\times6$ & 5.9848 & 0.1912 \\ \hline           
      1.2408 & $24^{3}\times6$ & 6.0230 & 0.1791 \\ \hline           
      1.3176 & $26^{3}\times6$ & 6.0594 & 0.1687 \\ \hline           
      1.3929 & $28^{3}\times6$ & 6.0941 & 0.1596 \\ \hline           
      1.4669 & $30^{3}\times6$ & 6.1274 & 0.1515 \\ \hline           
      1.5396 & $32^{3}\times6$ & 6.1592 & 0.1444 \\ \hline           
      1.6112 & $34^{3}\times6$ & 6.1897 & 0.1380 \\ \hline           
    \end{tabular}
  \end{center}
\end{table}

We first determine the lattice volume $V = 18^{3}\times6$ and parameter $\beta$ = 5.9019 so that the temperature of the lattice becomes the transition temperature of the standard configuration~(\ref{eq:trans_temp}). We then set the physical volume of the configurations at $V_{\text{phys}}$ = 5.3390 [fm$^{4}$] and the temporal direction of the configurations at $N_{t}/a = 6$. To change the temperature of the configurations, we vary the values of the parameter $\beta$ and the spatial components of the lattice volumes from $V_{s}$ = $14^{3}$, $16^{3}$, $18^{3}$, $\cdots$ to $34^{3}$; thus the temperatures of the configurations vary from $T/T_{c}^{\text{sta}}$ = 0.8282 to 1.6112.
\begin{figure}[htbp]
  \begin{center}
    \includegraphics[width=78mm]{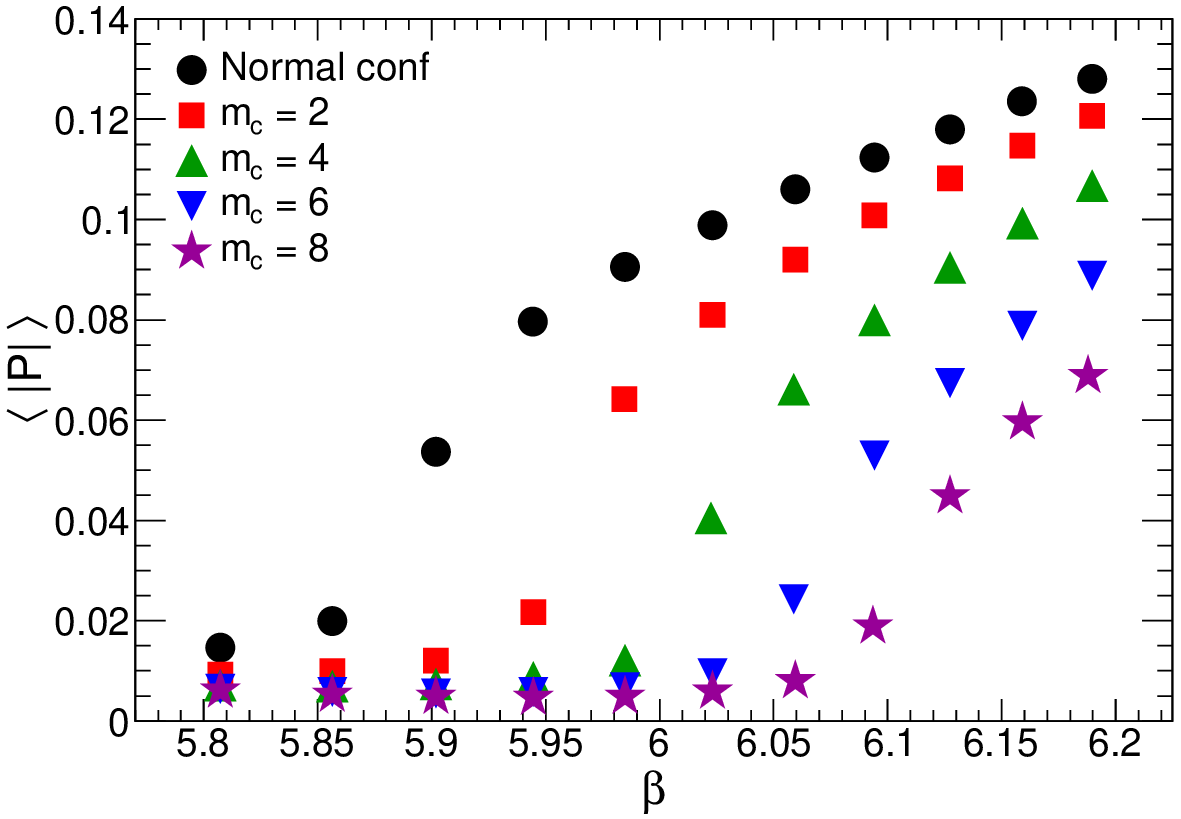}
  \end{center}
  \setlength\abovecaptionskip{-1pt}
  \caption{The phase transitions varying with increasing the magnetic charge $m_{c}$.}\label{fig:Ploop_vs_beta}
\end{figure}

The simulation parameters are presented in Table~\ref{tb:conf_info_tc_18xx3x6}. The numbers of configurations range from $N_{\text{conf}} = 3.0\times10^{3}$ to $7.4\times10^{3}$. We vary the magnetic charges of the additional monopoles and anti-monopoles from $m_{c}$ = 2, 4, 6, to 8. The distances between the additional monopole and anti-monopole are from 1.3 to 1.6 [fm].

In the quenched approximation, the expectation value of the Polyakov loop is an order parameter for the phase transition of color deconfinement. First, we compute the expectation value of the absolute value of the Polyakov loops $\langle |P| \rangle$ from the SU(3) non-Abelian link variables $U(x, k)$ as the order parameter, using the standard configurations and the configurations with the additional monopoles and anti-monopoles. The Polyakov loop is defined as follows:
\begin{equation}
  P(\vec{x}) = \text{Tr}\prod_{x_{0} = 0}^{N_{t}-1}U(x, 0)
\end{equation}
$U(x, 0)$ indicates the SU(3) non-Abelian link variables at site $x$ and direction $k = 0$.

We do not diagonalize the configurations under any condition to compute the Polyakov loops; therefore, this observable does not depend on the choice of the gauge condition\footnote{The problems of gauge dependence concerning the Abelian observables are discussed in references~\cite{Suzuki1,Hasegawa1}.}. 

Figure~\ref{fig:Ploop_vs_beta} demonstrates that the transition temperature from the color confinement phase to the deconfinement phase becomes higher than $\beta$ = 5.9019 with increasing the magnetic charges of the additional monopoles and anti-monopoles.
 \begin{figure}[htbp]
  \begin{center}
    \includegraphics[width=78mm]{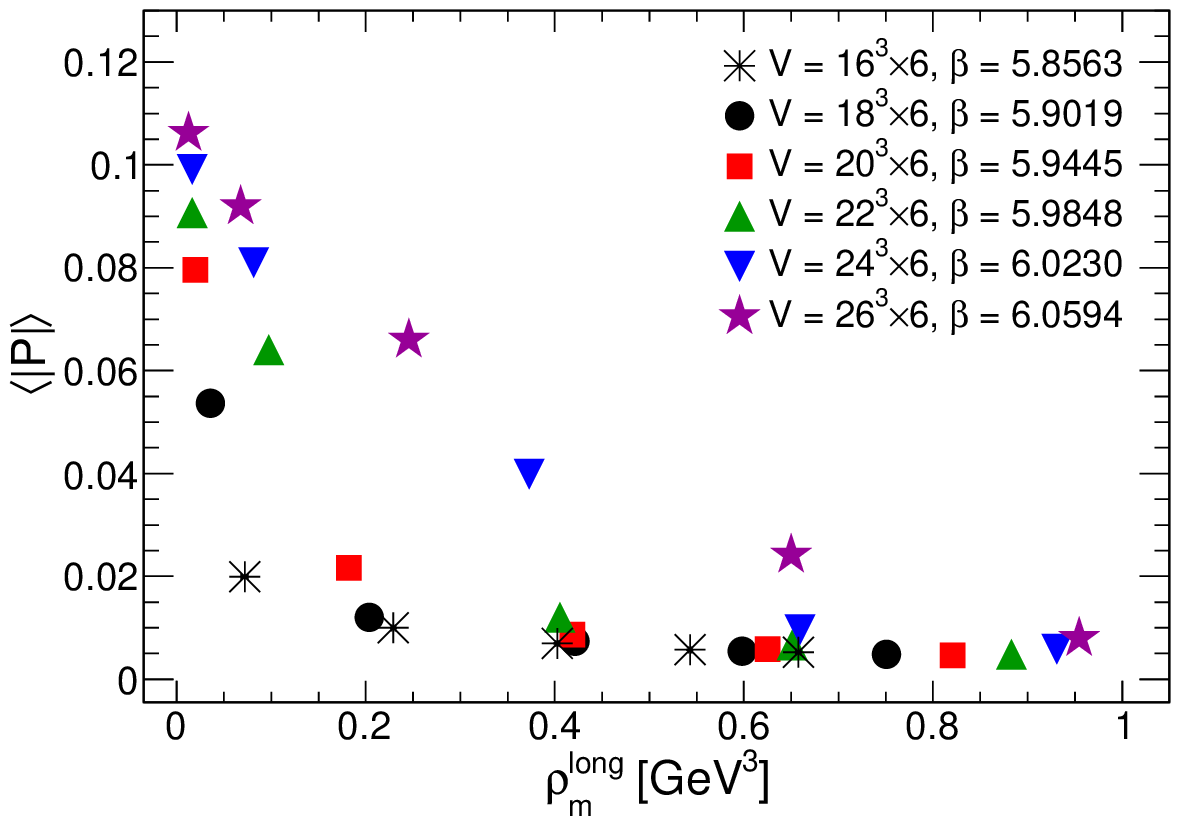}
  \end{center}
  \setlength\abovecaptionskip{-1pt}
  \caption{The relation between the order parameter $\langle |P| \rangle$ and the number density of the long monopole loops $\rho^{long}_{m}$ at finite temperatures.}\label{fig:Ploop_vs_mon_dens1}
 \end{figure}
\begin{table*}[htbp]
  \begin{center}
    \caption{The fitting results of the Polyakov loop susceptibility $\chi$ by the curve~(\ref{eq:gauss_func_tc}) and the analytical results of the transition temperature $T_{c}$. The errors of $T_{c}$ are analyzed including the errors of formula~(\ref{eq:chi_temp}).}\label{tb:Tc}
  \begin{tabular}{|c|c|c|c|c|c|c|c|}\hline
    $m_{c}$ & $p_{1}$ & $p_{2}$ & $p_{3}$ & $p_{4}$ & $T_{c}$  [MeV]  & FR: $\beta$ & $\frac{\chi^{2}}{\text{dof}}$ \\
     &  &  & $\times10^{-2}$ & & $\times10^{2}$ & & \\ \hline    
    N. C.  & 0.124(14) & 5.900(3) & 2.5(2) & 0.182(14) & 2.95(2)  & 5.79-6.19 & 3.5/7.0 \\  \hline
    2 & 0.104(14) & 5.960(5) & 2.7(2) & 0.115(10) & 3.30(2) & 5.84-6.19 & 4.0/6.0 \\  \hline 
    4 & 0.116(16) & 6.024(3) & 2.02(19) &  0.173(14) & 3.68(3) & 5.93-6.19 & 2.0/4.0 \\  \hline
    6 & 0.111(15) & 6.063(3) & 1.92(17) & 0.137(13) & 3.92(3) & 5.97-6.19 & 2.0/3.0 \\  \hline
    8 & 9.6(15)$\times10^{-2}$ & 6.099(4) & 1.91(18) & 0.132(15) & 4.16(3) & 6.01-6.19 & 1.3/2.0 \\ \hline
  \end{tabular}
  \end{center}
\end{table*}

Next, we diagonalize the configurations under the condition of the maximal Abelian gauge as in subsection~\ref{sec:monopole_ins}, calculate the number density of the long monopole loops $\rho^{long}_{m}$~(\ref{eq:mono_dens}), and compare them with the expectation value of the absolute value of the Polyakov loops $\langle| P |\rangle$ as shown in Fig.~\ref{fig:Ploop_vs_mon_dens1}. The computed results are presented in Table 5 in~\cite{Hasegawa5}. This figure shows that the expectation values of the absolute values of the Polyakov loops approach zero when the number densities of the long monopole loops increase.%; thus, the color deconfinement phase reaches the color confinement phase by expanding the long monopole loops.
\begin{figure}[htbp]
  \begin{center}
    \includegraphics[width=78mm]{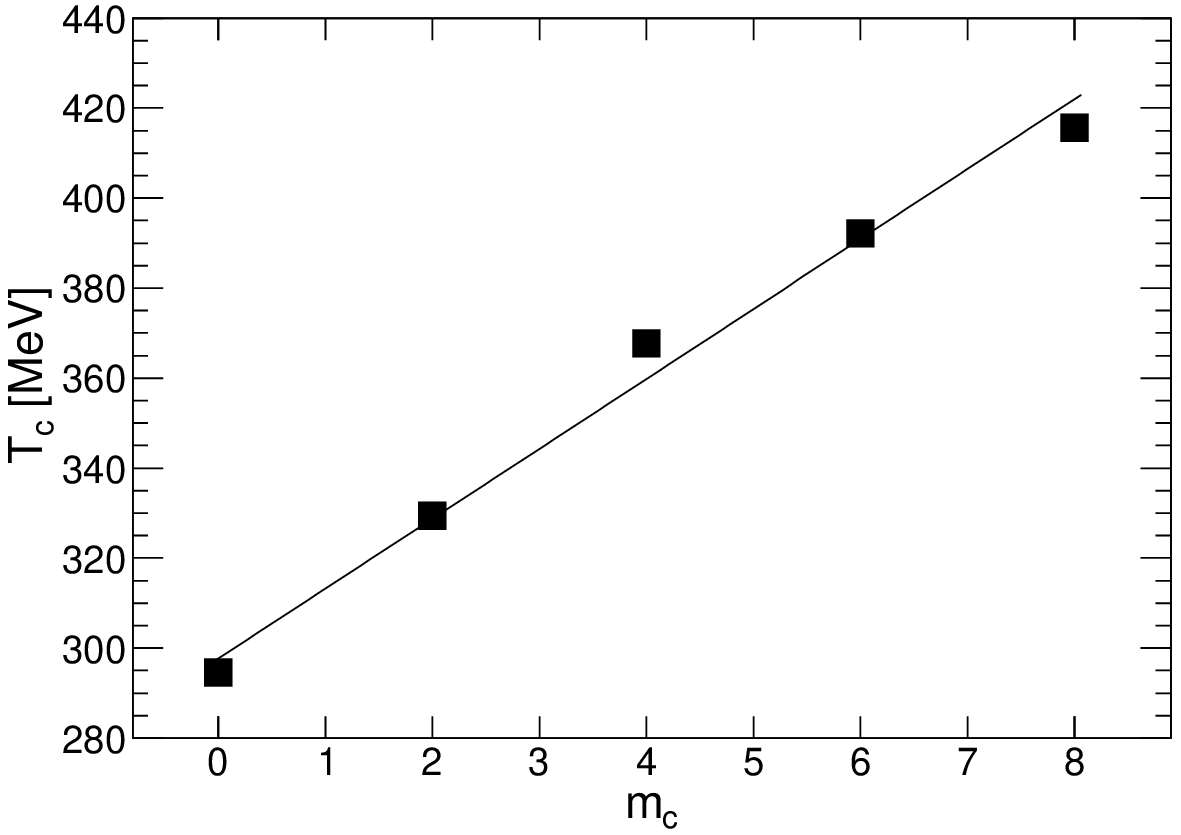}
  \end{center}
  \setlength\abovecaptionskip{-1pt}
  \caption{The rise of the critical temperatures $T_{c}$ with increasing the magnetic charges $m_{c}$. The outcome of the magnetic charge $m_{c} = 0$ is the result of the normal configuration.}\label{fig:Tc}
\end{figure}

Finally, we determine the transition temperatures from the color confinement phase to the color deconfinement phase by analyzing the Polyakov loop susceptibilities. The Polyakov loop susceptibility $\chi$ is calculated as follows:
\begin{equation}
  \chi = V_{s}\left(\langle |P^{2}| \rangle - \langle |P| \rangle^{2}\right)
\end{equation}
The numerical results are presented in Tables 6 and 7 in~\cite{Hasegawa5}. We fit the following Gaussian function to the numerical outcomes of the Polyakov loop susceptibilities and determine the transition temperature for each magnetic charge:
\begin{equation}
  \chi = \frac{p_{1}}{p_{3}\sqrt{2\pi}}\exp\left\{-\frac{1}{2}\left(\frac{x-p_{2}}{p_{3}}\right)^{2}\right\}+p_{4}, \ x = \beta.\label{eq:gauss_func_tc}
\end{equation}
The fitting results and outcomes of the critical temperatures are shown in Table~\ref{tb:Tc}. The outcome of $T_{c}$ of the normal configurations by fitting is consistent with the analytical result~(\ref{eq:trans_temp}); thus, this result demonstrates that we correctly determine the critical temperature in our calculations.

We plot the outcomes of the transition temperatures and fit the following linear function as shown in Fig.~\ref{fig:Tc}: $T_{c} = am_{c} + b$. The fitting results are $a$ = 15.5(4) [MeV], $b$ = $2.977(16)\times10^{2}$ [MeV], and $\frac{\chi^{2}}{\text{dof}} = 18/3$. This finding does not depend on the choice of the gauge condition and corroborates that the transition temperature from the color confinement phase to the color deconfinement phase linearly raises with increasing the magnetic charges of the additional monopoles and anti-monopoles; thus, the additional monopoles and anti-monopoles play a significant role in the color confinement mechanism.

Hereafter, we do not diagonalize the SU(3) matrices under any conditions. We do not perform the Abelian projection or smearing to the SU(3) gauge links. We directly calculate the overlap Dirac operator from the SU(3) non-Abelian gauge links in the sections below.%OK, 31 Aug 2022.
%%%%%%%%%%%%%%%%%%%%%%%%%%%%%%%%%%%%%%%%%%%%
%%%%%%%%%%%%%%%%%%%%%%%%%%%%%%%%%%%%%%%%%%%%
%%%%%%%%%%%%%%%%%%%%%%%%%%%%%%%%%%%%%%%%%%%%
%% END SEC 2 END
%%%%%%%%%%%%%%
\begin{figure*}[htbp]
  \begin{center}
    \includegraphics[width=160mm]{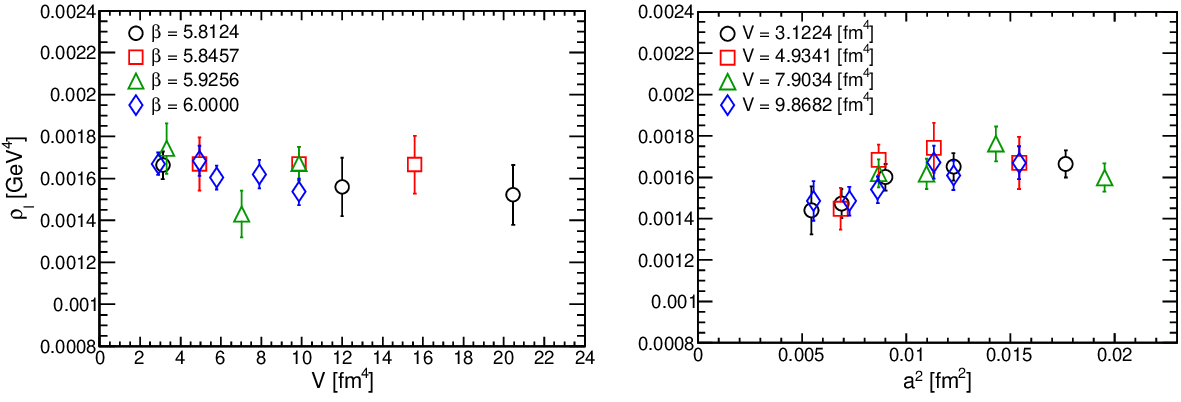}
  \end{center}
  \setlength\abovecaptionskip{-1pt}
  \caption{The number density of the instanton and anti-instantons of the standard configurations. The left panel shows the numerical results of the parameter $\beta$ = 5.8124, 5.8457, 5.9256, and 6.0000, and the right panel shows the numerical results of the physical volumes $V_{\text{phys}}$ = 3.1224, 4.9341, 7.9034, and 9.8682 [fm$^{4}$].}\label{fig:ins_dens_or}
\end{figure*}

\section{Instantons}\label{sec:3}

In this section, we detect the topological charges in the spectrum of the overlap Dirac operator and estimate the number density of the instantons and anti-instantons from the topological charges. We compare distributions of the topological charges with our prediction.

\subsection{The number density of the instantons and anti-instantons of standard configurations}

The overlap Dirac operator preserves exact chiral symmetry at the continuum limit, and there are fermion zero modes in its spectrum~\cite{Ginsparg1,Neuberger1,Neuberger2,Lusher1}.

We calculate the low-lying eigenvalues $\lambda_{i}$ and eigenvectors of the massless overlap Dirac operator to count the number of instantons and anti-instantons from the number of fermion zero modes. We use the numerical technique~\cite{Giusti6} and calculate approximately 100 pairs of low-lying eigenvalues and eigenvectors for each configuration by using the subroutines ({\small ARPACK}). The definition and parameters of the massless overlap Dirac operator are the same as those in our previous studies~\cite{DiGH3,Hasegawa2}.

To detect the zero modes in the spectrum, we first find eigenvalues with absolute values smaller than $1.0\times10^{-7}$, calculate their chiralities by multiplying $\gamma_{5}$ by their eigenvectors, and count the number of eigenvalues for each chirality. We define the number of zero modes as follows: when the chirality is a positive sign, the number of zero modes is $n_{+}$. When the chirality is a negative sign, the number of zero modes is $n_{-}$. Let the number of instantons be $n_{+}$ and the number of anti-instantons be $n_{-}$.

In our studies, we, however, have never observed zero modes of positive chirality and negative chirality from the same configuration. The observed zero modes are consistent with the topological charges $Q$, which are defined as follows:
\begin{equation}
Q = n_{+} - n_{-}\label{eq:top_q}
\end{equation}
Therefore, we estimate the number of instantons and anti-instantons $N_{I}$ from the topological charges $Q$ as follows~\cite{DiGH3}:
\begin{equation}
N_{I} = \langle Q^{2}\rangle
\end{equation}
The value of the brackets $\langle \mathcal{O} \rangle$ indicates that the average value provided by the sum of the numerical results is divided by the number of configurations. The number density of the instantons and anti-instantons $\rho_{I}$ is calculated as follows:
\begin{equation}
  \rho_{I} = \frac{N_{I}}{V_{\text{phys}}} \ [\text{GeV}^{4}]
\end{equation}
This formula indicates that the number density of the instantons and anti-instantons equals the topological susceptibility.

The phenomenological model~\cite{Shuryak2} predicts that the instanton (or anti-instanton) density is 8$\times10^{-4}$ [GeV$^{4}$]; accordingly, the number density of the instantons and anti-instantons of the standard configuration is
\begin{equation}
\rho_{I}^{\text{sta}} = 1.6\times10^{-3} \ [\text{GeV}^{4}]\label{eq:ins_dens},
\end{equation}
and the number of instantons and anti-instantons of the standard configuration in the physical volume $V_{\text{phys}}$ is
\begin{equation}
N_{I}^{\text{sta}} = \rho_{I}^{\text{sta}}V_{\text{phys}}\label{eq:num_ins}.
\end{equation}

Similar to the computations of the number density of the monopoles and anti-monopoles, we estimate the number density of the instantons and anti-instantons of the standard configurations. The calculated results and the simulation parameters are listed in Table 4 in~\cite{Hasegawa5}.
\begin{figure*}[htbp]
  \begin{center}
    \includegraphics[width=157mm]{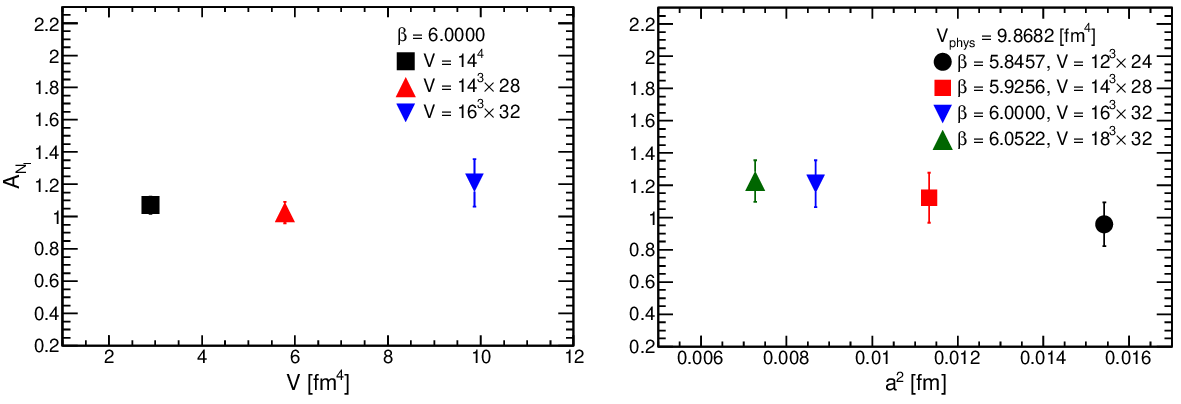}
  \end{center}
  \setlength\abovecaptionskip{-1pt}
  \caption{Examining the effects of the finite lattice volume (left) and the discretization (right) on the fitting results of the slope $A_{N_{I}}$. The outcomes of $\beta$ = 6.0000, $V = 14^{4}$ and $\beta$ = 6.0522, $V = 18^{3}\times32$ are reported in our previous research~\cite{DiGH3,Hasegawa2}}\label{fig:instanton_fitting_results}
\end{figure*}

First, to confirm the effects of the finite lattice volume on the calculated results, we set the value of the parameter $\beta$ to 5.8124, 5.8457, 5.9256, and 6.0000 and vary the lattice volumes as shown in the left panel of Fig.~\ref{fig:ins_dens_or}. This shows that the numerical results are consistent with the prediction~(\ref{eq:ins_dens}) and that the effects of the finite lattice volume are negligible. 

Second, to confirm the discretization effects, we set the values of the parameter $\beta$ and the lattice volumes $V$ so that the physical volumes are $V_{\text{phys}}$ = 3.1224, 4.9341, 7.9034, and 9.8682 [fm$^{4}$] and vary the values of the lattice spacing $a$ and the lattice volumes. The right panel of Fig.~\ref{fig:ins_dens_or} shows that the discretization effects are small and that the calculated results are reasonably consistent with the prediction~(\ref{eq:ins_dens}).

These results show that we can correctly estimate the number density of the instantons and anti-instantons, and the effects of the finite lattice volume and the discretization on the number density of the instantons and anti-instantons are small enough and negligible.
%%%%%%%%%%%%%%%%%%%%%%%%%%%%%%%%%%%%%%%%%%%%%%%%%%%%%%%%
%%%%%%%%%%%%%%%%%%%%%%%%%%%%%%%%%%%%%%%%%%%%%%%%%%%%%%%%

\subsection{The creation of instantons and anti-instantons}\label{sec:creation_ins}
\begin{figure*}[htbp]
  \begin{center}
    \includegraphics[width=162mm]{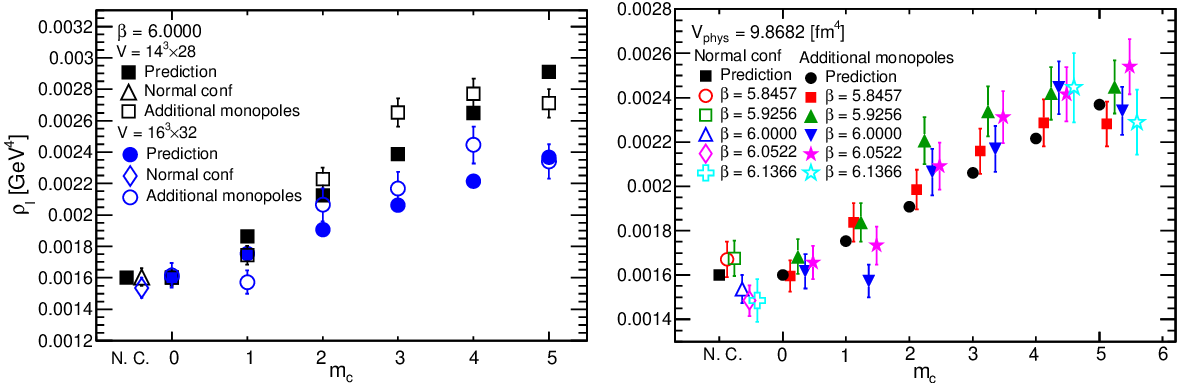}
  \end{center}
  \setlength\abovecaptionskip{-1pt}
  \caption{Comparing the calculated results of the number density of the instantons and anti-instantons $\rho_{I}$ using the configurations of the additional monopoles and anti-monopoles with the predictions. The left panel shows the results of the lattice volumes $V$ = $14^{3}\times28$ and $16^{3}\times32$, $\beta$ = 6.0000. The right panel shows the results of $V_{\text{phys}}$ = 9.8682 [fm$^{4}$]. The symbols in the right panel are slightly shifted along the horizontal axis to show the difference clearly.}\label{fig:ins_dens_add}
\end{figure*}

Our previous studies quantitatively demonstrate that a pair of additional monopole and anti-monopole of the magnetic charges $m_{c}$ = 1 makes one instanton or anti-instanton by fitting a linear curve. To confirm the influences of the finite lattice volume and discretization on this result, we quantitatively evaluate the increases in the number of instantons and anti-instantons $N_{I}$ to the increases in the number of magnetic charges $m_{c}$ from the results obtained by fitting the following linear curve:
\begin{equation}
  N_{I} = A_{N_{I}}m_{c} + B_{N_{I}}.
\end{equation}
The fitting results are listed in Table~\ref{tb:num_instanton_fitting_slopes}.
\begin{table}[htbp]
  \begin{footnotesize}
  \begin{center}
  \caption{The fitting results $A_{N_{I}}$, $B_{N_{I}}$, and $\chi^{2}/$dof obtained by fitting the curve $N_{I} = A_{N_{I}}m_{c} + B_{N_{I}}$ to the results of $N_{I}$.}\label{tb:num_instanton_fitting_slopes}
  \begin{tabular}{|c|c|c|c|c|c|}\hline
    $\beta$ & $V$ & $A_{N_{I}}$ & $B_{N_{I}}$ & FR:& $\frac{\chi^{2}}{\text{dof}}$\\
    &&&&$m_{c}$&\\ \hline
    5.8457 & $12^{3}\times24$ & 0.96(14) & 10.7(4) & 0-5 & 2.7/4.0 \\\hline
    5.9256 & $14^{3}\times28$ & 1.12(16) & 11.2(4) & 0-5 & 4.5/4.0 \\\hline
    6.0000 & $14^{3}\times28$ & 1.03(7) & 6.09(16) & 0-5 & 21.0/4.0 \\\cline{2-6}
    & $16^{3}\times32$ & 1.21(15) & 10.1(4) & 0-5 & 11.1/4.0 \\\hline
  \end{tabular}
  \end{center}
    \end{footnotesize}
\end{table}

The prediction of the slope $A_{N_{I}}^{\text{Pre}}$ is one because a pair of additional monopole and anti-monopole of the magnetic charges $m_{c}$ = 1 makes one instanton or anti-instanton. The predictions of the intercept $B_{N_{I}}^{\text{Pre}}$ are consistent with the numbers of instantons and anti-instantons of the normal configurations $N_{I}^{\text{sta}}$ = 6.1044 for $V_{\text{phys}}$ = 5.7845 [fm$^{4}$] ($V$ = $14^{3}\times28$, $\beta$ = 6.0000) and $N_{I}^{\text{sta}}$ = 10.414 for $V_{\text{phys}}$ = 9.8682 [fm$^{4}$] (apart from $V$ = $14^{3}\times28$, $\beta$ = 6.0000). These predictions are calculated from the function~(\ref{eq:num_ins}).

We compare the fitting results of the slope $A_{N_{I}}$ together with our previous outcomes of lattices $\beta$ = 6.0000, $V = 14^{4}$ and $\beta$ = 6.0522, $V = 18^{3}\times32$, as shown in Fig.~\ref{fig:instanton_fitting_results}. Figure~\ref{fig:instanton_fitting_results} and the fitting results in Table~\ref{tb:num_instanton_fitting_slopes} indicate that the slope $A_{N_{I}}$ and intercept $B_{N_{I}}$ are consistent with our previous outcomes and predictions $A_{N_{I}}^{\text{Pre}}$ and $B_{N_{I}}^{\text{Pre}}$, respectively.

Furthermore, Figure~\ref{fig:instanton_fitting_results} shows that the influences of the finite lattice volume and discretization on the increases in the numbers of instantons and anti-instantons are negligible and that one pair of additional monopole and anti-monopole of the magnetic charges $m_{c}$ = 1 makes one instanton or anti-instanton.

Therefore, we provide a prediction concerning the number of instantons and anti-instantons being in the configurations of the additional monopoles and anti-monopoles with the magnetic charges as follows:
\begin{equation}
N_{I}^{\text{Pre}}(m_{c}) = N_{I}^{\text{sta}} + m_{c}\label{eq:num_ins_add}
\end{equation}
Similarly, the prediction of the number density of the instantons and anti-instantons in the configurations of the additional monopoles and anti-monopoles with the magnetic charges is
\begin{equation}
\rho_{I}^{\text{Pre}}(m_{c}) = \left(\rho_{I}^{\text{sta}} + \frac{m_{c}}{V_{\text{phys}}}\right) \ [\text{GeV}^{4}]\label{eq:num_ins_dens_add}.
\end{equation}

Next, we investigate the effects of the finite lattice volume and discretization on the calculated results of the number density of the instantons and anti-instantons of the additional monopoles and anti-monopoles and compare the numerical results with the predictions.

Figure~\ref{fig:ins_dens_add} demonstrates that the number density of the instantons and anti-instantons increases with increasing the magnetic charges $m_{c}$, which shows that the numerical results are reasonably consistent with the predictions. Thus, the effects of the finite lattice volume and discretization on the calculated results are reasonably small.

The calculated results of the number of observed zero modes $N_{Z}$ = $|Q|$, number of instantons and anti-instantons $N_{I}$, number density of instantons and anti-instantons $\rho_{I}$, and number of configurations $N_{\text{conf}}$ together with the predictions are presented in Table~\ref{tb:Nzero_add_1} in~\ref{sec:nz_ni_niv}.
\begin{table}[htbp]
  \caption{The fitting results of the number density $\rho_{I}$ of the instantons and anti-instantons by interpolation.}\label{tb:inter_inst_density_add}
  \begin{center}
      \begin{footnotesize}
      \begin{tabular}{|c|c|c|c|c|} \hline
        $m_{c}$&$A_{\rho_{I}}$&$B_{\rho_{I}}$&FR: $a^{2}$& $\frac{\chi^{2}}{\text{dof}}$ \\
        &[GeV$^{4}\cdot\text{fm}^{-2}$]&[GeV$^{4}$]&[fm$^{2}$]& \\
        & $\times10^{-2}$ & $\times10^{-3}$ & $\times10^{-2}$ & \\\hline   
            N. C. & 2.2(1.0) &  1.35(11) & 5.5-15.5 & 1.1/4.0  \\ \hline     
            0 & -0.5(1.2) & 1.69(13) &7.2-15.5  & 0.6/2.0 \\ \hline        
            1 & 2.5(1.4) & 1.47(15) &7.2-15.5  & 4.2/2.0 \\ \hline        
            2 & -1.2(1.5) & 2.21(18) &7.2-15.5  & 2.0/2.0 \\ \hline       
            3 & -1.1(1.7) & 2.4(2)  &7.2-15.5  & 1.8/2.0 \\ \hline     
            4 & -0.3(3.1) & 2.5(3)  &5.5-11.4  & 0.0/2.0 \\ \hline     
            5 & 1.3(3.1)  & 2.3(3)  &5.5-11.4  &  2.1/2.0 \\ \hline   
      \end{tabular}
        \end{footnotesize}
  \end{center}
\end{table}
\begin{table}[htbp]
  \caption{The interpolated results of the number density of the instantons and anti-instantons $\rho_{I}$, square root of the number density $\rho_{I}^{\frac{1}{2}}$, and one-fourth root of the number density $\rho_{I}^{\frac{1}{4}}$. The fitting results of $\chi^{2}/$dof are for $\rho_{I}$.}\label{tb:final_Nzero_add_1}
  \begin{center}
      \begin{tabular}{|c|c|c|c|c|c|} \hline
               & $\rho_{I}^{\text{int}}$  & $\left(\rho_{I}^{\frac{1}{2}}\right)^{\text{int}}$& $\left(\rho_{I}^{\frac{1}{4}}\right)^{\text{int}}$  &FR: $a^{2}$ &  \\
 $m_{c}$   &  [GeV$^{4}$]  & [GeV$^{2}$] & [MeV] & [fm$^{2}$] & $\frac{\chi^{2}}{\text{dof}}$ \\
        &  $\times10^{-3}$ & $\times10^{-2}$ & $\times10^{2}$ & $\times10^{-3}$ &  \\\hline 
        N. C. & 1.57(3) & 3.97(4) & 1.993(10) & 5.5-15.5 & 6.1/5.0  \\ \hline 
        0 & 1.64(4) & 4.04(5) & 2.011(12) & 7.2-15.5 & 0.8/3.0  \\ \hline  
        1 & 1.73(4) & 4.17(5) & 2.042(12) & 7.2-15.5 & 7.5/3.0  \\ \hline  
        2 & 2.08(5) & 4.56(5) & 2.136(13) & 7.2-15.5 & 2.6/3.0  \\ \hline 
        3 & 2.24(5) & 4.73(6) & 2.175(13) & 7.2-15.5 & 2.2/3.0  \\ \hline 
        4 & 2.43(6) & 4.93(6) & 2.220(14) & 5.5-11.4 & 0.0/3.0  \\ \hline 
        5 & 2.41(7) & 4.91(6) & 2.216(14) & 5.5-11.4 & 2.3/3.0  \\ \hline
      \end{tabular}
  \end{center}
\end{table}

Finally, we interpolate the number density of the instantons and anti-instantons $\rho_{I}$ at the continuum limit for each magnetic charge $m_{c}$ by fitting the following curve to the calculated results:
\begin{equation}
  \rho_{I} = A_{\rho_{I}}x + B_{\rho_{I}}, \ x = a^{2} \ [\text{fm}^{2}].
\end{equation}
We do not include the numerical results obtained using the lattice $V = 12^{3}\times24$ of $\beta$ = 5.8457 of the magnetic charges 4 and 5 in the fitting ranges. The fitting results are presented in Table~\ref{tb:inter_inst_density_add}, which indicates that the errors of the slope $A_{\rho_{I}}$ of the normal configuration and configuration of the magnetic charge $m_{c} = 1$ are larger than 45$\%$. The fitting results of the slope apart from those results are zero because the errors are larger than the slope values. Moreover, we have shown that the influences of the finite lattice volume and discretization on the increases in the number of instantons and anti-instantons and their number density are small, as shown in Figs.~\ref{fig:instanton_fitting_results} and~\ref{fig:ins_dens_add}. Therefore, we fit a constant function to the calculated results of the number density $\rho_{I}$, the square root of the number density $\rho_{I}^{\frac{1}{2}}$, and the one-fourth root of the number density $\rho_{I}^{\frac{1}{4}}$. The fitting results are given in Table~\ref{tb:final_Nzero_add_1}.
\begin{figure*}[htbp]
  \begin{center}
    \includegraphics[width=133.5mm]{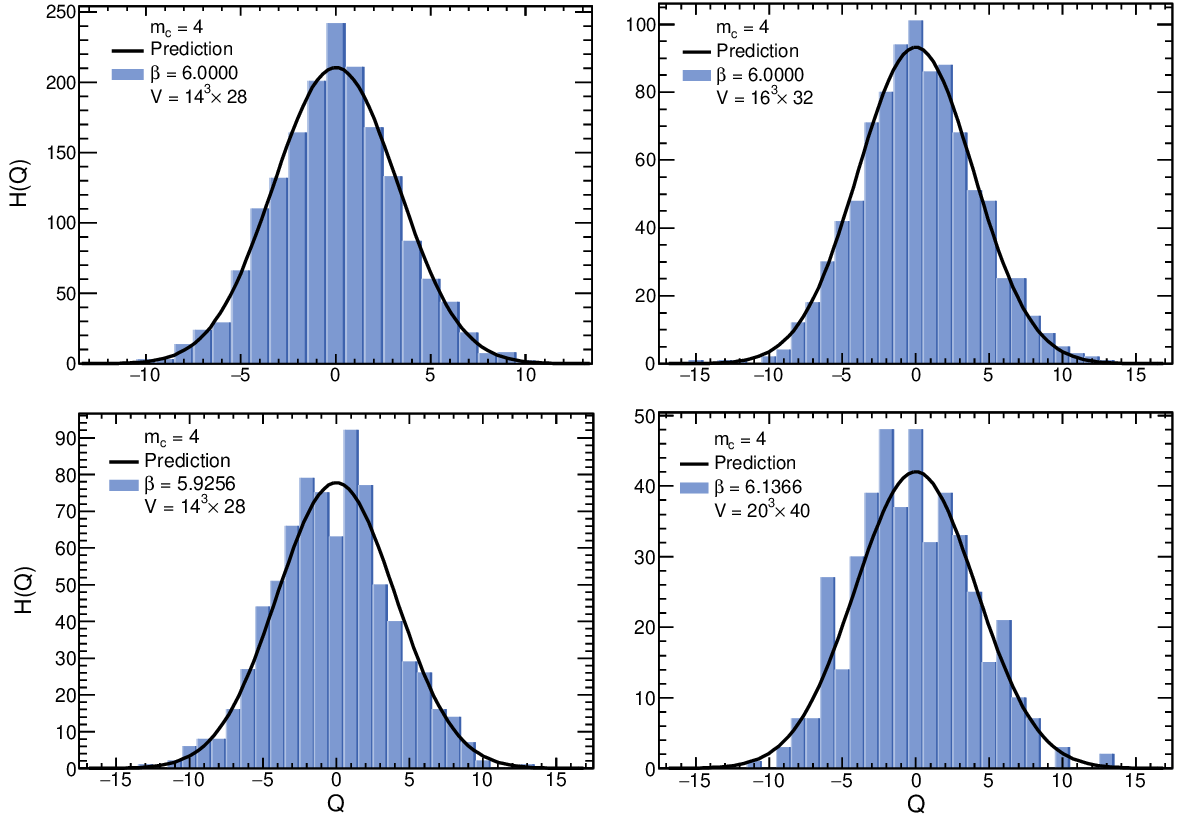}
  \end{center}
  \setlength\abovecaptionskip{-1pt}
  \caption{The histograms H($Q$) of the topological charges $Q$. The magnetic charges are $m_{c} =4$. The lattices are $\beta$ = 6.0000, $V$ = 14$^{3}\times$28 (upper left), $\beta$ = 5.9256, $V$ = 14$^{3}\times$28 (lower left), $\beta$ = 6.0000, $V$ = 16$^{3}\times$32 (upper right), and $\beta$ = 6.1366, $V$ = 20$^{3}\times$40 (lower right). The black lines indicate the fitting results according to the prediction function of the topological charges.}\label{fig:topological_q}
\end{figure*}
\begin{table}[htbp]
  \begin{center}
    \caption{The fitting results of the distributions of the topological charges by the prediction functions. The fitting range FR is -17 $\le Q \le$ 17. N. C. stands for the normal configuration.}\label{tb:dis_top_q}
    \begin{scriptsize}
  \begin{tabular}{|c|c|c|c|c|c|}\hline
    $\beta$ & $V$  & $m_{c}$& $\langle\delta^{2}\rangle$ & $\mathcal{O}(V^{-1})$ & $\frac{\chi^{2}}{\text{dof}}$ \\
    & & & & $\times10^{-2}$ &  \\\hline
   5.8457 & $12^{3}\times24$ &  N. C. & 10.3(5) & -2(3) & 24.4/20.0  \\\cline{3-6}    
    &                 & 0 & 10.0(5) & -2(3) & 21.8/20.0  \\\cline{3-6}
    &                 & 1 & 10.8(6) & -0.9(3) & 12.7/21.0  \\\cline{3-6}
    &                 & 2 & 10.3(6) & -2(3) & 19.0/23.0  \\\cline{3-6}
    &                 & 3 & 10.2(7) & -3(3) & 31.9/23.0  \\\cline{3-6}
    &                 & 4 & 10.3(8) & -2(3) & 21.7/22.0  \\\cline{3-6}
    &                 & 5 & 9.8(7) & -2(3) & 18.4/22.0  \\\hline
    5.9256 & $14^{3}\times28$ &  N. C. & 10.9(6) & -1(3) & 10.9/19.0  \\\cline{3-6}
    &                 & 0 & 10.8(6) & -2(3) & 15.1/19.0  \\\cline{3-6}
    &                 & 1 & 10.1(5) & -2(3) & 21.0/20.0  \\\cline{3-6}
    &                 & 2 & 11.9(8) & -3(3) & 25.3/21.0  \\\cline{3-6}
    &                 & 3 & 11.8(8) & -2(3) & 16.5/22.0  \\\cline{3-6}
    &                 & 4 & 12.0(9) & -2(4) & 21.1/22.0  \\\cline{3-6}
    &                 & 5 & 9.6(7) & -3(3) & 28.6/25.0  \\\hline
    6.0000 & $14^{3}\times28$ & N. C. & 5.9(2) & -1(2) & 18.9/16.0  \\\cline{3-6}
    &                 & 0 & 6.1(2) & -0.6(2.4) & 12.4/16.0  \\\cline{3-6}
    &                 & 1 & 5.5(2) & -0.8(2.4) & 14.1/16.0  \\\cline{3-6}
    &                 & 2 & 6.2(3) & -0.9(2.4) & 16.6/19.0  \\\cline{3-6}
    &                 & 3 & 7.1(4) & -0.5(2.4) & 8.7/20.0  \\\cline{3-6}
    &                 & 4 & 6.3(4) & -1(2) & 20.3/21.0  \\\cline{3-6}
    &                 & 5 & 5.3(4) & -0.6(2.4) & 10.9/20.0 \\ \cline{2-6} 
    & $16^{3}\times32$ & N. C. & 10.1(5) & -0.6(3.1) & 8.1/18.0 \\ \cline{3-6}
    &                 & 0 & 10.2(5) & -2(3) & 17.0/21.0 \\\cline{3-6}
    &                 & 1 & 8.9(5)   & -2(3) & 18.6/18.0 \\\cline{3-6}
    &                 & 2 & 10.6(7) & -2(3) & 20.1/22.0  \\\cline{3-6}
    &                 & 3 & 11.2(8) & -2(3) & 21.8/21.0  \\\cline{3-6}
    &                 & 4 & 11.3(8) & -1(3) & 13.6/25.0  \\\cline{3-6}
    &                 & 5 & 10.4(8) & -2(3) & 20.0/22.0  \\\hline
    6.1366 & $20^{3}\times40$ &  N. C. & 8.4(6) & -6(5) & 26.4/17.0 \\\cline{3-6}
    &                 & 4 & 12.8(1.5) &  -3(5) & 23.2/19.0  \\\cline{3-6}
    &                 & 5 & 9.0(1.1) & -4(5) & 18.3/20.0 \\\hline
  \end{tabular}
  \end{scriptsize}
 \end{center}
\end{table}

The interpolated result of the number density of the normal configuration is $\rho_{I}^{\text{int}}$ = 1.57(3)$\times10^{-3}$ [GeV$^{4}$], and the fitting result of $\chi^{2}/$dof is 1.2. This interpolated result is consistent with the outcome of the phenomenological model~(\ref{eq:ins_dens}). This indicates that we can adequately interpolate the numerical results of the number density of the instantons and anti-instantons to the continuum limit by fitting the constant function. Therefore, we use the interpolated results in Table~\ref{tb:final_Nzero_add_1} in the sections below. Hereafter, we do not include the numerical results obtained using the lattice $V = 12^{3}\times24$ of $\beta$ = 5.8457 of the additional monopoles and anti-monopoles with the magnetic charges 4 and 5 in the fitting ranges for the interpolations to the continuum limit because of the reason we mentioned in subsection~\ref{sec:monopole_ins}.%16 March 2022
%%%%%%%%%%%%%%%%%%%%%%%%%%%%%%%%%%%%%%%%%%%%%%%%%%%%%%%%
%%%%%%%%%%%%%%%%%%%%%%%%%%%%%%%%%%%%%%%%%%%%%%%%%%%%%%%

\subsection{Distributions of the topological charge $Q$ of additional monopoles and anti-monopoles}\label{sec:topolgical_charges}

We add monopoles and anti-monopoles to the configurations by the monopole creation operator, and then the added monopoles and anti-monopoles create the zero modes of the positive chirality and negative chirality with the same probability because we assume the CP invariance. We cannot directly count the number of zero modes; therefore, when we add the monopoles and anti-monopoles to the vacuum, the difference in the topological charges between the standard configurations and the configurations with the added monopoles and anti-monopoles arises.
\begin{figure*}[htbp]
  \begin{center}
    \includegraphics[width=168mm]{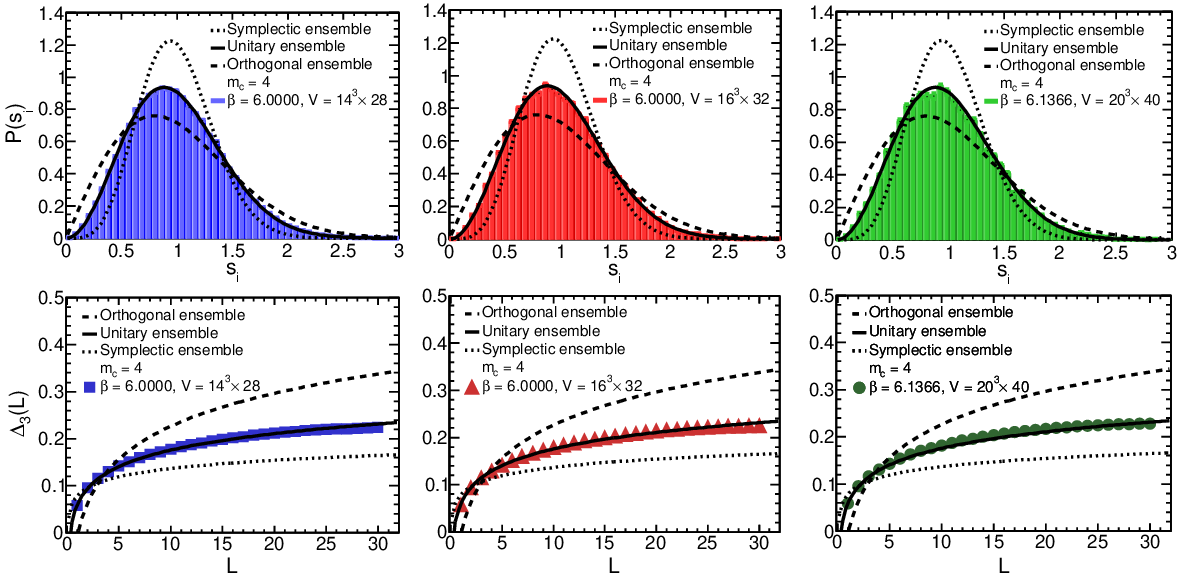}
  \end{center}
  \setlength\abovecaptionskip{-1pt}
  \caption{The distributions of the nearest neighbor spacing $P(s_{i})$ (upper figures) and the spectral rigidity $\Delta_{3}(L)$ (lower figures) compared with the GRMT. The magnetic charges are $m_{c} =4$. The lattices are $\beta$ = 5.8457, $V$ = 12$^{3}\times$24 (upper and lower left), $\beta$ = 6.0000, $V$ = 14$^{3}\times$28 (upper and lower middle), and $\beta$ = 6.0000, $V$ = 16$^{3}\times$32 (upper and lower right). The black dotted, full, and dashed lines represent the prediction functions of the GSE, GUE, and GOE of the GRMT, respectively. The colored lines and symbols indicate the numerical results.}\label{fig:ps_delta}
\end{figure*}

To confirm whether the added monopoles and anti-monopoles affect the vacuum structure, we predict the distribution functions of the topological charges for each magnetic charge and compare them with the histograms of the topological charges $H(Q)$, as shown in Fig.~\ref{fig:topological_q}. The distribution functions for each magnetic charge which comprise the Gaussian distributions and probabilities of the zero-mode creations are given in reference~\cite{DiGH3,Hasegawa2}. The fitting results of the normal configurations and the configurations with the added monopoles and anti-monopoles are listed in Table~\ref{tb:dis_top_q}.

If the prediction functions are consistent with the distributions of the topological charges that are numerically calculated, the fitting results $\langle\delta^{2}\rangle$ are consistent with the analytical results $N_{I}^{\text{sta}} =$ 6.1044 for $V_{\text{phys}}$ = 5.7845 [fm$^{4}$] and $N_{I}^{\text{sta}} =$ 10.414 for $V_{\text{phys}}$ = 9.8682 [fm$^{4}$], as presented in Table~\ref{tb:Nzero_add_1} in~\ref{sec:nz_ni_niv}. Moreover, the fitting results of $\mathcal{O}(V^{-1})$ that are the correction terms of the distribution functions are zero. 

The fitting results $\langle\delta^{2}\rangle$ in Table~\ref{tb:dis_top_q} are consistent with the predictions $N_{I}^{\text{sta}}$ and outcomes $N_{I}$ of the standard configurations in Table~\ref{tb:Nzero_add_1} in~\ref{sec:nz_ni_niv}. The fitting results of the correction terms $\mathcal{O}(V^{-1})$ are zero, and $\chi^{2}/$dof are small enough. Therefore, these results indicate that the effects of the finite lattice volume and discretization on detecting topological charges are negligible, and the added monopoles and anti-monopoles create instantons and anti-instantons without affecting the vacuum structure.
%%%%%%%%%%%%%%%%%%%%%%%%%%%%%%%%%%%%%%%%
%%%%%%%%%%%%%%%%%%%%%%%%%%%%%%%%%%%%%%%%
%%%%%%%%%%%%%%%%%%%%%%%%%%%%%%%%%%%%%%%%
%%% END SEC 3
%%%%%%%%%%%%%%%%%%%%%%%%%%%%%%%%%%%%%%%%

\section{Comparisons of the Dirac spectrum with random matrix theory}\label{sec:4}

We add the monopoles and anti-monopoles by applying the monopole creation operator to the vacuum. Therefore, to inspect whether the added monopoles and ant-monopoles affect the low-lying eigenvalues of the overlap Dirac operator, we compare the low-lying eigenvalues with the predictions of RMT.
\begin{figure*}[htbp]
  \begin{center}
    \includegraphics[width=133mm]{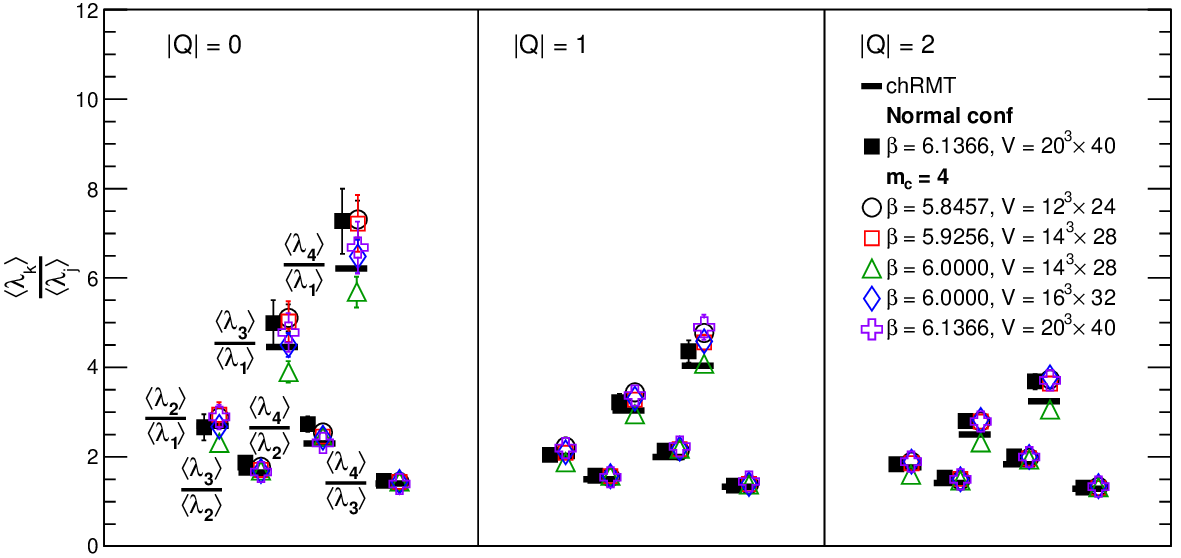}
  \end{center}
  \setlength\abovecaptionskip{-1pt}
  \caption{Comparing the ratios of the low-lying eigenvalues $\frac{\langle\bar{\lambda}_{k}^{|Q|}\rangle}{\langle\bar{\lambda}_{j}^{|Q|}\rangle}$ in each topological charge sector $|Q| =$ 0, 1, and 2 with the prediction of the chRMT and examining the effects of the finite lattice volume and the discretization. The black and colored symbols indicate the numerical results of the normal configuration and magnetic charge $m_{c}$ = 4. The black horizontal lines indicate the predictions of the chRMT.}\label{fig:rmt_test}
\end{figure*}
\begin{table*}[tbp]
  \centering
  \caption{Comparing the numerical results of $\frac{\langle\bar{\lambda}_{k}^{|Q|}\rangle}{\langle\bar{\lambda}_{j}^{|Q|}\rangle}$ with the prediction of the chRMT. The normal configurations and configurations of the magnetic charges $m_{c} = 4$ are used. The prediction of the chRMT is presented in Table 3~\cite{Giusti4}.}\label{tb:rmt_test_tb}
 \begin{scriptsize}
    \begin{tabular}{|c|c|c|c|c|c|c|c|c|c|c|c|c|} \hline
       &  & \multicolumn{2}{c|}{$\beta=5.8457$} & \multicolumn{2}{c|}{$\beta=5.9256$} & \multicolumn{2}{c|}{$\beta=6.0000$} & \multicolumn{2}{c|}{$\beta=6.0000$} &  \multicolumn{2}{c|}{$\beta=6.1366$} & \\ 
        $|Q|$ & $k/j$ & \multicolumn{2}{c|}{$V=12^{3}\times24$} & \multicolumn{2}{c|}{$V=14^{3}\times28$} & \multicolumn{2}{c|}{$V=14^{3}\times28$} & \multicolumn{2}{c|}{$V=16^{3}\times32$} & \multicolumn{2}{c|}{$V=20^{3}\times40$} & RMT \\ \cline{3-12}  
       &  & N. C. & $m_{c}=4$ & N. C. & $m_{c}=4$ & N. C. & $m_{c}=4$ & N. C. & $m_{c}=4$ & N. C. & $m_{c}=4$ & \\  \hline
                0 &2/1 &  2.90(18) &  2.88(19)   & 2.8(2)  &  2.9(3)   & 2.32(14)  & 2.31(15) & 3.1(2)   & 2.67(17)     & 2.7(3) &      2.9(3)        & 2.70\\\cline{2-13}  
                  &3/1 &  5.0(3)   &  5.1(3)     & 5.0(4)  &  5.0(4)   & 4.1(2)    & 3.9 (2)  & 5.3(3)   & 4.5(3)       & 5.0(5) &      4.8(4)                 & 4.46\\\cline{2-13}  
                  &4/1 &  7.3(4)   &  7.3(4)     & 7.1(5)  &  7.2(6)   & 6.0(3)    & 5.7 (3)  & 7.9(5)   & 6.5(4)       & 7.3(7) &      6.7(6)                 & 6.22\\\cline{2-13}  
                  &3/2 &  1.74(6)  &  1.77(8)    & 1.78(9) &  1.71(9)  & 1.79(12)  & 1.70 (10)  & 1.70(7)  & 1.68(7)    & 1.87(13) &    1.66(10)       & 1.65\\\cline{2-13}  
                  &4/2 &  2.53(8)  &  2.54(11)   & 2.56(12)&  2.45(12) & 2.57(16)  & 2.46(17) & 2.53(9)  & 2.43(9)      & 2.73(17) &    2.31(13)    & 2.30\\\cline{2-13}  
                  &4/3 &  1.46(4)  &  1.43(5)    & 1.44(6) &  1.44(5)  & 1.44(9)   & 1.45(10) & 1.49(5)  & 1.44(4)      & 1.46(8) &     1.40(7)       & 1.40\\\hline  
                1 &2/1 &  2.20(7)  &  2.23(8)    & 2.16(8) &  2.09(9)  & 1.95(7)   & 1.88(8)  & 2.18(8)  & 2.10(8)      & 2.04(12) &    2.19(14)        & 2.02\\\cline{2-13}  
                  &3/1 &  3.44(11) &  3.44(11)   & 3.46(13)&  3.28(13) & 2.99(11)  & 2.95(12) & 3.35(12) & 3.30(12)   & 3.22(19) &    3.4(2)  & 3.03\\\cline{2-13}  
                  &4/1 &  4.74(14) &  4.76(15)   & 4.81(17)&  4.56(18) & 4.00(14)  & 4.09(16) & 4.72(16) & 4.59(16)   & 4.4(2) &      4.9(3)   & 4.04\\\cline{2-13}  
                  &3/2 &  1.56(4)  &  1.55(4)    & 1.60(5) &  1.56(5)  & 1.53(6)   & 1.57(7)  & 1.54(4)  & 1.57(4)      & 1.58(6) &     1.54(7)         & 1.50\\\cline{2-13}  
                  &4/2 &  2.16(5)  &  2.14(5)    & 2.23(6) &  2.18(6)  & 2.05(8)   & 2.18(10) & 2.17(5)  & 2.18(6)      & 2.13(8) &     2.24(9)        & 2.00\\\cline{2-13}  
                  &4/3 &  1.38(2)  &  1.38(3)    & 1.39(3) &  1.39(3)  & 1.34(5)   & 1.39(6)  & 1.41(3)  & 1.39(3)      & 1.36(4) &     1.50(5)         & 1.33\\\hline
                2 &2/1 &  1.90(6)  &  1.89(6)    & 1.88(7) &  1.86(7)  & 1.75(6)   & 1.58(7)  & 1.90(7)  & 1.89(7)      & 1.84(9) &     1.87(10)       & 1.76\\\cline{2-13}  
                  &3/1 &  2.81(8)  &  2.79(8)    & 2.85(10)&  2.79(10) & 2.59(8)   & 2.32(11) & 2.83(10) & 2.82(10)     & 2.80(13) &    2.79(13)    & 2.50\\\cline{2-13}  
                  &4/1 &  3.76(11) &  3.74(11)   & 3.80(13)&  3.64(12) & 3.39(11)  & 3.06(14) & 3.77(12) & 3.78(13)   &  3.68(17) &    3.72(17)  & 3.24\\\cline{2-13}  
                  &3/2 &  1.48(4)  &  1.48(4)    & 1.52(4) &  1.51(4)  & 1.48(5)   & 1.47(8)  & 1.49(4)  & 1.49(4)      & 1.53(6) &     1.49(6)         & 1.42\\\cline{2-13}  
                  &4/2 &  1.98(5)  &  1.98(5)    & 2.03(5) &  1.96(5)  & 1.93(6)   & 1.94(11) & 1.99(5)  & 2.00(5)      & 2.01(7) &     1.99(8)        & 1.83\\\cline{2-13}  
                  &4/3 &  1.34(3)  &  1.34(3)    & 1.33(3) &  1.30(3)  & 1.31(4)   & 1.32(7)  & 1.33(3)  & 1.34(3)      & 1.31(4) &     1.33(5)         & 1.29\\\hline
              \end{tabular}
     \end{scriptsize}
\end{table*}

The Gaussian random matrix theory (GRMT) universally predicts the distributions of the short-range fluctuations and the mean-square deviations of the staircase function from the linear function. The predictions of the GRMT are classified into the following three Gaussian ensemble classes obeying the symmetries~\cite{Wigner1,Dyson2}: (i) Gaussian symplectic ensemble. (ii) Gaussian unitary ensemble. (iii) Gaussian orthogonal ensemble. 

We first improve the eigenvalues $\lambda_{k}$ of the massless overlap Dirac operator to approach the continuum limit~\cite{Capitani}. The improved eigenvalues are the pure imaginary numbers and come in positive and negative pairs. The subscript $k$ indicates the $k$th eigenvalue. We put the improved nonzero and positive eigenvalues in ascending order as follows: $a\bar{\lambda}_{1} < \cdots < a\bar{\lambda}_{k} < \cdots < a\bar{\lambda}_{n}$. 

We use the unfolding procedure techniques as follows~\cite{Guhr1}: (1) We plot the improved eigenvalues $a\bar{\lambda}_{k}$ on the horizontal axis against their numbers $k$ on the vertical axis for each configuration. (2) We fit the polynomial function of order d = 3 to the plot and obtain the smooth curve $N_{poly}$ using the fitting results of the coefficients for each configuration. (3) We obtain the following unfolded eigenvalues $E_{k}$ by substituting the improved eigenvalues $a\bar{\lambda}_{k}$ into the smooth curve $N_{poly}$ for each configuration:
\begin{equation}
  E_{k} = N_{poly}(a\bar{\lambda}_{k}).
\end{equation}

First, we calculate the nearest-neighbor spacing $s_{i}$ using the unfolded eigenvalues $E_{k}$ to inspect the influence of additional monopoles and anti-monopoles on short-range fluctuations as follows:
\begin{equation}
s_{i} = E_{i+1} - E_{i}.
\end{equation}
We make distributions of the nearest-neighbor spacing P($s_{i}$), which are normalized to unity, and show the comparisons with the predictions of the GRMT in the upper panels in Fig.~\ref{fig:ps_delta}. The distribution of the nearest-neighbor spacing for each Gaussian ensemble class is provided in reference~\cite{Guhr2}.

Next, we inspect the influence of additional monopoles and anti-monopoles on the fluctuations of the eigenvalues of the long interval of length $L$ on the unfolded scale by calculating the spectral rigidity $\Delta_{3} (L)$. The spectral rigidity $\Delta_{3} (L)$ is defined as the mean-square deviation of the staircase function from the linear function~\cite{Dyson2}. We use formula (I-39) of reference~\cite{Bohigas2} and almost all unfolded eigenvalues $E_{k}$ and evaluate the spectral rigidity by calculating the spectral and ensemble averages. The calculated results of the spectral rigidity compared with the predictions of three Gaussian ensemble classes are shown in the lower panels in Fig.~\ref{fig:ps_delta}. The predictions of the spectral rigidity are provided in reference~\cite{Guhr2}.

Figure~\ref{fig:ps_delta} shows that the numerical results of the configurations of $m_{c}$ = 4 of the different lattice volumes and values of the lattice spacing are consistent with the predictions of the Gaussian unitary ensemble. We have already shown that the outcomes of the different magnetic charges are compatible with the Gaussian unitary ensemble in our previous studies.

Therefore, additional monopoles and anti-monopoles, finite lattice volume, and discretization do not affect the fluctuations of the nearest-neighbor spacing and eigenvalues of the long interval of length $L$ on the unfolded scale.

Next, chiral random matrix theory (chRMT) universally predicts the distribution of low-lying eigenvalues of the Dirac operator for each eigenvalue number $k$~\cite{Nishigaki1,Damgaard1} of each topological charge sector $|Q|$. The following relation exists between the $k$th scaled eigenvalue of the chRMT $z_{k}^{|Q|}$ and the $k$th eigenvalue of the Dirac operator $a\lambda_{k}^{|Q|}$ of the topological charge sector $|Q|$:
\begin{equation}
  z_{k}^{|Q|} = \Sigma V \lambda_{k}^{|Q|}\label{eq:z_and_lam}.
\end{equation}
The parameter $\Sigma$ is the scale parameter of the eigenvalue distributions.

We compute the ratios of the eigenvalues $\frac{\langle z_{k}^{|Q|}\rangle}{\langle z_{j}^{|Q|}\rangle}$~\cite{Giusti4} using the improved eigenvalues $a\bar{\lambda}_{k}$ to remove the uncertainty coming from the scale parameter $\Sigma$ and compare the numerical results with the predictions of the chRMT for each topological charge sector as shown in Fig.~\ref{fig:rmt_test}. The numerical results and the predictions of the chRMT are listed in Table~\ref{tb:rmt_test_tb}. We have already shown that the ratios of the numerical results of the different magnetic charges using the lattice $V = 18^{3}\times32$, $\beta = 6.0522$ are consistent with the predictions of the chRMT in our previous study.

These results indicate that the numerical results are reasonably consistent with the predictions of the chRMT and that the finite lattice volume and discretization do not affect the ratios of the low-lying eigenvalues of the overlap Dirac operator.%15 march 2022
%%%%%%%%%%%%%%%%%%%%%%%%%%%%%%%%%%%%%%%%%%%%%%%%%%%%%%%%%
%%%%%%%%%%%%%%%%%%%%%%%%%%%%%%%%%%%%%%%%%%%%%%%%%%%%%%%%%
%%%%%%%%%%%%%%%%%%%%%%%%%%%%%%%%%%%%%%%%%%%%%%%%%%%%%%%%%
%%% END SEC4
%%%%%%%%%%%%%%%%%%%%%%%%%%%

\section{PCAC relation, renormalization constant $\hat{Z}_{S}$, and renormalized chiral condensate}\label{sec:5}

We calculate the correlations of the pseudoscalar and scalar densities using the eigenvalues and eigenvectors of the overlap Dirac operator and the renormalization constant for the scalar density. We then evaluate the renormalized chiral condensate in the chRMT using the analytic results of the scale parameter $\Sigma$ of the eigenvalue distributions of the overlap Dirac operator.
%%%%%%%%%%%%%%%%%%%%%%%%%%%%%%%%%%%%%%%%%%%%%%%%%%%%%%%%%
%%%%%%%%%%%%%%%%%%%%%%%%%%%%%%%%%%%%%%%%%%%%%%%%%%%%%%%%%

\subsection{PCAC relation}\label{sec:5_pcac}
  
This subsection briefly explains the calculations of the correlation functions. The computations of the correlation functions and the notation are the same as those in our previous study~\cite{Hasegawa2}.

We calculate the massive eigenvalues $\lambda_{k}^{\text{mass}}$ of the massive overlap Dirac operator using the eigenvalues $\lambda_{k}$ of the massless overlap Dirac operator $D$  as follows:
\begin{equation}
    a\lambda_{k}^{\text{mass}} = a\left(1 - \frac{\bar{m}_{ud}}{2\rho}\right)\lambda_{k} + a\bar{m}_{ud}
\end{equation}
In this study, we set the mass parameter $\rho$ for the massless overlap Dirac operator to $a\rho$ = 1.4. We define the input quark mass $\bar{m}_{ud}$ for the pion as follows:
\begin{equation}
   a\bar{m}_{ud} \equiv \frac{a(m_{u} + m_{d})}{2}\label{eq:qmass}
\end{equation}

The operators of the scalar density $\mathcal{O}_{S}$ and pseudoscalar density $\mathcal{O}_{P}$ are defined as follows:
\begin{align}
  &\mathcal{O}_{S} = a^{3}\bar{\psi}_{1}\left(1-\frac{D}{2\rho}\right)\psi_{2}\\
  &\mathcal{O}_{P} = a^{3}\bar{\psi}_{1}\gamma_{5}\left(1-\frac{D}{2\rho}\right)\psi_{2}
\end{align}
The quark propagator is defined as
\begin{equation}
  \mathcal{G}(\vec{y}, y^{0}; \vec{x}, x^{0})  \equiv a^{3}\sum_{k}\frac{\psi_{k}(\vec{x}, x^{0}) \psi_{k}^{\dagger}(\vec{y}, y^{0})}{\lambda_{k}^{\text{mass}}}\label{eq:qpro}.
\end{equation}
The correlation function of the scalar density is
\begin{equation}
  \mathcal{C}_{S}(\Delta t) = \frac{a^{3}}{V} \sum_{\vec{x}_{1}}\sum_{\vec{x}_{2}, t}\langle \mathcal{O}_{S}^{C}(\vec{x}_{2}, t) \mathcal{O}_{S}(\vec{x}_{1}, t + \Delta t)\rangle\label{eq:s_corre}.
\end{equation}
The correlation function of the pseudoscalar density is
\begin{equation}
 \mathcal{C}_{P}(\Delta t) = \frac{a^{3}}{V} \sum_{\vec{x}_{1}}\sum_{\vec{x}_{2}, t}\langle \mathcal{O}_{PS}^{C}(\vec{x}_{2}, t) \mathcal{O}_{PS}(\vec{x}_{1}, t + \Delta t)\rangle\label{eq:ps_corre}.
\end{equation}
The superscript $C$ stands for the Hermitian conjugate. We write the functions~(\ref{eq:qpro}),~(\ref{eq:s_corre}), and~(\ref{eq:ps_corre}) omitting the dimension $a$ of the lattice sites for convenience.

We want to precisely evaluate the effects of the monopoles and instantons on observables near the chiral limit; however, unphysical zero modes near the chiral limit affect the observables. Therefore, to remove undesirable influences caused by unphysical zero modes, we calculate the following subtracted correlation function~\cite{Blum1,Giusti3}:
\begin{equation}
 \mathcal{C}_{P-S} \equiv  \mathcal{C}_{P} - \mathcal{C}_{S}.\label{eq:sub_corre}
\end{equation}
Suppose that the following formula can approximate the subtracted correlation function~(\ref{eq:sub_corre}):
\begin{align}
  & \mathcal{C}_{P-S}(t)  =  \nonumber \\
  & \ \ \ \ \frac{a^{4}G_{P-S}}{am_{PS}} \exp\left( {-\frac{m_{PS}}{2}T}\right)\cosh\left[m_{PS} \left(\frac{T}{2} - t \right) \right].\label{eq:fiting_corre}
  \end{align}

We vary the input quark mass~(\ref{eq:qmass}) of the massive eigenvalues from 30 to 150 [MeV] and calculate the subtracted correlation function~(\ref{eq:sub_corre}). We fit formula~(\ref{eq:fiting_corre}) to the numerical results of the correlations, obtain the fitting results of the pseudoscalar mass $am_{PS}$ and coefficient $a^{4}G_{P-S}$, and calculate the observables using them in the later sections. The fitting ranges are determined so that the fitting results of $\chi^{2}/$dof are approximately 1.

We list the fitting results of $am_{PS}$, $a^{4}G_{P-S}$, and $\chi^{2}/$dof and the numerical results of $(am_{PS})^{2}$ and the decay constant of the pseudoscalar meson $aF_{PS}$ in the tables in~\cite{Hasegawa5} as follows: (i) The results of $V = 12^{3}\times24$ of $\beta = 5.8457$ are listed in Tables 8-10. (ii) The results of $V = 14^{3}\times28$ of $\beta = 5.9256$ are listed in Tables 11-14. (iii) The results of $V = 14^{3}\times28$ of $\beta = 6.0000$ are listed in Tables 15-18. (iv) The results of $V = 16^{3}\times32$ of $\beta = 6.0000$ are listed in Tables 19-22. (v) The results of $V = 20^{3}\times40$ of $\beta = 6.1366$ are listed in Tables 23-24. The fitting results and the numerical results of $V = 18^{3}\times32$ of $\beta = 6.0522$ are provided in reference~\cite{Hasegawa2}.

First, we confirm the effects of the additional monopoles and anti-monopoles on the partially conserved axial current (PCAC) relation between the numerical results of the square mass of the pseudoscalar meson $(am_{PS})^{2}$ and the input quark mass~(\ref{eq:qmass}). The PCAC relation is defined as follows:
\begin{equation}
  m_{PS}^{2} \equiv A\bar{m}_{ud}
\end{equation}
We fit the following linear function to the numerical results:
\begin{equation}
  (am_{PS})^{2} = a^{2}A_{\text{PCAC}}^{(1)}\bar{m}_{ud} + a^{2}B_{\text{PCAC}}.\label{eq:pcac_fit_func}
\end{equation}
The fitting ranges are determined so that the fitting results of $\chi^{2}/$dof are approximately 1. The fitting results of $a^{2}B_{\text{PCAC}}$ are nonzero; however, the values are small enough, as indicated in Table~\ref{tb:prop_v12xx3x24_b5p84572_fit1_2} in~\ref{sec:pcac_zs}.

The fitting results of the slope $aA_{\text{PCAC}}^{(1)}$ of each magnetic charge calculated using the same lattice are reasonably consistent, except that the fitting results of the lattice $V = 12^{3}\times24$ gradually decrease when the number of magnetic charges $m_{c}$ is larger than 3, as shown in the same table. The decrease comes by adding the monopoles and anti-monopoles with high magnetic charges to the small lattice volume, which we have already mentioned in subsection~\ref{sec:monopole_ins}. Moreover, we detect the influence of the finite lattice volume on the slope $aA_{\text{PCAC}}^{(1)}$ by comparing the fitting results of $V = 14^{3}\times28$ of $\beta$ = 6.0000 with $V = 16^{3}\times32$.

To reduce the errors of the renormalization constant $\hat{Z}_{S}$ for the scalar density, we remove the intercept from the fitting curve~(\ref{eq:pcac_fit_func}) and fit the following curve to the numerical results:
\begin{equation}
  (am_{PS})^{2} = a^{2}A_{\text{PCAC}}^{(2)}\bar{m}_{ud}.
\end{equation}
The fitting results of the slopes $aA_{\text{PCAC}}^{(2)}$ and parameter $\chi^{2}/$dof are shown in Table~\ref{tb:prop_v12xx3x24_b5p84572_fit1_2} in~\ref{sec:pcac_zs}. The fitting ranges are the same as fitting by the curve~(\ref{eq:pcac_fit_func}). In the sections below, we use the fitting results of the slope $A_{\text{PCAC}}^{(2)}$ for computing the renormalization constant $\hat{Z}_{S}$ for the scalar density and the average mass of the up and down quarks.%15 March 2022
%%%%%%%%%%%%%%%%%%%%%%%%%%%%%%%%%%%%%%%%%%%%%%%%
%%%%%%%%%%%%%%%%%%%%%%%%%%%%%%%%%%%%%%%%%%%%%%%%

\subsection{Renormalization constant $\hat{Z}_{S}$ for the scalar density}\label{sec:5_zs}

We obtain the renormalization constant $\hat{Z}_{S}$ for the scalar density by the nonperturbative calculations~\cite{Wennekers1,Hernandez2} using the fitting results of the slope $aA_{\text{PCAC}}^{(2)}$. We list the calculated results of $\hat{Z}_{S}$ in Table~\ref{tb:prop_v12xx3x24_b5p84572_fit1_2} in~\ref{sec:pcac_zs}. Comparing the renormalization constant $\hat{Z}_{S}$ computed using the lattice volume $V = 14^{3}\times28$ of $\beta = 6.0000$ with the result of the lattice volume $V = 16^{3}\times32$ of $\beta = 6.0000$ indicates that there is an influence of the finite lattice volume.
\begin{figure}[htbp]
  \begin{center}
    \includegraphics[width=80mm]{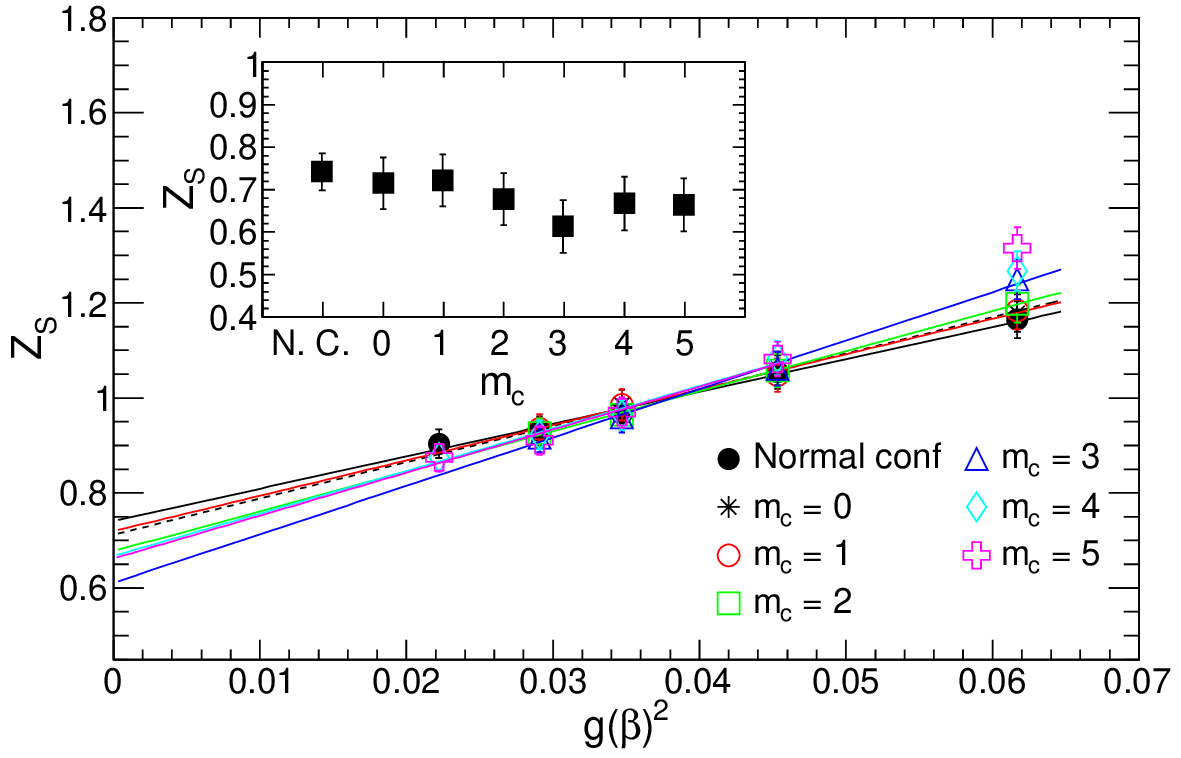}
  \end{center}
  \setlength\abovecaptionskip{-1pt}
  \caption{The interpolations at the continuum limit of the numerical results of the renormalization constant $\hat{Z}_{S}$ for the scalar density. The colored symbols and the colored lines indicate the numerical results and the fitting results, respectively. The small window in the figure shows the interpolated results of $\hat{Z}_{S}$.}\label{fig:inter_zs}
\end{figure}

The renormalization constant $\hat{Z}_{S}$ depends on the bare coupling $g(\beta)$~\cite{Wennekers1}; therefore, we interpolate the calculated results of the renormalization constant $\hat{Z}_{S}$ to the continuum limit by fitting the following linear function as shown in Fig.~\ref{fig:inter_zs}:
\begin{equation}
  Z_{S} = A_{\hat{Z}_{S}}x + \hat{Z}_{S}^{\text{int}}, \ x = g(\beta)^{2}.
\end{equation}
The fitting results are presented in Table~\ref{tb:interp_res_Zs}. The small window in Fig.~\ref{fig:inter_zs} demonstrates that the interpolated results $\hat{Z}_{S}^{\text{int}}$ are consistent except that the result of the magnetic charge $m_{c}$ = 3 shows a minor effect caused by the finite lattice volume. This indicates that the additional monopoles and anti-monopoles do not affect the renormalization constant for the scalar density at the continuum limit.%16 March 2022
\begin{table}[htbp]
  \caption{The fitting results of the slope $A_{\hat{Z}_{S}}$ and the interpolated results $\hat{Z}_{S}^{\text{int}}$ of the renormalization constant.}\label{tb:interp_res_Zs}
  \begin{center}
      \begin{tabular}{|c|c|c|c|c|c|c|c|} \hline
        $m_{c}$ & $A_{\hat{Z}_{S}}$ & $\hat{Z}_{S}^{\text{int}}$ & FR: $g(\beta)^{2}$ & $\frac{\chi^{2}}{\text{dof}}$\\
        & & & $\times10^{-2}$ &  \\\hline 
        N. C.
           & 6.8(1.1)  & 0.74(4) & 2.2-6.2 & 0.4/3.0 \\ \hline   
        0 & 7.6(1.4)  & 0.71(6) & 2.8-6.2 & 0.2/2.0 \\ \hline   
        1 & 7.4(1.4)  & 0.72(6) & 2.8-6.2 & 0.1/2.0 \\ \hline   
        2 & 8.4(1.5)  & 0.68(6) & 2.8-6.2 & 0.0/2.0 \\ \hline 
        3 & 10.2(1.5) & 0.61(6) & 2.8-6.2 & 0.3/2.0 \\ \hline   
        4 & 8.9(1.9)  & 0.67(6) & 2.2-4.6 & 0.4/2.0 \\ \hline   
        5 & 9.0(1.9)  & 0.66(6) & 2.2-4.6 & 0.5/2.0 \\ \hline
      \end{tabular}
  \end{center}
\end{table}
%%%%%%%%%%%%%%%%%%%%%%%%%%%%%%%%%%%%%%%%%%%%%%%%%%%%%%%%%%%%%%%%%%
%%%%%%%%%%%%%%%%%%%%%%%%%%%%%%%%%%%%%%%%%%%%%%%%%%%%%%%%%%%%%%%%%%

\subsection{The renormalized chiral condensate in the chRMT}\label{sec:sec4_chiral_cond_rmt}
\begin{table}[htbp]
  \begin{center}
    \caption{Comparing the prediction $\langle\bar{\psi}\psi\rangle^{\text{Pre}}$ with the interpolated results $\langle\bar{\psi}\psi\rangle_{\text{RMT}}^{\overline{\text{MS}}, \text{int}}$ at the continuum limit, and their fitting results.}\label{tb:interp_sigma_RMT}
    \begin{tabular}{|c|c|c|c|c|}\hline
       & $\langle\bar{\psi}\psi\rangle^{\text{Pre}}$ & $\langle\bar{\psi}\psi\rangle_{\text{RMT}}^{\overline{\text{MS}}, \text{int}}$  & FR: $a^{2}$  & \\
      $m_{c}$ & [GeV$^{3}$] & [GeV$^{3}$] & [fm$^{2}$] & $\frac{\chi^{2}}{\text{dof}}$ \\
      & $\times10^{-2}$ & $\times10^{-2}$ & $\times10^{-2}$  &  \\ \hline
      N. C. &  -2.0280 & -2.52(4) & 5.5-15.5 & 2.1/4.0 \\\hline
      0 &  -2.0280 & -2.52(4) & 7.2-15.5 & 1.6/3.0 \\\hline     
      1 &  -2.1231 & -2.66(5) & 7.2-15.5 & 1.0/3.0 \\\hline      
      2 &  -2.2142 & -2.89(5) & 7.2-15.5 & 0.6/3.0 \\\hline       
      3 &  -2.3016 & -3.01(5) & 7.2-15.5 & 1.1/3.0 \\\hline       
      4 &  -2.3859 & -3.11(5) & 5.5-11.4 & 0.9/3.0 \\\hline
      5 &  -2.4672 & -3.16(6) & 5.5-11.4 & 4.8/3.0 \\ \hline
    \end{tabular}
  \end{center}
\end{table}
\begin{figure*}[htbp]
  \begin{center}
    \includegraphics[width=160mm]{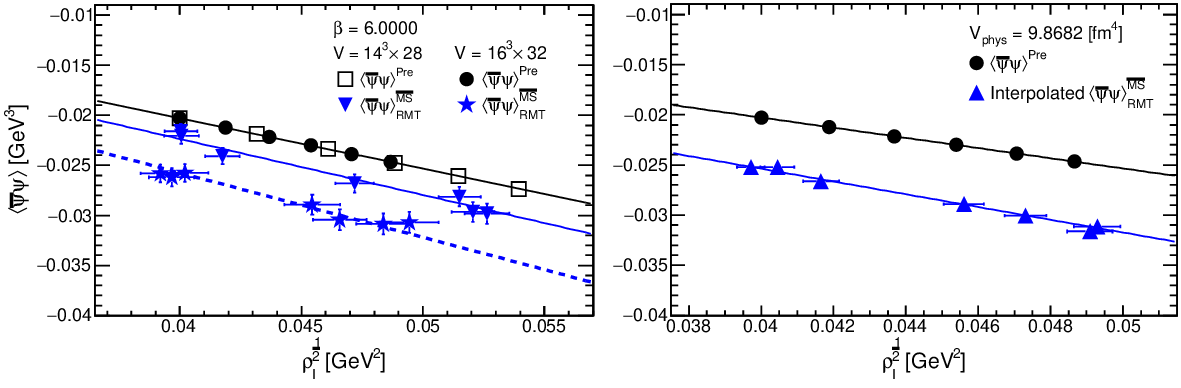}
  \end{center}
  \setlength\abovecaptionskip{-1pt}
  \caption{Comparisons of the renormalized chiral condensate $\langle\bar{\psi}\psi\rangle_{\text{RMT}}^{\overline{\text{MS}}}$ in the $\overline{\text{MS}}$-scheme at 2 [GeV] of the configurations of $\beta = 6.0000$, $V= 14^{3}\times28$ and $V= 16^{3}\times32$ (left) and the interpolated results at the continuum limit (right) with the predictions $\langle\bar{\psi}\psi\rangle^{\text{Pre}}$. In the left figure, the black full, the blue full, and the blue dashed lines indicate the fitting results of the predictions, the numerical results of $V = 14^{3}\times28$, and $V = 16^{3}\times32$, respectively. In the right figure, the black full line and the blue full line indicate the fitting results of the prediction and the interpolated results, respectively.}\label{fig:sigma_RMT_14xx3x28_16xx3x32_inter}
\end{figure*}

Suppose that the following relationship in $N_{f}$ = 3 between the chiral condensate and the scale parameter of the eigenvalue distribution in the chRMT holds~\cite{Wennekers1}:
\begin{equation}
  \langle \bar{\psi}\psi\rangle_{\text{RMT}} = -\Sigma\label{eq:z_lambda}.
\end{equation}
We evaluate the renormalized chiral condensate $\langle\bar{\psi}\psi\rangle_{\text{RMT}}^{\overline{\text{MS}}}$ in the $\overline{\text{MS}}$-scheme at 2 [GeV] using the outcomes of the scale parameter $\Sigma$ and renormalization constant $\hat{Z}_{S}$ for the scalar density.

First, we derive the scale parameter by substituting the scaled eigenvalues $z_{k}^{|Q|}$ that are analytically calculated and the numerical results of the improved eigenvalues $\lambda_{k}^{|Q|}$ into formula~(\ref{eq:z_and_lam})~\cite{Giusti4}. We presume that the scale parameter computed for each eigenvalue number $k$ of each topological charge sector $|Q|$ is independent, and then calculate the average of the scale parameter using the results of the eigenvalue numbers $k$ from 1 to 4 of each topological charge sector $|Q|$ from 0 to 2.

Finally, the renormalized chiral condensate in the $\overline{\text{MS}}$-scheme at 2 [GeV] is evaluated as follows:
\begin{equation}
  \langle \bar{\psi}\psi\rangle_{\text{RMT}}^{\overline{\text{MS}}} \ \ (2 \ [\text{GeV}]) = \frac{\hat{Z}_{S}}{0.72076}\langle \bar{\psi}\psi\rangle_{\text{RMT}}\label{eq:chiral_cond_rmt}
\end{equation}
We use each numerical result of the renormalization constant calculated for each lattice and configuration type (for each magnetic charge or standard configuration). The numerical results of the scale parameter and renormalized chiral condensate in the $\overline{\text{MS}}$-scheme at 2 [GeV] are presented in Table~\ref{tb:computed_sigma_rmt} in~\ref{sec:rmt_Sigma_conds}.

We first interpolate the calculated results of the renormalized chiral condensate in the $\overline{\text{MS}}$-scheme at 2 [GeV] to the continuum limit by fitting a linear function. The fitting results of the slope are zero considering their errors, except that the error of the slope value of the magnetic charge $m_{c}$ = 5 is 52$\%$. Therefore, we fit a constant function and list the interpolated results of $\langle\bar{\psi}\psi\rangle_{\text{RMT}}^{\overline{\text{MS}}}$ in Table~\ref{tb:interp_sigma_RMT}. The fitting results of $\chi^{2}/$dof are less than 1.6. This shows that the discretization influence on the numerical results is insignificant.

We have already reported that the renormalized chiral condensate in the $\overline{\text{MS}}$-scheme at 2 [GeV] decreases in direct proportion to the square root of the number density of the instantons and anti-instantons by comparing with the following prediction concerning the chiral condensate~\cite{Hasegawa2}:
\begin{equation}
\langle{\bar\psi} \psi \rangle^{\text{Pre}}(m_{c}) = -\frac{1}{{\bar\rho}}\left( \frac{\pi N_{c}}{13.2} \right)^{\frac{1}{2}}\left(\rho_{I}^{\text{sta}} + \frac{m_{c}}{V_{\text{phys}}}\right)^{\frac{1}{2}}\label{eq:chiral_ins_prediction1}.
\end{equation}
Here the predictions of the number density of the instantons and anti-instantons $\rho_{I}^{\text{Pre}}(m_{c})^{\frac{1}{2}}$ are given in Table~\ref{tb:Nzero_add_1} in~\ref{sec:nz_ni_niv}. The number of colors is $N_{c}$ = 3. The inverse of the average size of the instanton is
\begin{equation}
\frac{1}{\bar{\rho}} = 6 \times 10^{2} \ [\text{MeV}]\label{eq:inv_ins_size}.
\end{equation}
The predictions of the chiral condensate $\langle\bar{\psi}\psi\rangle^{\text{Pre}}$ are given in Table~\ref{tb:interp_sigma_RMT} and Table~\ref{tb:computed_ops} in~\ref{sec:comp_results_all}.

We then fit the following linear function to the numerical results to quantitatively evaluate the decreases in the chiral condensate by comparing the fitting results with the prediction shown in Fig.~\ref{fig:sigma_RMT_14xx3x28_16xx3x32_inter}:
\begin{equation}
 \langle\bar{\psi}\psi\rangle = -A_{\langle \bar{\psi}\psi\rangle}^{\text{RMT}}x, \  x = \rho_{I}^{\frac{1}{2}} \ [\text{GeV}^{2}].
\end{equation}
The slope of the prediction is $A_{\langle \bar{\psi}\psi\rangle}^{\text{Pre}}$ =  0.5070 [GeV]. The fitting results are listed in Table~\ref{tb:rmt_sigma_fit}.
\begin{table}[htbp]
  \begin{center}
    \caption{The fitting results obtained by the linear curves $\langle\bar{\psi}\psi\rangle = -A_{\langle \bar{\psi}\psi\rangle}^{\text{RMT}}x$, ($x = \rho_{I}^{\frac{1}{2}}$, $[\text{GeV}^{2}]$). The configurations of the lattice volumes $V = 14^{3}\times28$ and $V = 16^{3}\times32$ of $\beta = 6.0000$ and the interpolated results are used.}\label{tb:rmt_sigma_fit}
    \begin{tabular}{|c|c|c|c|}\hline
      Conf & $A_{\langle \bar{\psi}\psi\rangle}^{\text{RMT}}$& FR: $\rho_{I}^{\frac{1}{2}}$ [GeV$^{2}$] &$\frac{\chi^{2}}{\text{dof}}$\\
      &  [GeV] & $\times10^{-2}$ & \\\hline
      $14^{3}\times28$ & 0.559(8)  & 3.99-5.27 & 2.7/6.0 \\\cline{1-4}
      $16^{3}\times32$ & 0.644(10) & 3.92-4.95 & 1.7/6.0 \\\cline{1-4}
      Interp. & 0.635(5) & 3.97-4.93 & 1.4/6.0 \\\hline
    \end{tabular}
  \end{center}
\end{table}

Figure~\ref{fig:sigma_RMT_14xx3x28_16xx3x32_inter} and the fitting results of the slope in Table~\ref{tb:rmt_sigma_fit} demonstrate that the renormalized chiral condensate that is calculated from the scale parameter $\Sigma$ decreases in direct proportion to the square root of the number density of the instantons and anti-instantons; however, the fitting results of the slope indicate that the effects of the finite lattice volume on the slope values appear and the interpolated results are not consistent with the prediction~(\ref{eq:chiral_ins_prediction1}). Therefore, in the sections below, we improve the calculations of the observables by matching the numerical results with the experimental results.%16 Mar 2022
%%%%%%%%%%%%%%%%%%%%%%%%%%%%%%%%%%%%%%%%%%%%%%%%%%%%%%
%%%%%%%%%%%%%%%%%%%%%%%%%%%%%%%%%%%%%%%%%%%%%%%%%%%%%%
%%%%%%%%%%%%%%%%%%%%%%%%%%%%%%%%%%%%%%%%%%%%%%%%%%%%%%
%%% END SEC 5

\section{Instanton effects}\label{sec:6}

We perform simulations in the quenched approximation of QCD. Therefore, to compare the observables that are numerically computed with the experimental results, we determine the normalization factor by matching the experimental results with the numerical results. We quantitatively evaluate the effects of the instanton and anti-instantons created by the additional monopoles and anti-monopoles on the renormalized chiral condensate, the renormalized average mass of the light quarks, decay constants, and pion mass. Finally, the catalytic effect of the pion decay is estimated with the interpolated outcomes at the continuum limit of the pion mass and pion decay constant as input values.
%%%%%%%%%%%%%%%%%%%%%%%%%%%%%%%%%%%%%%%%%%%%%%%%%%%
%%%%%%%%%%%%%%%%%%%%%%%%%%%%%%%%%%%%%%%%%%%%%%%%%%%

\subsection{Matching the decay constant and mass of the pseudoscalar meson with the experimental outcomes}\label{sec:6_match_exp}

We cannot determine the physical quantities or directly compare the numerical results with the experiments without using the outcomes of the chiral perturbation theory or experiments because our calculations are performed in the quenched approximation. Therefore, we improve the computations in reference~\cite{Giusti3,Alton1} to remove the ambiguities coming from the determinations of the lattice scales. We use the calculated results of the lattice spacing by the analytic formula in Table~\ref{tb:lattice} as lattice scales, determine the normalization factor $Z_{\pi}$ using the experimental outcomes, and calculate the observables using the normalization factor.

In this study, we calculate the observables of the configurations with the additional monopoles and anti-monopoles using the normalization factor $Z_{\pi}$ calculated for each standard configuration.

First, we calculate the decay constant of the pseudoscalar meson $aF_{PS}$ using the fitting results of the coefficient $a^{4}G_{P-S}$ and the results of the square mass of the pseudoscalar meson $(am_{PS})^{2}$ as follows:
\begin{equation}
  aF_{PS} = \frac{2a\bar{m}_{ud}\sqrt{a^{4}G_{P-S}}}{(am_{PS})^{2}}\label{eq:fps}
\end{equation}
The pion decay constant is $F_{\pi}$ = 93 [MeV] in this notation. The calculated results of the decay constant of $aF_{PS}$ are listed in the same tables as the fitting results of $a^{4}G_{P-S}$ and $am_{PS}$ mentioned in subsection~\ref{sec:5_pcac}. We define the decay constant of the pseudoscalar meson at the chiral limit $f_{0}$ as follows:
\begin{equation}
  af_{0} \equiv \lim_{a\bar{m}_{ud}\rightarrow 0}aF_{PS} =  \lim_{(am_{PS})^{2} \rightarrow 0}aF_{PS}
\end{equation}
The third term holds because the PCAC relation holds even if we add the monopoles and anti-monopoles with the magnetic charges to the configurations.

In the studies using the overlap Dirac operator in the quenched SU(3), there is a linear relationship between the decay constant and square mass of the pseudoscalar meson. This behavior corresponds to the features of the SU(2) Lagrangian in the quenched chiral perturbation theory~\cite{Colangelo1}, which was reported by other groups~\cite{Giusti2,Giusti1}.

Therefore, we plot the numerical results of the decay constant and square mass of the pseudoscalar meson, confirm the linear relationship between them without divergence near the chiral limit, and then fit the following linear function to the numerical results:
\begin{equation}
  aF_{PS} = a^{-1}A_{PS}(am_{PS})^{2} + aB_{PS}.
\end{equation}
The fitting results of the slope $a^{-1}A_{PS}$, intercept $aB_{PS}$, and $\chi^{2}/$dof are presented in Table~\ref{tb:fitresults_aps_ampi2_inter} in~\ref{sec:fitting_res_fps_intersec_fpi_mpi}. The fitting results of $\chi^{2}/$dof are less than 1; thus, these results indicate that the decay constant of the pseudoscalar meson $aF_{PS}$ linearly increases with its square mass $(am_{PS})^{2}$ without divergence near the chiral limit. 

To determine the normalization factor $Z_{\pi}$, we first make a curve using the experimental outcomes of the pion decay constant and pion mass as follows:
\begin{equation}
  aF_{PS} = am_{PS}\mathcal{C}_{\pi}^{\text{Exp}}, \ \mathcal{C}_{\pi}^{\text{Exp}} = \frac{F_{\pi^{-}}^{\text{Exp}}}{m_{\pi^{\pm}}^{\text{Exp}}\sqrt{2}}\label{eq:fpi_mpi_exp1}.
\end{equation}
Here we use the following outcomes of the pion decay constant and pion mass reported in reference~\cite{PDG_2017}: 
\begin{align}
  &\frac{F_{\pi^{-}}^{\text{Exp}}}{\sqrt{2}} = \frac{130.50(12)}{\sqrt{2}} \ [\text{MeV}]\label{eq:exp_fpi},\\
  &m_{\pi^{\pm}}^{\text{Exp}} = 139.5706(2) \ [\text{MeV}]\label{eq:exp_mpi}.
\end{align}
\begin{table}[htbp]
  \begin{center}
  \caption{The calculated results of the normalization factor $Z_{\pi}$ using the standard configurations.}\label{tb:z_pi}
  \begin{tabular}{|c|c|c|}\hline
    $\beta$ & $V$ & $Z_{\pi}$  \\\hline
    5.8457 & $12^{3}\times$24 &  1.24(3) \\\hline
    5.9256 & $14^{3}\times$28 &  1.26(2) \\\hline
    6.0000 & $14^{3}\times$28 &  1.31(2) \\\cline{2-3}
     & $16^{3}\times$32 &  1.294(19) \\\hline
    6.1366 & $20^{3}\times$40 & 1.270(18) \\\hline
  \end{tabular}
  \end{center}
\end{table}
\begin{table}[htbp]
  \begin{center}
    \caption{Comparing the interpolated results $F_{0}^{\text{int}}$ of the decay constant at the continuum limit with the prediction $F_{0}^{\text{Pre}}$.}\label{tb:interp_ops_f0}
      \begin{tabular}{|c|c|c|c|c|}\hline
        $m_{c}$ & $F_{0}^{\text{Pre}}$ & $F_{0}^{\text{int}}$ & FR: $a^{2}$ [fm$^{2}$] &$\frac{\chi^{2}}{\text{dof}}$\\
        &  [MeV] & [MeV]  &  $\times10^{-2}$ &\\\hline
        N. C.
        & 85.366  & 90.6(9)    & 5.5-15.5  & 0.0/4.0 \\ \hline
        0 & 85.366  & 90.8(1.1)   & 7.2-15.5 & 0.2/3.0 \\ \hline
        1 & 87.345  & 92.6(1.1)   & 7.2-15.5 & 0.0/3.0 \\ \hline
        2 & 89.199  & 96.3(1.2)   & 7.2-15.5 & 0.4/3.0 \\ \hline
        3 & 90.943  & 97.5(1.2)   & 7.2-15.5 & 0.7/3.0 \\ \hline
        4 & 92.593  & 99.2(1.1)   & 5.5-11.4 & 2.3/3.0 \\ \hline
        5 & 94.159  & 100.0(1.1)  & 5.5-11.4 & 1.2/3.0 \\ \hline
      \end{tabular}
  \end{center}
\end{table}
\begin{figure*}[htbp]
  \begin{center}
    \includegraphics[width=160mm]{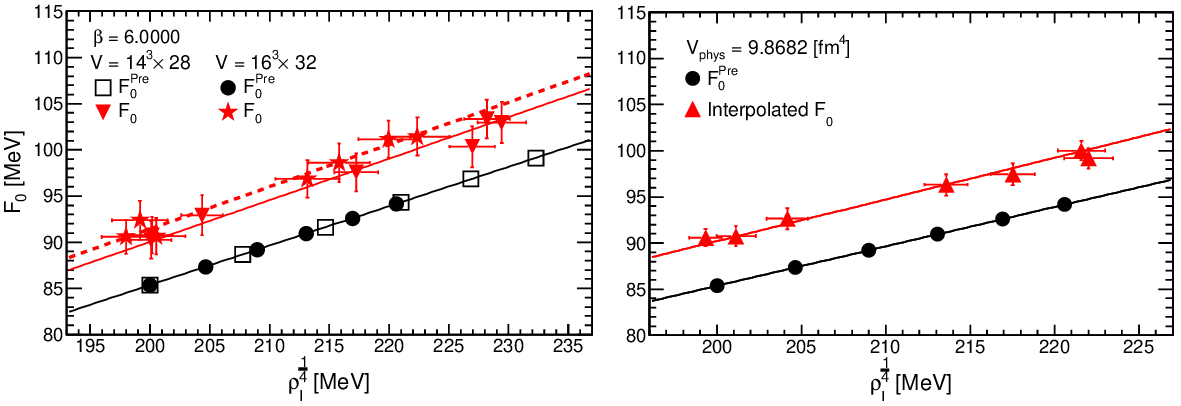}
  \end{center}
  \setlength\abovecaptionskip{-1pt}
  \caption{Comparisons of the decay constant $F_{0}$ at the chiral limit of the configurations of $\beta = 6.0000$, $V = 14^{3}\times28$ and $V= 16^{3}\times32$ (left) and the interpolated results at the continuum limit (right) with the predictions $F_{0}^{\text{Pre}}$. In the left figure, the black full, the red full, and the red dashed lines indicate the fitting results of the predictions, the numerical results of $V = 14^{3}\times28$, and $V = 16^{3}\times32$, respectively. In the right figure, the black full line and the red full line indicate the fitting results of the prediction and the interpolated results, respectively.}\label{fig:decay_constant_rescale_f0}
\end{figure*}

Next, we calculate the intersections $aF_{PS}^{\pi}$ and $am_{PS}^{\pi}$ between the curve~(\ref{eq:fpi_mpi_exp1}) and the linear functions made using the fitting results of the slope $a^{-1}A_{PS}$ and intercept $aB_{PS}$. The normalization factor $Z_{\pi}$ is determined for each lattice using the intersections of the standard configuration and the experimental outcomes of the pion decay constant~(\ref{eq:exp_fpi}) or pion mass~(\ref{eq:exp_mpi}) as follows:
\begin{equation}
Z_{\pi} = \frac{92.277}{F_{PS}^{\pi}} = \frac{139.5706}{m_{PS}^{\pi}}\label{eq:zpi}\\
\end{equation}
Here we do not consider the errors of the experimental outcomes because their errors are substantially smaller than the errors of the numerical results. The computed results of the intersections $aF_{PS}^{\pi}$ and $am_{PS}^{\pi}$ are presented in Table~\ref{tb:fitresults_aps_ampi2_inter} in~\ref{sec:fitting_res_fps_intersec_fpi_mpi}, and the normalization factor $Z_{\pi}$ of the standard configurations are listed in Table~\ref{tb:z_pi}.

Next, we define the decay constant at the chiral limit $aF_{0}$ using the normalization factor $Z_{\pi}$ of the standard configuration and the fitting results of the intercept $aB_{PS}$ as follows:
\begin{equation}
  aF_{0} \equiv Z_{\pi}af_{0} = Z_{\pi}aB_{PS}\label{eq:f0}
\end{equation}
The numerical results of the decay constant at the chiral limit in Table~\ref{tb:computed_ops} in~\ref{sec:comp_results_all} show that there are no influences of the discretization or finite lattice volume on the numerical results. Therefore, we can compute the observables using the normalization factor $Z_{\pi}$ without harmful influences of the finite lattice volume or the discretization on the observables. We then fit the constant function to interpolate the results to the continuum limit and list the fitting results in Table~\ref{tb:interp_ops_f0}.

The prediction of the decay constant $F_{0}^{\text{Pre}}$ at the chiral limit using the Gell-Mann-Oakes-Renner (GMOR) relation~\cite{Gellmann1} and the prediction of the chiral condensate~(\ref{eq:chiral_ins_prediction1}) is given in~\cite{Hasegawa2} as follows:
\begin{align}
  & F_{0}^{\text{Pre}}(m_{c}) = \nonumber\\
 & \ \ \ \ \ \ \ \ \  \frac{1}{m_{\pi}}\left( \frac{2\bar{m}_{ud}}{{\bar\rho}} \right)^{\frac{1}{2}} \left( \frac{\pi N_{c}}{13.2} \right)^{\frac{1}{4}} \left(\rho_{I}^{\text{sta}} + \frac{m_{c}}{V_{\text{phys}}}\right)^{\frac{1}{4}}.\label{eq:pion_decay}
\end{align}
The prediction of the normal configuration calculated by this formula is $F_{0}^{\text{Pre}}(0)$ = 85.366 [MeV]. This outcome is consistent with the prediction in the chiral perturbation theory $F_{0}^{\chi PT} = 86.2(5)$ [MeV]~\cite{Colangelo3}. The prediction $F_{0}^{\text{Pre}}$ of the decay constant at the chiral limit for each magnetic charge is given in Table~\ref{tb:interp_ops_f0} and Table~\ref{tb:computed_ops} in~\ref{sec:comp_results_all}. 
\begin{equation}
  F_{0} = A_{F_{0}}x,\ x = \rho_{I}^{\frac{1}{4}} \ [\text{MeV}].
\end{equation}

Last, we have already reported that the decay constant at the chiral limit increases in direct proportion to the one-fourth root of the number density of the instantons and anti-instantons in our previous study~\cite{Hasegawa2}; accordingly, to compare the computed results of the decay constant at the chiral limit with the predictions, we fit the following linear function as shown in Fig.~\ref{fig:decay_constant_rescale_f0}:
\begin{table}[htbp]
  \begin{center}
     \caption{Comparisons of the fitting results. The fitting results are obtained by the linear curves $F_{0} = A_{F_{0}}x$, ($x = \rho_{I}^{\frac{1}{4}}$ [MeV]). The configurations of $\beta = 6.0000$, $V = 14^{3}\times28$ and $V = 16^{3}\times32$ and the interpolated results are used.}\label{tb:rescaled_all_fitting_f0}
    \begin{tabular}{|c|c|c|c|}\hline
      Conf&$A_{F_{0}}$&FR: $\rho_{I}^{\frac{1}{4}}$ [MeV]&$\frac{\chi^{2}}{\text{dof}}$\\
      &  & $\times10^{2}$ & \\\hline
      $14^{3}\times28$ & 0.450(4) & 1.99-2.30 & 0.9/6.0 \\\hline
      $16^{3}\times32$ & 0.457(4) & 1.97-2.23 & 0.7/6.0 \\\hline
      Interp. & 0.451(2) & 1.99-2.23 & 1.4/6.0 \\\hline
      \end{tabular}
  \end{center}
\end{table}

The prediction of the slope is $A_{F_{0}}^{Pre}$ = 0.4268. The fitting results of the slope of the numerical results are 5$\sim$7$\%$ steeper than the prediction indicated in Table~\ref{tb:rescaled_all_fitting_f0}; however, there is no influence of the finite lattice volume on the slope. Therefore, the results demonstrate that the decay constant at the chiral limit increases in direct proportion to the one-fourth root of the number density of the instanton and anti-instantons.
\begin{figure*}[htbp]
  \begin{center}
    \includegraphics[width=160mm]{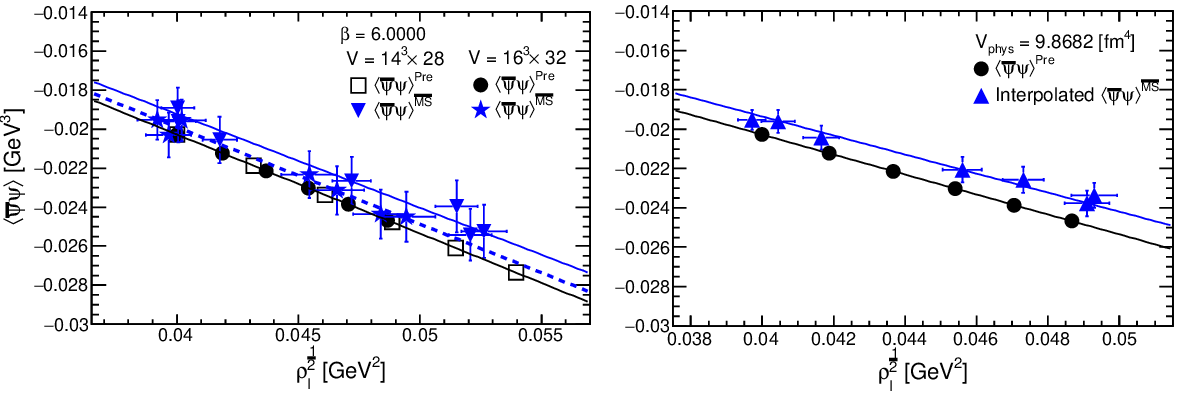}
  \end{center}
  \setlength\abovecaptionskip{-1pt}
  \caption{Comparisons of the renormalized chiral condensate $\langle\bar{\psi}\psi\rangle^{\overline{\text{MS}}}$ in the $\overline{\text{MS}}$-scheme at 2 [GeV] of the configurations of $\beta = 6.0000$, $V= 14^{3}\times28$ and $V= 16^{3}\times32$ (left) and the interpolated results at the continuum limit (right) with the predictions $\langle\bar{\psi}\psi\rangle^{\text{Pre}}$. In the left figure, the black full, the blue full, and the blue dashed lines indicate the fitting results of the predictions, the numerical results of $V = 14^{3}\times28$, and $V = 16^{3}\times32$, respectively. In the right figure, the black full line and the blue full line indicate the fitting results of the prediction and the interpolated results, respectively.}\label{fig:chiral_condensates}
\end{figure*}
%%%%%%%%%%%%%%%%%%%%%%%%%%%%%%%%%%%%%%%%%%%%%%%%%%%%%%%%%%
%%%%%%%%%%%%%%%%%%%%%%%%%%%%%%%%%%%%%%%%%%%%%%%%%%%%%%%%%%

\subsection{Instanton effects on chiral symmetry breaking}\label{sec:sec4_chi_cond_gmor}

The chiral condensate at the chiral limit is derived using the fitting results of the slope $aA_{\text{PCAC}}^{(2)}$ of the PCAC relation and the decay constant of the pseudoscalar meson at the chiral limit~(\ref{eq:f0}) as follows:
\begin{align}
    a^{3}\langle \bar{\psi}\psi \rangle = - \frac{aA_{\text{PCAC}}^{(2)}}{2}{(aF_{0})^{2}}\label{eq:gmor_new2}
\end{align}
The renormalized chiral condensate in the $\overline{\text{MS}}$-scheme at 2 [GeV] is evaluated as follows:
\begin{equation}
  \langle \bar{\psi}\psi\rangle^{\overline{\text{MS}}} \ \ (2 \ [\text{GeV}]) = \frac{\hat{Z}_{S}}{0.72076}\langle \bar{\psi}\psi\rangle\label{eq:chiral_cond_gmor}
\end{equation}
We use each numerical result of the renormalization constant $\hat{Z}_{S}$ calculated for each lattice and configuration type, as mentioned in subsection~\ref{sec:sec4_chiral_cond_rmt}. The numerical results of the renormalized chiral condensate in the $\overline{\text{MS}}$-scheme at 2 [GeV] are listed in Table~\ref{tb:computed_ops} in~\ref{sec:comp_results_all}.
\begin{table}[htbp]
  \begin{center}
    \caption{The prediction $\langle\bar{\psi}\psi\rangle^{\text{Pre}}$, interpolated results $\langle\bar{\psi}\psi\rangle_{\overline{\text{MS}}}^{\text{int}}$ at the continuum limit by the constant function, and the fitting results.}\label{tb:interp_ops_cond}
    \begin{tabular}{|c|c|c|c|c|}\hline
      & $\langle\bar{\psi}\psi\rangle^{\text{Pre}}$ & $\langle\bar{\psi}\psi\rangle_{\overline{\text{MS}}}^{\text{int}}$  & FR: $a^{2}$  &  \\
      $m_{c}$ & [GeV$^{3}$]  & [GeV$^{3}$]  &  [fm$^{2}$] &$\frac{\chi^{2}}{\text{dof}}$\\
      &   $\times10^{-2}$ &  $\times10^{-2}$ &  $\times10^{-2}$ & \\\hline
       N. C.
          &  -2.0280 &  -1.95(5) & 5.5-15.5 & 0.0/4.0 \\ \hline
       0 &  -2.0280 &  -1.96(6)  & 7.2-15.5 & 0.2/3.0 \\ \hline
       1 &  -2.1231 &  -2.04(6)  & 7.2-15.5 & 0.0/3.0 \\ \hline
       2 &  -2.2142 &  -2.21(6)  & 7.2-15.5 & 0.2/3.0 \\ \hline
       3 &  -2.3016 &  -2.26(7)  & 7.2-15.5 & 0.5/3.0 \\ \hline
       4 &  -2.3859 &  -2.34(6)  & 5.5-11.4 & 1.5/3.0 \\ \hline
       5 &  -2.4672 &  -2.38(6)  & 5.5-11.4 & 0.8/3.0 \\ \hline
    \end{tabular}
  \end{center}
\end{table}

Table~\ref{tb:computed_ops} in~\ref{sec:comp_results_all} shows that the numerical results of the renormalized chiral condensate are consistent without the discretization influence by using the normalization factor; therefore, we interpolate the calculated results to the continuum limit by fitting the constant function. The fitting results are given in Table~\ref{tb:interp_ops_cond}. This table shows that the fitting results of $\chi^{2}$/dof are less than 0.5 and that the interpolated results are reasonably consistent with the prediction; therefore, the influence of the discretization is negligible.

To confirm that the influence of the finite lattice volume is improved, we compare the slope values of the numerical results with the prediction by fitting the following linear function as shown in the left panel of Fig.~\ref{fig:chiral_condensates}:
\begin{equation}
\langle \bar{\psi}\psi\rangle = -A_{\langle \bar{\psi}\psi\rangle}x, \ x = \rho_{I}^{\frac{1}{2}} \ [\text{GeV}^{2}].
\end{equation}
The fitting results of the slope $A_{\langle \bar{\psi}\psi\rangle}$ of the lattice volumes $V = 14^{3}\times28$ and $V = 16^{3}\times32$ of $\beta$ = 6.0000 in Table~\ref{tb:rescaled_all_fitting_cond} are consistent with the prediction $A_{\langle \bar{\psi}\psi\rangle}^{\text{Pre}}$ = 0.5070 [GeV]; thus, there is no influence of the finite lattice volume on the renormalized chiral condensate in the $\overline{\text{MS}}$-scheme at 2 [GeV].
\begin{table}[htbp]
  \begin{center}
    \caption{The fitting results of the slope $A_{\langle \bar{\psi}\psi\rangle}$ obtained by the linear curves $\langle \bar{\psi}\psi\rangle = -A_{\langle \bar{\psi}\psi\rangle}x$, ($x = \rho_{I}^{\frac{1}{2}}$ $[\text{GeV}^{2}]$). The configurations of the lattice volumes $V = 14^{3}\times28$ and $V = 16^{3}\times32$ of $\beta = 6.0000$ and the interpolated results are used.}\label{tb:rescaled_all_fitting_cond}
    \begin{tabular}{|c|c|c|c|}\hline
      Conf & $A_{\langle \bar{\psi}\psi\rangle}$ [GeV] & FR: $\rho_{I}^{\frac{1}{2}}$  [GeV$^{2}$] &$\frac{\chi^{2}}{\text{dof}}$\\
        &  & $\times10^{-2}$ & \\\hline
      $14^{3}\times28$ & 0.481(10) & 3.99-5.27 & 0.7/6.0 \\\hline
      $16^{3}\times32$ & 0.497(11) & 3.92-4.95 & 0.5/6.0 \\\hline
      Interp. & 0.484(6) & 3.97-4.93 & 1.0/6.0 \\\hline
    \end{tabular}
  \end{center}
\end{table}

Next, we compare the slope value of the interpolated results with the prediction as shown in the right panel of Fig.~\ref{fig:chiral_condensates}. The slope value of the interpolated results that is presented in Table~\ref{tb:rescaled_all_fitting_cond} is consistent with the prediction of the slope $A_{\langle \bar{\psi}\psi\rangle}^{\text{Pre}}$ = 0.5070 [GeV]. 

These results demonstrate that the renormalized chiral condensate in the $\overline{\text{MS}}$-scheme at 2 [GeV] decreases in direct proportion to the square root of the number density of the instantons and anti-instantons, with neither the influences of the finite lattice volume nor the discretization.
     
In addition, we estimate the inverse of the average size of the instanton from the slope value of the interpolated result as follows:
\begin{equation}
\frac{1}{\bar{\rho}} = 5.73(7) \times10^{2} \ [\text{MeV}]
\end{equation}
This result is compatible with the outcome of the phenomenological model~(\ref{eq:inv_ins_size}) and indicates that the additional monopoles and anti-monopoles do not change the average size of the instanton or anti-instanton. Moreover, we have already mentioned that the interpolated result of the number density of the instantons and anti-instantons computed using the standard configurations is consistent with the outcome~(\ref{eq:ins_dens}) of the phenomenological model in subsection~\ref{sec:creation_ins}.

Finally, these results demonstrate that the instantons and anti-instantons created by the additional monopoles and anti-monopoles induce chiral symmetry breaking and the features are consistent with the phenomenological models concerning the instantons~\cite{Dyakonov6,Shuryak2}.
\begin{figure*}[htbp]
  \begin{center}
   \includegraphics[width=160mm]{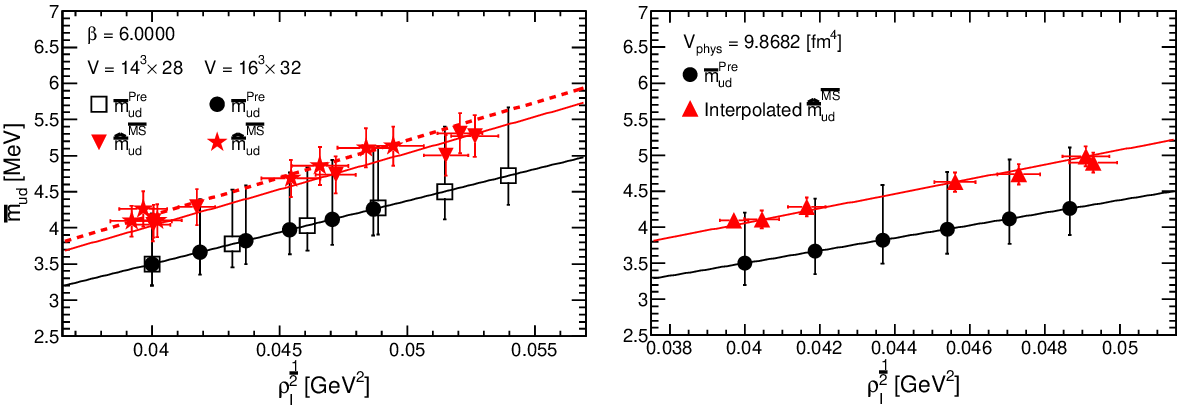}
  \end{center}
  \setlength\abovecaptionskip{-1pt}
  \caption{Comparisons of the renormalized average quark mass $\hat{\bar{m}}_{ud}^{\overline{\text{MS}}}$ in the $\overline{\text{MS}}$-scheme at 2 [GeV] of the configurations of $V= 14^{3}\times28$ and $V= 16^{3}\times32$ of $\beta = 6.0000$ (left) and the interpolated results at the continuum limit (right) with the predictions $\bar{m}_{ud}^{\text{Pre}}$. In the left figure, the black full, the red full, and the red dashed lines indicate the fitting results of the predictions, the numerical results of $V = 14^{3}\times28$, and $V = 16^{3}\times32$, respectively. In the right figure, the black full line and the red full line indicate the fitting results of the prediction and the interpolated results, respectively.}\label{fig:quark_masses}
\end{figure*}%16 March 2022
%%%%%%%%%%%%%%%%%%%%%%%%%%%%%%%%%%%%%%%%%%%%%%%%%%%%%%
%%%%%%%%%%%%%%%%%%%%%%%%%%%%%%%%%%%%%%%%%%%%%%%%%%%%%%

\subsection{Instanton effects on light quark masses}\label{sec:sec4_quark_mass}

We have quantitatively demonstrated that the absolute value of the renormalized chiral condensate in the $\overline{\text{MS}}$-scheme at 2 [GeV] increases in direct proportion to the square root of the number density of the instantons and anti-instantons. Quark-mass generation is closely related to chiral symmetry breaking; therefore, suppose that the quark masses become heavy in direct proportion to the square root of the number density of the instantons and anti-instantons~\cite{Hasegawa2}.

In this study, we estimate the average mass of the up and down quarks using the computed results of the intersection $am_{PS}^{\pi}$ in Table~\ref{tb:fitresults_aps_ampi2_inter} in~\ref{sec:fitting_res_fps_intersec_fpi_mpi}, normalization factor $Z_{\pi}$ in Table~\ref{tb:z_pi}, and the fitting results of the slope value $aA^{(2)}$ of the PCAC relation in Table~\ref{tb:prop_v12xx3x24_b5p84572_fit1_2} in~\ref{sec:pcac_zs} as follows:
\begin{equation}
  a\bar{m}_{ud} = \frac{(Z_{\pi}am_{PS}^{\pi})^{2}}{aA^{(2)}}
\end{equation}

The renormalized average mass of the up and down quarks is evaluated in the $\overline{\text{MS}}$-scheme at 2 [GeV] using the renormalization constant $\hat{Z_{S}}$ for the scalar density in Table~\ref{tb:prop_v12xx3x24_b5p84572_fit1_2} in~\ref{sec:pcac_zs} as follows:
\begin{equation}
 \hat{\bar{m}}_{ud}^{\overline{\text{MS}}} = \frac{0.72076}{\hat{Z}_{S}}\bar{m}_{ud}
\end{equation}
We use each result of the renormalization constant for the scalar density calculated for each type of configuration. The calculated results of the renormalized average mass $\hat{\bar{m}}_{ud}^{\overline{\text{MS}}}$ in the $\overline{\text{MS}}$-scheme at 2 [GeV] are presented in Table~\ref{tb:computed_ops} in~\ref{sec:comp_results_all} and they show no influence of the discretization. Therefore, we fit the calculated results by the constant function, obtain the interpolated results at the continuum limit, and list them in Table~\ref{tb:interp_ops_quark_mass}.
\begin{table}[htbp]
  \begin{center}
    \caption{Comparing the interpolated results $\hat{\bar{m}}_{ud}^{\overline{\text{MS}}, \text{int}}$ at the continuum limit with the prediction $\bar{m}_{ud}^{\text{Pre}}$ together with the fitting results.}\label{tb:interp_ops_quark_mass}
    \begin{tabular}{|c|c|c|c|c|}\hline
       $m_{c}$ & $\bar{m}_{ud}^{\text{Pre}}$ & $\hat{\bar{m}}_{ud}^{\overline{\text{MS}}, \text{int}}$ & FR: $a^{2}$ [fm$^{2}$] &$\frac{\chi^{2}}{\text{dof}}$\\
        &  [MeV]    &  [MeV]   &  $\times10^{-2}$ &\\\hline
       N. C.
         &  3.5$^{+0.7}_{-0.3}$ & 4.09(10) & 5.5-15.5 & 0.0/4.0 \\ \hline
       0 &  3.5$^{+0.7}_{-0.3}$ & 4.11(12) & 7.2-15.5 & 0.2/3.0  \\ \hline
       1 &  3.7$^{+0.7}_{-0.3}$ & 4.28(13) & 7.2-15.5 & 0.0/3.0  \\ \hline
       2 &  3.8$^{+0.8}_{-0.3}$ & 4.63(14) & 7.2-15.5 & 0.2/3.0  \\ \hline
       3 &  4.0$^{+0.8}_{-0.3}$ & 4.74(14) & 7.2-15.5 & 0.5/3.0  \\ \hline
       4 &  4.1$^{+0.8}_{-0.4}$ & 4.90(14) & 5.5-11.4 & 1.4/3.0  \\ \hline
       5 &  4.3$^{+0.9}_{-0.4}$ & 4.98(14) & 5.5-11.4 & 0.7/3.0  \\ \hline
    \end{tabular}
  \end{center}
\end{table}

We assume that the average mass of the light quarks becomes heavy in direct proportion to the square root of the number density of the instantons and anti-instantons; accordingly, we provide the following prediction concerning the average mass of the light quarks using the predictions~(\ref{eq:ins_dens}),~(\ref{eq:num_ins_dens_add}), and experimental outcome $\bar{m}_{ud}^{\text{\text{Exp}}}$ of the average mass of the light quarks: 
\begin{equation}
  \bar{m}_{ud}^{\text{Pre}}(m_{c}) = \bar{m}_{ud}^{\text{\text{Exp}}}\left[\frac{\rho_{I}^{\text{Pre}}(m_{c})}{\rho_{I}^{\text{sta}}}\right]^{\frac{1}{2}}\label{eq:mq_ins_pred}.
\end{equation}
Here, the experimental outcome is $\bar{m}_{ud}^{\text{\text{Exp}}} = 3.5_{-0.3}^{+0.7}$ [MeV]\\~\cite{PDG_2017}. The predictions are given in Table~\ref{tb:interp_ops_quark_mass} and Table~\ref{tb:computed_ops} in~\ref{sec:comp_results_all}.

In order to quantitatively demonstrate the increase in the average mass of the light quarks to the increase in the square root of the number density of the instantons and anti-instantons, we plot the numerical results and predictions and fit them by the following linear function, as shown in Fig.~\ref{fig:quark_masses}:
\begin{equation}
  \bar{m}_{ud} = A_{\bar{m}_{ud}}x, \ x = \rho_{I}^{\frac{1}{2}} \ [\text{GeV}^{2}].
\end{equation}

The slope values $A_{\bar{m}_{ud}}$ of the lattice volumes $V = 14^{3}\times28$ and $V = 16^{3}\times32$ and the interpolated results are 15$\%$ steeper than the prediction, as shown in Table~\ref{tb:rescaled_all_fitting_quark_mass}; however, there are no influences of the finite lattice volume on the slope values.
\begin{table}[htbp]
  \begin{center}
    \caption{The fitting results of the numerical results and prediction. The linear curve is $\bar{m}_{ud} = A_{\bar{m}_{ud}}x$, ($x = \rho_{I}^{\frac{1}{2}}$ $[\text{GeV}^{2}]$). The configurations of the lattice volumes $V = 14^{3}\times28$ and $V = 16^{3}\times32$ of $\beta = 6.0000$ and the interpolated results are used.}\label{tb:rescaled_all_fitting_quark_mass}
    \begin{tabular}{|c|c|c|c|}\hline
      Conf &$A_{\bar{m}_{ud}}$ [MeV$^{-1}$]&FR: $\rho_{I}^{\frac{1}{2}}$&$\frac{\chi^{2}}{\text{dof}}$\\
       & $\times10^{-4}$ &[GeV$^{2}$]& \\\hline
      Pred. & 0.88(4) & 3.99-5.40 & 0.0/5.0 \\ \hline
      $14^{3}\times28$ & 1.01(2) & 3.99-5.27 & 0.6/6.0 \\\hline
      $16^{3}\times32$ & 1.04(2) & 3.92-4.95 & 0.5/6.0 \\\hline
      Interp. & 1.014(12) & 3.97-4.93 & 1.1/6.0 \\\hline
    \end{tabular}
  \end{center}
\end{table}

Thus, the renormalized average mass of the up and down quarks in the $\overline{\text{MS}}$-scheme at 2 [GeV] becomes heavy in direct proportion to the square root of the number density of the instantons and anti-instantons, without any influences of the finite lattice volume or discretization. The increase in the average mass of the quarks is almost the same as the increase in the absolute value of the chiral condensate. We will quantitatively evaluate the decrease and increases in the observables in subsection~\ref{sec:6_ratios}.
%%%%%%%%%%%%%%%%%%%%%%%%%%%%%%%%%%%%%%%%%%%%%%%%%%%%%%%
%%%%%%%%%%%%%%%%%%%%%%%%%%%%%%%%%%%%%%%%%%%%%%%%%%%%%%%

\subsection{Instanton effects on the pion mass and pion decay constant}\label{sec:sec5_qmass}

The pion mass $am_{\pi}$ and pion decay constant $aF_{\pi}$ are estimated using the normalization factor $Z_{\pi}$ in Table~\ref{tb:z_pi} and the intersections $am_{PS}^{\pi}$ and $aF_{PS}^{\pi}$ in Table~\ref{tb:fitresults_aps_ampi2_inter} in~\ref{sec:fitting_res_fps_intersec_fpi_mpi} as follows:
\begin{align}
  & am_{\pi} = Z_{\pi}am_{PS}^{\pi}\\
  & aF_{\pi} = Z_{\pi}aF_{PS}^{\pi}
\end{align}

The discretization does not affect the numerical results of the pion mass, pion decay constant, and ratio of the decay constants $\frac{F_{\pi}}{F_{0}}$, as shown in Table~\ref{tb:computed_ops} in~\ref{sec:comp_results_all}. Moreover, the table shows that the calculated results of the ratio $\frac{F_{\pi}}{F_{0}}$ do not change even if we add the monopoles and anti-monopoles to the configurations; therefore, we fit the constant function to the numerical results of the pion mass and pion decay constant. The fitting results in Tables~\ref{tb:interp_ops_masses} and~\ref{tb:interp_ops_decay} show that the values of $\chi^{2}$/dof are small enough and the interpolated results of the pion mass and decay constant are reasonably consistent with their predictions.
\begin{table}[htbp]
  \begin{center}
    \caption{The prediction of the pion mass $m_{\pi}^{\text{Pre}}$, interpolated results of the pion mass $m_{\pi}^{\text{int}}$, and fitting results.}\label{tb:interp_ops_masses}
    \begin{tabular}{|c|c|c|c|c|}\hline
       &$m_{\pi}^{\text{Pre}}$&$m_{\pi}^{\text{int}}$& FR: $a^{2}$&\\
      $m_{c}$ &[MeV]&[MeV]&[fm$^{2}$]&$\frac{\chi^{2}}{\text{dof}}$\\  
       & $\times10^{2}$ & $\times10^{2}$ & $\times10^{-2}$ & \\ \hline 
       N. C.
         &  1.395706(2) & 1.396(14) & 5.5-15.5 & 0.0/4.0  \\ \hline
       0 &  1.395706(2) & 1.399(17)  & 7.2-15.5 & 0.3/3.0  \\ \hline
       1 &  1.428069(2) & 1.428(18)  & 7.2-15.5 & 0.0/3.0  \\ \hline
       2 &  1.458370(2) & 1.485(18)  & 7.2-15.5 & 0.4/3.0  \\ \hline
       3 &  1.486892(2) & 1.503(18)  & 7.2-15.5 & 0.7/3.0  \\ \hline
       4 &  1.513862(2) & 1.528(16)  & 5.5-11.4 & 2.2/3.0  \\ \hline
       5 &  1.539462(2) & 1.541(17)  & 5.5-11.4 & 1.2/3.0  \\ \hline
  \end{tabular}
  \end{center}
\end{table}
\begin{table}[htbp]
  \begin{center}
    \caption{The prediction of the pion decay constant $F_{\pi}^{\text{Pre}}$, interpolated results of the pion decay constant $F_{\pi}^{\text{int}}$, and fitting results.}\label{tb:interp_ops_decay}
    \begin{tabular}{|c|c|c|c|c|}\hline       
       &$F_{\pi}^{\text{Pre}}$&$F_{\pi}^{\text{int}}$&FR: $a^{2}$&\\
       $m_{c}$&[MeV]&[MeV]&[fm$^{2}$]&$\frac{\chi^{2}}{\text{dof}}$\\
       &&&$\times10^{-2}$&\\\hline
       N. C.
         &  92.28(9)  & 92.3(1.0)  & 5.5-15.5 & 0.0/4.0 \\ \hline
       0 &  92.28(9)  & 92.5(1.1)  & 7.2-15.5 & 0.3/3.0 \\ \hline
       1 &  94.42(9)  & 94.4(1.2)  & 7.2-15.5 & 0.0/3.0 \\ \hline
       2 &  96.42(9)  & 98.2(1.2)  & 7.2-15.5 & 0.4/3.0 \\ \hline
       3 &  98.31(9)  & 99.3(1.2)  & 7.2-15.5 & 0.7/3.0 \\ \hline
       4 &  100.09(9) & 101.1(1.1) & 5.5-11.4 & 2.2/3.0 \\ \hline
       5 &  101.78(9) & 101.9(1.1) & 5.5-11.4 & 1.2/3.0 \\ \hline
    \end{tabular}
  \end{center}
\end{table}

We have shown that the average mass of the up and down quarks becomes heavy in direct proportion to the square root of the number density of the instantons and anti-instantons. Moreover, the PCAC relation holds even if the monopoles and anti-monopoles are added to the configurations. These outcomes predict that the pion mass becomes heavy in direct proportion to the one-fourth root of the number density of the instantons and anti-instantons. Therefore, similar to the prediction concerning the average mass of the light quarks, we provide the following prediction concerning the pion mass using the experimental outcome~(\ref{eq:exp_mpi}):
\begin{equation}
m_{\pi}^{\text{Pre}}(m_{c}) = m_{\pi^{\pm}}^{\text{Exp}}\left[\frac{\rho_{I}^{\text{Pre}}(m_{c})}{\rho_{I}^{\text{sta}}}\right]^{\frac{1}{4}}\label{eq:mpi_ins_pred}.
\end{equation}

In addition, we have demonstrated that the decay constant of the pseudoscalar meson linearly increases with the increase in the square of the pseudoscalar mass. The decay constant of the pseudoscalar meson at the chiral limit increases in direct proportion to the one-fourth root of the number density of the instantons and anti-instantons. These quantitative relations indicate that the pion decay constant also increases in direct proportion to the one-fourth root of the number density of the instantons and anti-instantons. Therefore, we provide the following prediction concerning the pion decay constant using the experimental outcome~(\ref{eq:exp_fpi}):
\begin{equation}
  F_{\pi}^{\text{Pre}}(m_{c}) = \frac{F_{\pi^{-}}^{\text{Exp}}}{\sqrt{2}}\left[\frac{\rho_{I}^{\text{Pre}}(m_{c})}{\rho_{I}^{\text{sta}}}\right]^{\frac{1}{4}}\label{eq:fpi_ins_pred}.
\end{equation}
The predictions concerning $m_{\pi}^{\text{Pre}}$ and $F_{\pi}^{\text{Pre}}$ are listed in Table~\ref{tb:interp_ops_masses} and Table~\ref{tb:computed_ops} in~\ref{sec:comp_results_all}.
\begin{figure*}[htbp]
  \begin{center}
    \includegraphics[width=160mm]{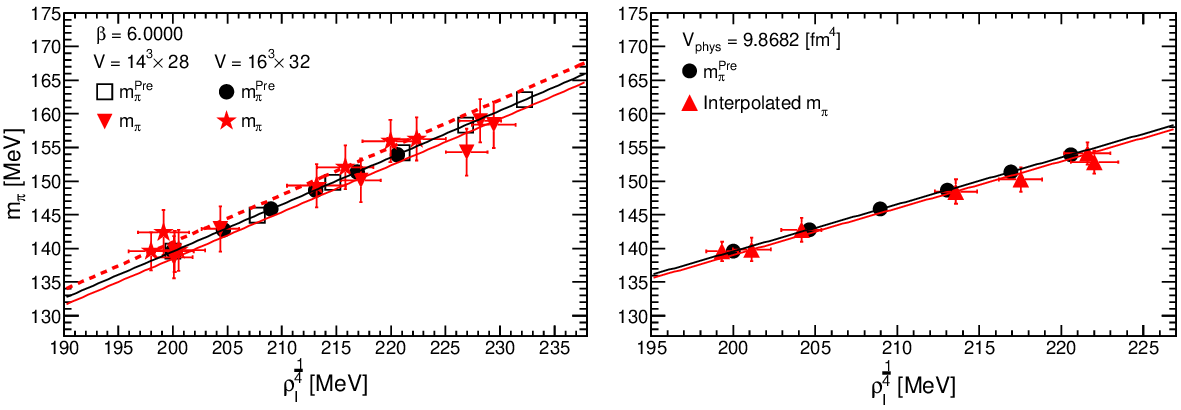}
  \end{center}
  \setlength\abovecaptionskip{-1pt}
  \caption{Comparisons of the pion mass $m_{\pi}$ of the configurations of $\beta = 6.0000$, $V= 14^{3}\times28$ and $V= 16^{3}\times32$ (left) and the interpolated results at the continuum limit (right) with the predictions $m_{\pi}^{\text{Pre}}$. In the left figure, the black full line, the red full line, and the red dashed line indicate the fitting results of the predictions, the numerical results of $V = 14^{3}\times28$, and $V = 16^{3}\times32$, respectively. In the right figure, the black full line and the red full line indicate the fitting results of the prediction and the interpolated results, respectively.}\label{fig:pion_masses}
\end{figure*}
\begin{figure*}[htbp]
  \begin{center}
  \includegraphics[width=160mm]{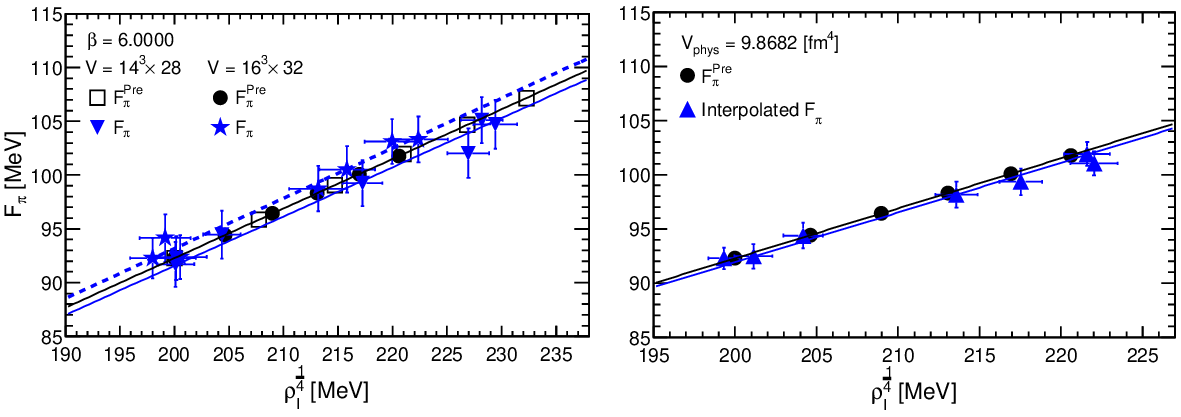}
  \end{center}
  \setlength\abovecaptionskip{-1pt}
  \caption{Comparisons of the pion decay constant $F_{\pi}$ of the configurations of $\beta = 6.0000$, $V= 14^{3}\times28$ and $V= 16^{3}\times32$ (left) and the interpolated results at the continuum limit (right) with the predictions $F_{\pi}^{\text{Pre}}$. In the left figure, the black full, the blue full, and the blue dashed lines indicate the fitting results of the predictions, the numerical results of $V = 14^{3}\times28$, and $V = 16^{3}\times32$, respectively. In the right figure, the black full line and the blue full line indicate the fitting results of the prediction and the interpolated results, respectively.}\label{fig:pion_decay_constants}
\end{figure*}

Finally, we fit the following linear functions to the numerical results of the pion mass and pion decay constant and their predictions, as shown in Figs.~\ref{fig:pion_masses} and~\ref{fig:pion_decay_constants}:
\begin{align}
  & m_{\pi} = A_{m_{\pi}}x, \ x = \rho_{I}^{\frac{1}{4}} \ [\text{MeV}],\\
  & F_{\pi} = A_{F_{\pi}}x, \ x = \rho_{I}^{\frac{1}{4}} \ [\text{MeV}].
\end{align}
The fitting results concerning the pion mass and the pion decay constant are presented in Tables~\ref{tb:fitting_mpi} and~\ref{tb:fitting_fpi}, respectively.

The fitting results concening the pion mass and the pion decay constant of the lattice volumes $V = 14^{3}\times28$ and $V = 16^{3}\times32$ of $\beta$ = 6.0000 in Tables~\ref{tb:fitting_mpi} and~\ref{tb:fitting_fpi} indicate that there is no influence of the finite lattice volume and that the slope values are consistent with the predictions. Furthermore, the fitting results of the slopes of the interpolated results in Tables~\ref{tb:fitting_mpi} and~\ref{tb:fitting_fpi} are consistent with the slope values of the predictions. These results show that the pion mass and pion decay constant increase in direct proportion to the one-fourth root of the number density of the instantons and anti-instantons without the influence of the finite lattice volume or discretization.
\begin{table}[htbp]
  \begin{center}
    \caption{Comparisons of the fitting results of the slope $A_{m_{\pi}}$ obtained by the linear function $m_{\pi} = A_{m_{\pi}}x$, ($x = \rho_{I}^{\frac{1}{4}}$ [MeV]). The configurations of the lattice volumes $V = 14^{3}\times28$ and $V = 16^{3}\times32$ of $\beta = 6.0000$ and the interpolated results are used.}\label{tb:fitting_mpi}
    \begin{tabular}{|c|c|c|c|}\hline
      Conf & $A_{m_{\pi}}$ & FR: $\rho_{I}^{\frac{1}{4}}$ [MeV] &$\frac{\chi^{2}}{\text{dof}}$\\
      &  & $\times10^{2}$ & \\\hline
      Pred. & 0.697853(4) & 1.99-2.33 & 0.0/5.0 \\ \hline
      $14^{3}\times28$ & 0.693(6) & 1.99-2.30 & 0.9/6.0 \\\hline
      $16^{3}\times32$ & 0.705(6) & 1.97-2.23 & 0.6/6.0 \\\hline
      Interp. & 0.695(3)                   & 1.99-2.23 & 1.4/6.0 \\\hline
    \end{tabular}
  \end{center}
\end{table}
\begin{table}[htbp]
  \begin{center}
    \caption{Comparisons of the fitting results of the slope $A_{F_{\pi}}$ obtained by the linear function $F_{\pi} = A_{F_{\pi}}x$, ($x = \rho_{I}^{\frac{1}{4}}$ [MeV]). The configurations of the lattice volumes $V = 14^{3}\times28$ and $V = 16^{3}\times32$ of $\beta = 6.0000$ and the interpolated results are used.}\label{tb:fitting_fpi}
    \begin{tabular}{|c|c|c|c|}\hline
      Conf & $A_{F_{\pi}}$ & FR: $\rho_{I}^{\frac{1}{4}}$  [MeV] &$\frac{\chi^{2}}{\text{dof}}$\\
      &  & $\times10^{2}$ & \\\hline
      Pred. & 0.46140(17) & 1.99-2.33 & 0.0/5.0 \\ \hline
      $14^{3}\times28$ & 0.458(4) & 1.99-2.30 & 0.9/6.0 \\\hline
      $16^{3}\times32$ & 0.466(4) & 1.97-2.23 & 0.6/6.0 \\\hline
      Interp. & 0.460(2)                   & 1.99-2.23 & 1.4/6.0 \\\hline
    \end{tabular}
  \end{center}
\end{table}
%%%%%%%%%%%%%%%%%%%%%%%%%%%%%%%%%%%%%%%%%%%%%%%%%%%%%%%
%%%%%%%%%%%%%%%%%%%%%%%%%%%%%%%%%%%%%%%%%%%%%%%%%%%%%%%

\subsection{Evaluations of the proportions}\label{sec:6_ratios}

We calculate the ratios of the observables to quantitatively evaluate the decrease and increase in the observables to the increase in the number density of the instantons and anti-instantons without suffering uncertainties that come from the renormalization constant, normalization factor, and lattice scale. The ratios are defined as follows:
\begin{equation}
  R_{\text{obs}}(m_{c}) = \frac{\langle \mathcal{O}(m_{c})\rangle}{\langle \mathcal{O}^{\text{sta}}\rangle}\label{eq:dif_ratios}
\end{equation}
The numerator $\langle \mathcal{O}(m_{c}) \rangle$ inidicates the observables that are calculated using the configurations with the additional monopoles and anti-monopoles of the magnetic charges $m_{c}$. The denominator $\langle \mathcal{O}^{\text{sta}}\rangle$ refers to the observables that are calculated using the standard configurations.
\begin{figure*}[htbp]
  \begin{center}
    \includegraphics[width=160mm]{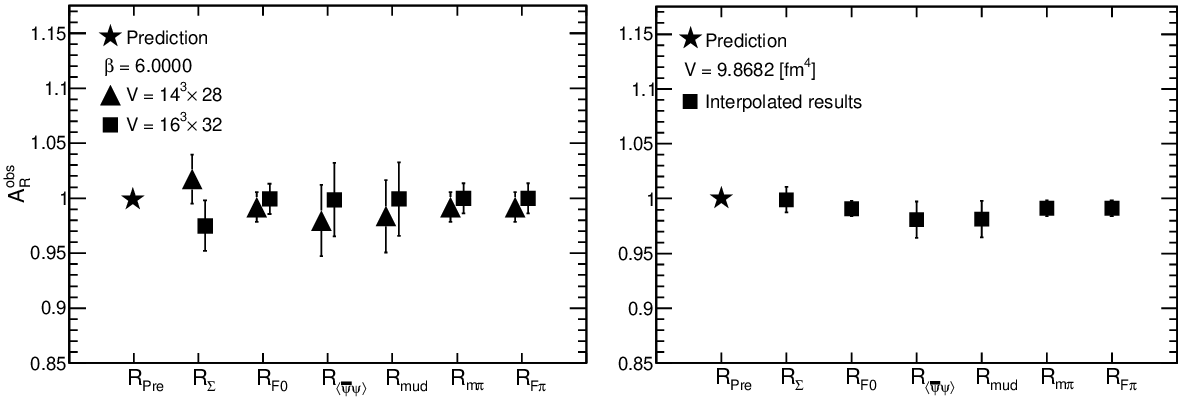}
  \end{center}
  \setlength\abovecaptionskip{-1pt}
  \caption{Comparing the prediction with the fitting results of the slopes $A_{R}^{\text{obs}}$ to the ratios $R_{\text{obs}}$. The configurations of $V= 14^{3}\times28$ and $V= 16^{3}\times32$ of $\beta = 6.0000$ (left) and the interpolated results (right).}\label{fig:R_fitting_results}
\end{figure*}
\begin{table}[htbp]
  \begin{center}
    \caption{The fitting results of the slopes $A_{R}^{\text{\text{obs}}}$ obtained by the linear functions $R_{\text{obs}} = A_{R}^{\text{obs}}R_{N_{I}}^{\frac{1}{2}}$ and $R_{\text{\text{obs}}} = A_{R}^{\text{obs}}R_{N_{I}}^{\frac{1}{4}}$. The ratios are computed using the results of the configurations of the lattice volumes $V = 14^{3}\times28$ and $V = 16^{3}\times32$ of $\beta = 6.0000$ and the interpolated results. The prediction of the slope values is $A_{R}^{\text{Pre}}$ = 1.}\label{tb:Ratios_fitting}
    \begin{tabular}{|c|c|c|c|c|}\hline
      Conf & Obs. & $A_{R}^{\text{obs}}$ & FR &$\frac{\chi^{2}}{\text{dof}}$\\\hline      
      & $\Sigma_{\text{RMT}}$                  & 1.02(2)     & $R_{N_{I}}^{\frac{1}{2}}$: 0.99-1.32 & 1.2/5.0 \\\cline{2-5}
      & $F_{0}$                         & 0.992(13)   &  $R_{N_{I}}^{\frac{1}{4}}$: 0.99-1.15 & 0.4/5.0 \\\cline{2-5}
      $14^{3}\times28$ & $\langle\bar{\psi}\psi\rangle$ & 0.98(3)     & $R_{N_{I}}^{\frac{1}{2}}$: 0.99-1.32 & 0.3/5.0 \\\cline{2-5}
       & $\hat{\bar{m}}_{ud}$            & 0.98(3)     & $R_{N_{I}}^{\frac{1}{2}}$: 0.99-1.32 & 0.3/5.0 \\\cline{2-5}
      & $m_{\pi}$                       & 0.992(14)   & $R_{N_{I}}^{\frac{1}{4}}$: 0.99-1.15 & 0.4/5.0 \\\cline{2-5}
     & $F_{\pi}$                       & 0.992(14)   & $R_{N_{I}}^{\frac{1}{4}}$: 0.99-1.15 & 0.4/5.0 \\\hline
      & $\Sigma_{\text{RMT}}$                  & 0.97(2)      & $R_{N_{I}}^{\frac{1}{2}}$: 1.02-1.27 & 0.7/5.0 \\\cline{2-5}
      & $F_{0}$                          & 0.999(14)   & $R_{N_{I}}^{\frac{1}{4}}$: 1.00-1.13 & 0.4/5.0 \\\cline{2-5}
     $16^{3}\times32$ & $\langle\bar{\psi}\psi\rangle$ & 1.00(3)    & $R_{N_{I}}^{\frac{1}{2}}$: 1.02-1.27 & 0.2/5.0 \\\cline{2-5}
         & $\bar{m}_{ud}$                   & 1.00(3)     & $R_{N_{I}}^{\frac{1}{2}}$: 1.02-1.27 & 0.2/5.0 \\\cline{2-5}
      & $m_{\pi}$                        & 1.000(14)   & $R_{N_{I}}^{\frac{1}{4}}$: 1.00-1.13 & 0.4/5.0 \\\cline{2-5}
     & $F_{\pi}$                        & 1.000(14)   & $R_{N_{I}}^{\frac{1}{4}}$: 1.00-1.13 & 0.4/5.0 \\ \hline
       & $\Sigma_{\text{RMT}}$           & 0.999(12)   & $R_{N_{I}}^{\frac{1}{2}}$: 1.01-1.25 & 0.8/5.0 \\\cline{2-5}
      & $F_{0}$                         & 0.991(7)     & $R_{N_{I}}^{\frac{1}{4}}$: 1.00-1.12 & 0.5/5.0 \\\cline{2-5}
      Interp. & $\langle\bar{\psi}\psi\rangle$ & 0.981(17)   & $R_{N_{I}}^{\frac{1}{2}}$: 1.01-1.25 & 0.4/5.0 \\\cline{2-5}
       & $\bar{m}_{ud}$                  & 0.981(17)    & $R_{N_{I}}^{\frac{1}{2}}$: 1.01-1.25 & 0.4/5.0 \\\cline{2-5}
      & $m_{\pi}$                        & 0.991(7)    & $R_{N_{I}}^{\frac{1}{4}}$: 1.00-1.12 & 0.5/5.0 \\\cline{2-5}
      & $F_{\pi}$                        & 0.991(7)    & $R_{N_{I}}^{\frac{1}{4}}$: 1.00-1.12 & 0.5/5.0 \\\hline
   \end{tabular}
  \end{center}
\end{table}

The ratios are predicted using the predictions concerning the chiral condensate~(\ref{eq:chiral_ins_prediction1}) and the decay constant of the pseudoscalar at the chiral limit~(\ref{eq:pion_decay}) as follows:
\begin{enumerate}
\item When the observable increases in direct proportion to the square root of the number density of the instantons and anti-instantons, the ratio~(\ref{eq:dif_ratios}) equals
\begin{equation}
  \left[\frac{N_{I}(m_{c})}{N_{I}^{\text{sta}}}\right]^{\frac{1}{2}} \equiv R_{N_{I}}^{\frac{1}{2}}\label{eq:ni_pred_1ov2}.
\end{equation}
\item When the observable increases in direct proportion to the one-fourth root of the number density of the instantons and anti-instantons, the ratio~(\ref{eq:dif_ratios}) equals
\begin{equation}
  \left[\frac{N_{I}(m_{c})}{N_{I}^{\text{sta}}}\right]^{\frac{1}{4}} \equiv R_{N_{I}}^{\frac{1}{4}}\label{eq:ni_pred_1ov4}.
\end{equation}
\end{enumerate}
We plot the ratios concerning the numbers of instantons and anti-instantons along the horizontal axis and the ratio of the observables to the vertical axis, and fit the linear function
\begin{equation}
  R_{\text{obs}} = A_{R}^{\text{obs}}R_{N_{I}}^{\frac{1}{2}}
\end{equation}
to the numerical results of case 1. Similarly, we fit the linear function
\begin{equation}
  R_{\text{obs}} = A_{R}^{\text{obs}}R_{N_{I}}^{\frac{1}{4}}
\end{equation}
to the numerical results of case 2. The prediction of the slopes is $A_{R}^{\text{Pre}} = 1$. We compare the fitting results of the slope values with the predictions, as shown in Table~\ref{tb:Ratios_fitting} and Fig.~\ref{fig:R_fitting_results}. Figure~\ref{fig:R_fitting_results} shows no influence of the finite lattice volume on the slope values, and the slope values are consistent with the prediction. Therefore, the observables decrease or increase in direct proportion to the square root or the one-fourth root of the number density of the instantons and anti-instantons.
%%%%%%%%%%%%%%%%%%%%%%%%%%%%%%%%%%%%%%%%%%%%%%%%%%%%%%%
%%%%%%%%%%%%%%%%%%%%%%%%%%%%%%%%%%%%%%%%%%%%%%%%%%%%%%%

\subsection{Catalytic effect on the pion decay}\label{sec:sec4_catalytic_effects}
%&&&&&&&&&&&&&&&&&&&&&&&&&&&&&&&&&&&&&&&&&&&&&&&&
\begin{table*}[htbp]
  \begin{center}
    \caption{Comparisons of the computed results of the decay width $\Gamma^{14^{3}28}$, $\Gamma^{16^{3}32}$, and $\Gamma^{\text{int}}$ with the predictions $\Gamma^{\text{Pre1}}$ and $\Gamma^{\text{Pre2}}$. The outcomes of $\Gamma^{14^{3}28}$, $\Gamma^{16^{3}32}$, and $\Gamma^{\text{int}}$ calculated using the lattice $V = 14^{3}\times28$,  $V = 16^{3}\times32$, and interpolated results, respectively. The predictions $\Gamma^{\text{Pre1}}$ and $\Gamma^{\text{Pre2}}$ are for the lattice volumes $V = 14^{3}\times28$ ($V_{\text{phys}} = 5.7845 $ [fm$^{4}$]) and $V = 16^{3}\times32$ ($V_{\text{phys}} = $ 9.8682 [fm$^{4}$]), respectively.}\label{tb:interp_Gamma}
    \begin{tabular}{|c|c|c|c|c|c|}\hline
      $m_{c}$  & $\Gamma^{\text{Pre1}}$ [sec$^{-1}$]  & $\Gamma^{14^{3}28}$  [sec$^{-1}$]  & $\Gamma^{\text{Pre2}}$ [sec$^{-1}$]  & $\Gamma^{16^{3}32}$ [sec$^{-1}$]  & $\Gamma^{\text{int}}$ [sec$^{-1}$]  \\
      & $\times10^{7}$ & $\times10^{7}$ &  $\times10^{7}$ & $\times10^{7}$ & $\times10^{7}$  \\\hline
       N. C. 
         &  3.774(7)  &  3.8(3) & 3.774(7)    & 3.8(3) & 3.77(14)  \\\hline
       0  &  3.774(7)  &  3.6(3) & 3.774(7)    & 3.8(3) & 3.84(17) \\\hline
       1  &  5.099(9)  &  4.6(4) & 4.544(8)    & 4.4(4) & 4.5(2)    \\\hline
       2  &  6.471(12) & 6.6(5) & 5.333(10)  & 6.3(5)  &  6.1(3)    \\\hline
       3  &  7.873(14) & 7.9(6) & 6.136(11)  & 7.2(6)  &  6.6(3)     \\\hline
       4  &  9.294(17) & 9.3(7) & 6.951(13)  & 8.6(6)  &  7.4(3)     \\\hline
       5  &  10.73(2)   & 9.5(7) & 7.775(14)  & 8.5(6)  &  7.8(3)    \\\hline
    \end{tabular}
  \end{center}
\end{table*}
%&&&&&&&&&&&&&&&&&&&&&&&&&&&&&&&&&&&&&&&&&&&&&&&&
\begin{figure*}[htbp]
  \begin{center}
    \includegraphics[width=160mm]{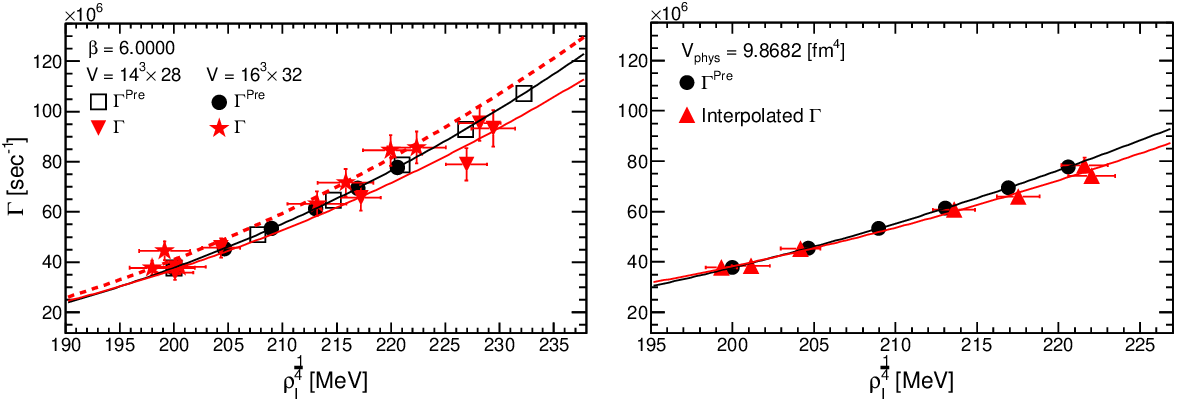}
  \end{center}
  \setlength\abovecaptionskip{-1pt}
  \caption{Comparing the partial decay width of the charged pion $\Gamma$ of the configurations $\beta = 6.0000$, $V= 14^{3}\times28$ and $V= 16^{3}\times32$ (left) and the interpolated results (right) with the predictions $\Gamma^{\text{Pre}}$. In the left figure, the black full line, the red full line, and the red dashed line indicate the fitting results of the predictions, the numerical results of $V = 14^{3}\times28$, and $V = 16^{3}\times32$, respectively. In the right figure, the full black line and red line indicate the fitting results of the prediction and the interpolated results, respectively.}\label{fig:Gamma}
\end{figure*}
%&&&&&&&&&&&&&&&&&&&&&&&&&&&&&&&&&&&&&&&&&&&&&&&&

We have shown that the pion mass becomes heavy, and the pion decay constant increases. These are caused by increasing the number density of the instantons and anti-instantons created by the additional monopoles and anti-monopoles. We have quantitatively demonstrated that the increases in the pion mass and pion decay constant are directly proportional to the one-fourth root of the number density of the instantons and anti-instantons. This relation is confirmed by comparing the ratios of the predictions with the ratios of the numerical results.

The instanton and monopole creations increase the mass and decay constant of the pion, and the increases affect the pion decay. Therefore, we estimate the effects of the instanton and monopole creations on the pion decay using the numerical results and predictions of the pion mass and pion decay constant as the input values.

This study focuses on the following decay of the charged pions $\pi^{\pm}$, that is, more than 99.99$\%$~\cite{PDG_2017} of the charged pions decay into muons and neutrinos.
\begin{equation}
    \pi^{+} \rightarrow \mu^{+} + \nu_{\mu}, \ \ \pi^{-} \rightarrow \mu^{-} + \overline{\nu}_{\mu}\nonumber
\end{equation}
The partial decay width of the charged pions is estimated as follows~\cite{Kugo1}:
\begin{equation}
  \Gamma(\pi^{-} \rightarrow \mu^{-} + \overline{\nu}_{\mu}) = \frac{(G_{F}F_{\pi}\cos\theta_{c})^{2}}{4\pi m_{\pi}^{3}}m_{\mu}^{2}(m_{\pi}^{2} - m_{\mu}^{2})^{2}\label{eq:decay_width_1}.
\end{equation}
Here, we use the experimental outcomes as follows: The Dirac constant is $\hbar = 6.582119514 (40) \times10^{-16}$ [eV$\cdot s$], and the Fermi constant is $G_{F} = 1.1663787(6) \times 10^{-5}$ [GeV$^{-2}$]~\cite{PDG_2017}. We do not consider the errors of the experimental outcomes because they are negligibly more minuscule than the errors of the numerical results. When we matched the numerical results of the pion mass and pion decay constant with the experimental outcomes in subsection~\ref{sec:6_match_exp}, we used the experimental outcomes of $\pi^{-}$; therefore, suppose that the experimental outcomes of the decay of the charged pion $\pi^{-}$ are the same as $\pi^{+}$.
%&&&&&&&&&&&&&&&&&&&&&&&&&&&&&&&&&&&&&&&&&&&&&&&&
\begin{table*}[htbp]
  \begin{center}
    \caption{The fitting results of the partial decay width $\Gamma$ by the curve $\Gamma = p_{1}x^{3} - p_{2}x + \frac{p_{3}}{x}$, ($x$ = $\rho_{I}^{\frac{1}{4}}$ [MeV]). Pre1 stands for the fitting results of the prediction $\Gamma^{\text{Pre1}}$.}\label{tb:Gamma_fitting1}
    \begin{tabular}{|c|c|c|c|c|c|}\hline
      Conf & $p_{1}$ [sec$^{-1}\cdot$MeV$^{-3}$] & $p_{2}$  [sec$^{-1}\cdot$MeV$^{-1}$] & $p_{3}$ [sec$^{-1}\cdot$MeV] & FR: $\rho^{\frac{1}{4}}$ [MeV]  &$\frac{\chi^{2}}{\text{dof}}$\\
      &  &  $\times10^{6}$ & $\times10^{10}$ & $\times10^{2}$ & \\\hline
      Pre1  & 25.89(4) & 1.187(3) & 1.360(8) & 1.99-2.33 & 0.0/3.0\\  \hline               
      $14^{3}\times28$ & 22.5(1.8) & 1.00(14) & 1.1(4) & 1.99-2.30 & 1.9/4.0  \\\hline
      $16^{3}\times32$ & 26(2) & 1.11(17) & 1.1(5) & 1.97-2.23 & 1.7/4.0 \\\hline
      Interp. & 24.2(1.1) & 1.16(8) & 1.5(2) & 1.97-2.23 & 1.5/4.0 \\\hline
    \end{tabular}
  \end{center}
\end{table*}
%&&&&&&&&&&&&&&&&&&&&&&&&&&&&&&&&&&&&&&&&&&&&&&&&

To check the consistency of formula~(\ref{eq:decay_width_1}) with the experimental result of the partial decay width, first, we substitute the interpolated results of the pion mass and pion decay constant of the normal configurations in Tables~\ref{tb:interp_ops_decay} and~\ref{tb:interp_ops_masses} and estimate the partial decay width. The result is $\Gamma^{\text{int}}$ = 3.77(14)$\times10^{7}$ [sec$^{-1}$].

The experimental result of the lifetime of the charged pion~\cite{PDG_2017} is 
\begin{equation}
\tau^{\text{Exp}} = 2.6033(5)\times10^{-8} \ [\text{sec}]\label{eq:exp_lifet}.
\end{equation}
The decay width of the charged pion, which is estimated using the experimental outcome of the lifetime of the charged pion is $\Gamma^{\text{Exp}} = 3.8413(7)\times10^{7} \ [\text{sec}^{-1}]$. The estimation coming from the numerical result is 1.9$\%$ larger than the experimental outcome; however, this result verifies that we can adequately estimate the partial decay width of the charged pion with formula~(\ref{eq:decay_width_1}) and the numerical results.
%%%%%%%%%%%%%%%%%%%%%%%%%
\begin{table*}[htbp]
  \begin{center}
    \caption{Comparisons of the lifetimes $\tau^{14^{3}28}$, $\tau^{16^{3}32}$, and $\tau^{\text{int}}$ with the predictions $\tau^{\text{Pre1}}$ and $\tau^{\text{Pre2}}$. The outcomes of $\tau^{14^{3}28}$, $\tau^{16^{3}32}$, and $\tau^{\text{int}}$ calculated using the lattice $V = 14^{3}\times28$, $V = 16^{3}\times32$, and the interpolated results, respectively. The predictions $\tau^{\text{Pre1}}$ and $\tau^{\text{Pre2}}$ are for the lattice volumes $V = 14^{3}\times28$ ($V_{\text{phys}} = 5.7845 $ [fm$^{4}$]) and $V = 16^{3}\times32$ ($V_{\text{phys}} = $ 9.8682 [fm$^{4}$]), respectively.}\label{tb:interp_tau}
    \begin{tabular}{|c|c|c|c|c|c|}\hline
      $m_{c}$ & $\tau^{\text{Pre1}}$  [sec] & $\tau^{14^{3}28}$  [sec] & $\tau^{\text{Pre2}}$  [sec] & $\tau^{16^{3}32}$  [sec] & $\tau^{\text{int}}$ [sec]\\\
      & $\times10^{-8}$ & $\times10^{-8}$ &  $\times10^{-8}$ & $\times10^{-8}$ & $\times10^{-8}$ \\\hline
       N. C. 
         & 2.649(5)    & 2.6(2)      & 2.649(5) & 2.65(19) & 2.65(10) \\\hline
       0  & 2.649(5)    & 2.8(2)      & 2.649(5) & 2.6(2)     & 2.60(11) \\\hline
       1  & 1.961(4)    & 2.19(19)   & 2.201(4) &  2.25(19)  & 2.21(10) \\\hline
       2  & 1.545(3)    & 1.52(12)   & 1.875(3) &  1.58(12)   & 1.65(7)  \\\hline
       3  & 1.270(2)    & 1.27(10)   & 1.630(3) &  1.40(11)  & 1.51(7)  \\\hline
       4  & 1.076(2)    & 1.07(8)     & 1.439(3) & 1.17(9)    & 1.35(5)  \\\hline
       5  & 0.9321(17) & 1.05(7)    &   1.286(2) & 1.18(9)    & 1.28(5)  \\\hline
    \end{tabular}
  \end{center}
\end{table*}
%%%%%%%%%%%%%%%%%%%%%%%%%%%%
\begin{figure*}[htbp]
  \begin{center}
    \includegraphics[width=160mm]{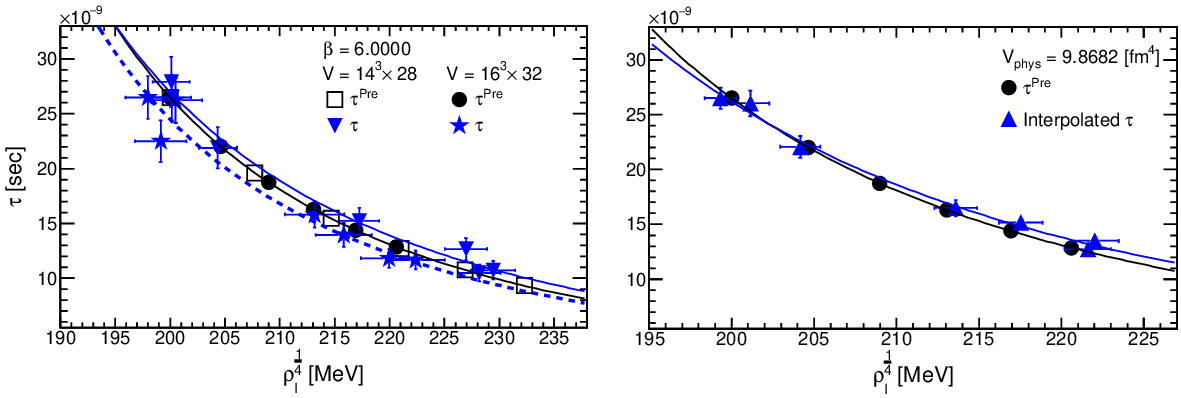}
  \end{center}
  \setlength\abovecaptionskip{-1pt}
  \caption{Comparing the lifetime of the charged pion $\tau$ of the configurations of $\beta = 6.0000$, $V= 14^{3}\times28$ and $V= 16^{3}\times32$ (left) and the continuum limit (right) with the prediction $\tau^{\text{Pre}}$. In the left figure, the black full line, the blue full line, and the blue dashed line indicate the fitting results of the predictions, the numerical results of $V = 14^{3}\times28$, and $V = 16^{3}\times32$, respectively. In the right figure, the full black line and blue line indicate the fitting results of the prediction and the interpolated results, respectively.}\label{fig:tau}
\end{figure*}
%%%%%%%%%%%%%%%%%%%%%%%%%%%%

Next, we estimate the partial decay width by substituting the following numerical results and predictions into formula~(\ref{eq:decay_width_1}): (i) The numerical results of the lattice volumes $V = 14^{3}\times28$ and $V = 16^{3}\times32$ of $\beta = 6.0000$ and the predictions are presented in Table~\ref{tb:computed_ops} in~\ref{sec:comp_results_all}. (ii) The interpolated results of the pion mass and pion decay constant are shown in Tables~\ref{tb:interp_ops_masses} and~\ref{tb:interp_ops_decay}. The calculated results of the partial decay width compared with the predictions are listed in Table~\ref{tb:interp_Gamma}.

We have confirmed that the increases in the pion mass and pion decay constant are directly proportional to the one-fourth root of the number density of the instantons and anti-instantons. Therefore, we make the following curve to fit the calculated results of the decay width:
\begin{equation}
  \Gamma = p_{1}x^{3} - p_{2}x + \frac{p_{3}}{x}, \ x = \rho_{I}^{\frac{1}{4}} \ [\text{MeV}]\label{eq:fitting_width}.
\end{equation}
We fit this curve to the calculated results, as shown in Fig.~\ref{fig:Gamma}, and list the fitting results in Table~\ref{tb:Gamma_fitting1}. The fitting results of the parameters $p_{1}$, $p_{2}$, and $p_{3}$ are resonably consistent with the fitting results of the prediction, as indicated in Table~\ref{tb:Gamma_fitting1}; thus, the finite lattice volume does not affect the results of the partial decay width of the charged pion. Figure~\ref{fig:Gamma} shows that the partial decay width becomes wider with increasing one-fourth root of number density of the instantons and anti-instantons, and the increases are consistent with the predictions.
%%%%%%%%%%%%%%%%%%%%%%%%%%%%
\begin{table*}[htbp]
  \begin{center}
    \caption{The fitting results of the lifetime $\tau$ by the curve $\tau = \left(p_{1}x^{3} - p_{2}x + \frac{p_{3}}{x}\right)^{-1}$, ($x$ = $\rho_{I}^{\frac{1}{4}}$ [MeV]). Pre1 stands for the fitting results of the prediction $\tau^{\text{Pre1}}$.}\label{tb:tau_fitting1}
    \begin{tabular}{|c|c|c|c|c|c|}\hline
   Conf & $p_{1}$ [sec$^{-1}\cdot$MeV$^{-3}$] & $p_{2}$  [sec$^{-1}\cdot$MeV$^{-1}$] & $p_{3}$ [sec$^{-1}\cdot$MeV] & FR: $\rho^{\frac{1}{4}}$ [MeV]  &$\frac{\chi^{2}}{\text{dof}}$\\
      &  &  $\times10^{6}$ & $\times10^{10}$ & $\times10^{2}$ & \\\hline      
      Pre1 & 25.89(4) & 1.187(3) & 1.360(8) & 1.99-2.33 & 0.0/3.0\\\hline            
      $14^{3}\times28$ & 22.4(1.4) & 0.98(1.4) & 1.1(5) & 1.99-2.30 & 1.8/4.0 \\\hline
      $16^{3}\times32$ & 25(2) & 1.04(17) & 1.0(5) & 1.97-2.23 & 1.7/4.0 \\\hline
      Interp. & 23.1(1.1) & 1.05(8) & 1.3(2) & 1.97-2.23 & 1.5/4.0 \\\hline
    \end{tabular}
  \end{center}
\end{table*}

The partial decay width of the charged pion into the muon and neutrino is almost 100$\%$; therefore, finally, we estimate the lifetime $\tau$ of the charged pion from its inverse of the partial decay width and quantitatively demonstrate the catalytic effect caused by the monopole and instanton creations on the pion decay.

Similar to the decay width, we estimate the lifetime using the numerical results of the lattice volumes $V = 14^{3}\times28$ and $V = 16^{3}\times32$, interpolated results, and predictions. We compare the numerical results with the predictions in Table~\ref{tb:interp_tau} and plot them in Fig.~\ref{fig:tau}. To quantitatively evaluate the catalytic effect on the pion decay, we fit the inverse of the fitting curve~(\ref{eq:fitting_width})
\begin{equation}
  \tau = \left(p_{1}x^{3} - p_{2}x + \frac{p_{3}}{x}\right)^{-1}, \ x = \rho_{I}^{\frac{1}{4}} \ [\text{MeV}]\label{eq:fitting_tau}
\end{equation}
to the computed results, as shown in Fig.~\ref{fig:tau}. The fitting results of the lifetime are given in Table~\ref{tb:tau_fitting1}. The fitting results of the parameters $p_{1}$, $p_{2}$, and $p_{3}$ indicate no influence of the finite lattice volume, and they are reasonably consistent with the fitting results of the predictions and the partial decay width $\Gamma$.

The results demonstrate that the lifetime of the charged pion becomes shorter than the experimental result~(\ref{eq:exp_lifet}) by increasing the one-fourth root of the number density of the instantons and anti-instantons without the finite lattice volume influence. This is the catalytic effect caused by the monopole and instanton creations on the pion decay. % 16 March 2022 
%%%%%%%%%%%%%%%%%%%%%%%%%%%%%%%%%%%%%%%%%%%%%%%%%%%%%%%
%%%%%%%%%%%%%%%%%%%%%%%%%%%%%%%%%%%%%%%%%%%%%%%%%%%%%%%
%%%%%%%%%%%%%%%%%%%%%%%%%%%%%%%%%%%%%%%%%%%%%%%%%%%%%%%

\section{Summary and conclusions}\label{sec:7}

We performed the simulations to find the clues to observe the effects of the magnetic monopoles and instantons on the observables by experiments. The primary purposes of this study were to inspect the influences of the finite lattice volume and discretization on the observables and quantitative relations, which we obtained in our previous research, and to obtain the interpolated results at the continuum limit.

For these purposes, we prepared the standard configurations and the configurations of the following lattice volumes and parameter values to which the monopoles and anti-monopoles were added, in the quenched approximation of QCD: (i) To inspect the influence of the finite lattice volume, $V = 14^{3}\times28$ and $V = 16^{3}\times32$ of the parameter value $\beta$ = 6.0000. (ii) To inspect the influence of the discretization, $V = 12^{3}\times24$ of $\beta = 5.8124$, $V = 14^{3}\times28$ of $\beta = 5.9256$, and $V = 20^{3}\times40$ of $\beta = 6.1366$. We added the monopoles and anti-monopoles varying the magnetic charges from 0 to 5 to these configurations. We calculated the low-lying eigenvalues and eigenvectors of the overlap Dirac operator using these configurations, evaluated the observables using the low-lying eigenvalues and eigenvectors, and demonstrated the quantitative relations by comparing the observables with the predictions. We interpolated the results to the continuum limit using the outcomes in this study and the outcomes of the lattice of $V = 18^{3}\times32$ of $\beta = 6.0522$ in the previous study.

First, we confirmed that the additional monopoles and anti-monopoles do not affect the numerical results of the lattice spacing by computing the static potential. We did not observe significant effects of the additional monopoles and anti-monopoles on Abelian dominance or monopole dominance.

Then, we found minor effects of the finite lattice volume on the number density of the monopoles and anti-monopoles that are computed using the standard configurations.

We analyzed the effects of the additional monopoles and anti-monopoles. We found that the influences of the finite lattice volume appear when we add the monopoles and anti-monopoles with the magnetic charges $m_{c}$ higher than 3 to the small lattice volume $V = 12^{3}\times24$ of the coarse lattice spacing $\beta = 5.8457$. Therefore, we calculated the numerical results using the lattices $V = 20^{3}\times40$ of $\beta = 6.1366$ of the magnetic charges 4 and 5 and interpolated the numerical results to the continuum limit instead of using the numerical results of the lattices $V = 12^{3}\times24$ of $\beta = 5.8457$ of the magnetic charges 4 and 5.

Incidentally, to reveal the effects of the added monopoles and anti-monopoles on color confinement, we calculated the average values of the absolute values of the Polyakov loops as the order parameter of the color deconfinement phase transition, number density of the long monopole loops, and Polyakov loop susceptibility at finite temperatures.

At finite temperatures, we demonstrated that the order parameter of the color deconfinement phase transition becomes approximately zero even in the deconfinement phase by increasing the number density of the long monopole loops; thus, the color deconfinement phase reaches the color confinement phase by lengthening the long monopole loops. We then demonstrated that the transition temperature from the color confinement phase to the color deconfinement phase linearly rises by increasing the magnetic charges of the monopole and anti-monopole, without using the diagonalized configurations under certain gauge conditions. This result indicated that a quark is confined even if the temperature exceeds the transition temperature of color deconfinement by increasing the number of monopoles and anti-monopoles composing the long loops; thus, the additional monopoles and anti-monopoles induce color confinement.

Next, we calculated the number density of the instantons and anti-instantons using the standard configurations. We confirmed that the outcomes of the number density of the instantons and anti-instantons do not vary even if we change the physical volumes fixing the values of the lattice spacing parameter, or we change the lattice volumes and values of the lattice spacing parameter fixing the physical volumes. Furthermore, these outcomes are reasonably consistent with the prediction of the phenomenological model concerning the instantons.

In previous studies, we have already shown that one pair of the additional monopole and anti-monopole of the magnetic charge $m_{c}$ = 1 makes one instanton or anti-instanton. To inspect the influences of the finite lattice volume and the discretization on this quantitative relationship, we plotted the number of magnetic charges $m_{c}$ of the additional monopole and anti-monopole to the horizontal axis and the number of instantons and anti-instantons $N_{I}$ to the vertical axis, and we fitted the linear function to the numerical results. We confirmed that the fitting results are reasonably consistent with our previous outcomes and predictions, without the influence of the finite lattice volume or discretization. Therefore, we obtained the number density of the instantons and anti-instantons at the continuum limit by the interpolation.

This study analytically estimated the number of instantons and anti-instantons from the topological charges; therefore, we inspected the influence of the added monopoles and anti-monopoles, finite lattice volume, and discretization on the distributions of the topological charge by comparing them with the predicted distribution functions. We confirmed that the additional monopoles and anti-monopoles add the topological charges without changing the vacuum structures; in addition, the finite lattice volume and discretization do not affect the distributions of the topological charges.

Next, to inspect the influences of the added monopoles and anti-monopoles, finite lattice volume, and discretization on the low-lying eigenvalues of the overlap Dirac operator, we calculated the distributions of the nearest-neighbor spacing of the eigenvalues, the spectral rigidity, and ratios of the low-lying eigenvalues. We demonstrated that the numerical results are consistent with the predictions of the RMT and chRMT. This result indicated that the additional monopoles and anti-monopoles, finite lattice volume, and discretization do not affect the fluctuations of the eigenvalues from the short to long-range or the ratios of the low-lying eigenvalues.

Our previous research has already demonstrated the effects caused by the instanton and anti-instanton creations on the renormalized chiral condensate and renormalized quark masses in the $\overline{\text{MS}}$-scheme at 2 [GeV], light meson masses, and decay constants. In addition, the previous research revealed the quantitative relations among the number density of the instantons and anti-instantons and the observables; however, we showed only the outcomes that were obtained using the lattice $V = 18^{3}\times32$ of $\beta = 6.0522$. Therefore, we needed to inspect the influences of the finite lattice volume and discretization on these outcomes using the lattices (i) and (ii) that we mentioned above.

We first computed the correlation functions of the scalar density and the pseudoscalar density using the eigenvalues and eigenvectors. We then confirmed that the additional monopoles and anti-monopoles, finite lattice volume, and discretization do not affect the PCAC relation. We demonstrated that the interpolated results of the renormalization constant for the scalar density at the continuum limit are reasonably consistent even if we add the monopoles and anti-monopoles.

It is well known that instantons and anti-instantons are closely related to chiral symmetry breaking; therefore, we evaluated the renormalized chiral condensate in the $\overline{\text{MS}}$-scheme at 2 [GeV] using the outcomes of the scale parameter of the eigenvalue distribution in the chRMT and the renormalization constant for the scalar density. We demonstrated that the renormalized chiral condensate decreases in direct proportion to the square root of the number density of the instanrtons and anti-instantons. However, we found that the finite lattice volume influenced the renormalized chiral condensate and the interpolated results are not consistent with the prediction.

Therefore, we obtained the normalization factor by matching the numerical results with the experimental results and improved the computation method by using the normalization factor without suffering from uncertainties that come from the determinations of the lattice scales. By this computation, we could calculate the observables without any discretization influences. The higher precision calculations than our previous study revealed the quantitative relations among the observables and the number density of the instantons and anti-instantons.

First, to show that we could adequately calculate the observables using the normalization factor, we estimated the decay constant of the pseudoscalar at the chiral limit. We showed that the decay constant of the pseudoscalar meson at the chiral limit linearly increases with the one-fourth root of the number density of the instantons and anti-instantons without any influences of the finite lattice volume or the discretization.

We then recalculated the renormalized chiral condensate in the $\overline{\text{MS}}$-scheme at 2 [GeV], which is derived from the GMOR relation, using the normalization factor, and showed that the discretization does not influence the outcomes. As a result, we demonstrated that the renormalized chiral condensate in the $\overline{\text{MS}}$-scheme at 2 [GeV] decreases in direct proportion to the square root of the number density of the instantons and anti-instantons. Furthermore, the slope values are consistent with the predictions without the influences of the finite lattice volume. The phenomenological model cannot predict the average size of the instanton or anti-instanton; therefore, we estimated the inverse of the average size of the instanton or anti-instanton using the slope value of the interpolated results and confirmed the consistency with the phenomenological model. 

These results indicated that the instantons and anti-instantons created by the additional monopoles and anti-monopoles induce chiral symmetry breaking, as explained in the phenomenological model. Therefore, we supposed that chiral symmetry breaking which is induced by instanton and anti-instanton creations acts on quark-mass generation.

In order to verify this assumption, we estimated the renormalized average mass of the up and down quarks in the $\overline{\text{MS}}$-scheme at 2 [GeV] and revealed that the renormalized average mass of the light quarks increases in direct proportion to the square root of the number density of the instantons and anti-instantons. We compared the fitting results of the slope value showing that the quark mass increases with the predictions to quantitatively evaluate the increases. The fitting results showed that the finite lattice volume does not affect the slope values; however, the fitting results of the slope are slightly steeper than the prediction.

We supposed that the pion mass increases with increasing the number density of the instantons and anti-instantons because the pion mass is estimated from the up and down quark masses. 

Incidentally, we indicated that the slope of the PCAC relation linearly increases without the logarithmic divergence near the chiral limit and that the slope values are not affected by increasing the magnetic charges of the additional monopoles and anti-monopoles.

Furthermore, we demonstrated that the linear relationship between the decay constant of the pseudoscalar and the square mass of the pseudoscalar holds and that the logarithmic divergence near the chiral limit does not appear. We showed that the additional monopoles and anti-monopoles push the intercept value of this linear relationship, that is the decay constant at the chiral limit, up.

Therefore, we supposed that the pion mass and pion decay constant increase in direct proportion to the one-fourth root of the number density of the instantons and anti-instantons. We evaluated these increases by comparing the fitting results of the slope values with the predictions and confirmed that the finite lattice volume does not affect the outcomes. Finally, we revealed that the numerical results are consistent with these assumptions.

However, the decrease and increases in the observables include uncertainties that come from the renormalization constant, normalization factor, and lattice scales; therefore, to remove these uncertainties, we calculated the ratios among the observables computed using the standard configurations and the observables computed using the configurations with the additional monopoles and anti-monopoles. We showed that the ratios of the numerical results are consistent with the prediction, without any influences of the finite lattice volume.

The pion mass and pion decay constant are increased by increasing the number density of the instantons and anti-instantons. Accordingly, we predicted that the pion decay is affected by the increases in the pion mass and pion decay constant. We focused on the charged pion decay and estimated the effects of the instanton and anti-instanton creations on the partial decay width of the charged pion. We found that the partial decay width of the charged pion becomes wider than the experimental outcome when the one-fourth root of the number density of the instantons and anti-instantons increases. We quantitatively demonstrated that the increases in the partial decay width are consistent with the prediction by fitting the curve. Furthermore, we did not observe any influence of the finite lattice volume on these outcomes.

We estimated the lifetime of the charged pion from the inverse of the decay width and found the catalytic effect: the lifetime of the charged pion becomes shorter than the experimental outcome by increasing the one-fourth root of the number density of the instantons and anti-instantons. To quantitatively demonstrate the catalytic effect, we fitted the curve to the numerical results and ascertained that the fitting results of the numerical results are consistent with the fitting results of the prediction without the influence of the finite lattice volume.

Finally, we provide the following conclusion: when the monopoles and anti-monopoles are added, they form long monopole loops that are closely related to color confinement and create instantons and anti-instantons. The instantons and anti-instantons induce chiral symmetry breaking and increase the light quark masses, pion mass, and pion decay constant. These effects result in the lifetime of the charged pion becoming shorter than the experimental result.
%%%%%%%%%%%%%%%%%%%%%%%%%
%%%%%%%%%%%%%%%%%%%%%%%%%
%%%%%%%%%%%%%%%%%%%%%%%%%
%%% END SEC 7
%%%%%%%%%%%%%%%%%%%%%

  \section*{Acknowledgements}
  The author has started this research project with A. Di Giacomo and F. Pucci and appreciates the helpful discussion and advice. The author would like to thank M. D'Elia for the helpful discussions. Furthermore, the author received financial support for visiting the University of Pisa from the Istituto Nazionale di Fisica Nucleare at the University of Pisa and the Joint Institute for Nuclear Research. This research project was performed using the SX-series, computer clusters, and XC40 at the Research Center for Nuclear Physics and the Cybermedia Center at Osaka University and the Yukawa Institute for Theoretical Physics at Kyoto University. In addition, we have used the storage elements from the Japan Lattice Data Grid at the Research Center for Nuclear Physics at Osaka University. We sincerely appreciate the computer resources that they provided to us and their technical support.
%%%%%%%%%%%%%%%%%%%%%%%%%
%%%%%%%%%%%%%%%%%%%%%%%%%
%%%%%%%%%%%%%%%%%%%%%%%%%

\appendix
\onecolumn
\section{The number of observed zero modes $N_{Z}$, number of instantons $N_{I}$, and number density of instantons and anti-instantons $\frac{N_{I}}{V}$}\label{sec:nz_ni_niv}
\begin{table*}[htbp]
  \caption{The number of observed zero modes $N_{Z}$ = $|Q|$, number of instantons $N_{I}$, and number density of instantons and anti-instantons $\rho_{I}$, $\rho_{I}^{\frac{1}{2}}$, $\rho_{I}^{\frac{1}{4}}$, and their predictions. The predictions are indicated as Pre1 for the physical volume $V_{\text{phys}}$ = 5.7845 [fm$^{4}$] ($\beta$ = 6.0000, $V = 14^{3}\times28$) and Pre2 for the physical volume $V_{\text{phys}}$ = 9.8582 [fm$^{4}$] (apart from $\beta$ = 6.0000, $V = 16^{3}\times32$) in the column of $\beta$. These are calculated with equations~(\ref{eq:num_ins_add}) and~(\ref{eq:num_ins_dens_add}).}\label{tb:Nzero_add_1}
  \begin{center}
      \begin{tabular}{|c|c|c|c|c|c|c|c|c|} \hline
        $\beta$ & $V$ &  $m_{c}$ & $N_{Z}$ & $N_{I}$ & $\rho_{I}$ [GeV$^{4}$] & $\rho_{I}^{\frac{1}{2}}$ [GeV$^{2}$]  & $\rho_{I}^{\frac{1}{4}}$ [MeV] & $N_{\text{conf}}$ \\
        & & & & & $\times10^{-3}$ & $\times10^{-2}$ & $\times10^{2}$ &  \\ \hline
        Pre1 & - & Normal conf
             & 1.9713 & 6.1044 & 1.6000 & 4.0000 & 2.0000 & - \\ \cline{3-9} 
      & &  1 & 2.1306 & 7.1044 & 1.8621 & 4.3152 & 2.0773 & - \\ \cline{3-9}
      & &  2 & 2.2777 & 8.1044 & 2.1242 & 4.6089 & 2.1468 & - \\ \cline{3-9} 
      & &  3 & 2.4153 & 9.1044 & 2.3863 & 4.8850 & 2.2102 & - \\ \cline{3-9} 
      & &  4 & 2.5450 & 10.104 & 2.6484 & 5.1463 & 2.2685 & - \\ \cline{3-9} 
      & &  5 & 2.6682 & 11.104 & 2.9105 & 5.3949 & 2.3227 & - \\ \hline          
      Pre2 & - & Normal conf
             & 2.5748 & 10.414 & 1.6000 & 4.0000 & 2.0000 & - \\ \cline{3-9} 
      & &  1 & 2.6975 & 11.414 & 1.7536 & 4.1877 & 2.0464 & - \\ \cline{3-9}  
      & &  2 & 2.8144 & 12.414 & 1.9073 & 4.3672 & 2.0898 & - \\ \cline{3-9} 
      & &  3 & 2.9265 & 13.414 & 2.0609 & 4.5397 & 2.1307 & - \\ \cline{3-9} 
      & &  4 & 3.0343 & 14.414 & 2.2146 & 4.7059 & 2.1693 & - \\ \cline{3-9} 
      & &  5 & 3.1383 & 15.414 & 2.3682 & 4.8664 & 2.2060 & - \\ \hline
      5.8457 & $12^{3}\times24$ &  Normal conf      
                & 2.61(7) & 10.9(5) & 1.67(8) & 4.09(10) & 2.02(2) & 900 \\ \cline{3-9}
      & & 0 & 2.51(6) & 10.4(5) & 1.60(7)  & 4.00(9)  & 2.00(2) & 1100 \\ \cline{3-9}
      & & 1 & 2.75(7) & 12.0(6) & 1.84(9) & 4.29(10) & 2.07(2) &  950 \\ \cline{3-9}
      & & 2 & 2.84(7) & 12.9(6) & 1.99(9)  & 4.46(10) & 2.11(2) & 1050 \\ \cline{3-9}
      & & 3 & 2.96(7) & 14.1(7) & 2.16(10) & 4.65(11) & 2.16(3) & 950 \\ \cline{3-9}
      & & 4 & 3.05(8) & 14.9(7) & 2.29(11) & 4.78(11) & 2.19(3) & 950 \\ \cline{3-9}
      & & 5 & 3.09(7) & 14.8(7) & 2.28(10) & 4.78(10) & 2.19(2) & 973 \\ \hline
      5.9256 & $14^{3}\times28$ &  Normal conf                    
            & 2.64(7) & 10.9(5) & 1.67(8)  & 4.09(9)  & 2.02(2) & 850 \\ \cline{3-9}
      & & 0 & 2.62(7) & 11.0(5) & 1.68(8)  & 4.10(10) & 2.03(2) & 868 \\ \cline{3-9}
      & & 1 & 2.73(7) & 11.9(6) & 1.84(9)  & 4.28(10) & 2.07(2) & 950 \\ \cline{3-9}
      & & 2 & 3.00(8) & 14.4(7) & 2.21(11) & 4.70(11) & 2.17(3) & 852 \\ \cline{3-9}
      & & 3 & 3.14(8) & 15.2(7) & 2.34(11) & 4.83(12) & 2.20(3) & 802 \\ \cline{3-9}
      & & 4 & 3.17(8) & 15.7(8) & 2.42(12) & 4.92(12) & 2.22(3) & 800 \\ \cline{3-9}
      & & 5 & 3.17(8) & 15.9(8) & 2.45(12) & 4.95(12) & 2.22(3) & 910 \\ \hline
      6.0000 & $14^{3}\times28$ & Normal conf
            & 1.93(4) & 6.1(2)  & 1.60(6)  & 4.01(7) & 2.002(1.7) & 1720  \\ \cline{3-9}
      & & 0 & 1.93(4) & 6.1(2)  & 1.60(5)  & 4.00(7) & 2.001(1.7) & 1800 \\ \cline{3-9}
      & & 1 & 2.05(4) & 6.7(2)  & 1.74(6)  & 4.18(7) & 2.043(1.7) & 1710 \\ \cline{3-9}
      & & 2 & 2.33(4) & 8.5(3)  & 2.23(7)  & 4.72(8) & 2.172(1.8) & 1710 \\ \cline{3-9}
      & & 3 & 2.52(5) & 10.1(3) & 2.65(9)  & 5.15(9) & 2.270(1.9) & 1720 \\ \cline{3-9}
      & & 4 & 2.53(5) & 10.6(4) & 2.77(10) & 5.26(9) & 2.29(2)    & 1732 \\ \cline{3-9}
      & & 5 & 2.53(5) & 10.3(5) & 2.71(9)  & 5.21(9) & 2.282(1.9) & 1743 \\  \cline{2-9}
       & $16^{3}\times32$ &  Normal conf          
            & 2.54(6) & 10.0(4) & 1.54(6)  & 3.92(8)  & 1.98(2) & 1040 \\ \cline{3-9}
      & & 0 & 2.55(7) & 10.5(5) & 1.62(8)  & 4.02(10) & 2.01(2) & 880 \\ \cline{3-9}
      & & 1 & 2.54(7) & 10.2(5) & 1.57(7)  & 3.97(9)  & 1.99(2) & 880 \\ \cline{3-9}
      & & 2 & 2.88(8) & 13.4(7) & 2.07(10) & 4.54(11) & 2.13(3) & 880 \\ \cline{3-9}
      & & 3 & 2.95(8) & 14.1(7) & 2.17(10) & 4.66(11) & 2.16(3) & 880 \\ \cline{3-9}
      & & 4 & 3.13(8) & 15.9(8) & 2.44(12) & 4.94(12) & 2.22(3) & 930 \\ \cline{3-9}
      & & 5 & 3.13(8) & 15.2(7) & 2.34(11) & 4.84(11) & 2.20(3) & 861 \\ \hline
      6.1366 & $20^{3}\times40$ &  Normal conf
                & 2.49(9) &  9.7(6) & 1.49(10) & 3.85(12) & 1.96(3) &  440 \\ \cline{3-9}
      & & 4 & 3.21(11) & 15.9(1.0) & 2.44(16) & 4.94(16) & 2.22(4) & 448 \\ \cline{3-9}
      & & 5 & 3.10(11) & 14.9(1.0) & 2.29(15) & 4.78(15) & 2.19(4) & 450 \\ \hline
    \end{tabular}
  \end{center}
\end{table*}
\clearpage

\section{The numerical results of the scale parameter $\Sigma_{\text{RMT}}$ and renormalized chiral condensate $\langle\bar{\psi}\psi\rangle_{\text{RMT}}^{\overline{\text{MS}}}$ in the $\overline{\text{MS}}$-scheme at 2 [GeV]}\label{sec:rmt_Sigma_conds}
  
\begin{table}[htbp]
  \caption{The scale parameter $\Sigma_{\text{RMT}}$ of the distribution of the eigenvalues and renormalized chiral condensate $\langle\bar{\psi}\psi\rangle_{\text{RMT}}^{\overline{\text{MS}}}$ in the $\overline{\text{MS}}$-scheme at 2 [GeV]. The renormalized chiral condensate of $\beta =$ 5.9044, $V = 16^{4}$ is estimated with $\hat{Z}_{S} =$ 1.08(7) which is obtained by the interpolation of the numerical results of $\hat{Z}_{S}$.}\label{tb:computed_sigma_rmt}
  \begin{center}
      \begin{tabular}{|c|c|c|c|c|} \hline
        $\beta$ & $V$ & $m_{c}$ & $a^{3}\Sigma_{\text{RMT}}$ & $\langle\bar{\psi}\psi\rangle_{\text{RMT}}^{\overline{\text{MS}}}$ [GeV$^{3}$] \\
       &  &  &  $\times10^{-3}$  & $\times10^{-2}$   \\\hline
        5.8457 & $12^{3}\times24$ &  N. C.
& 3.90(3) & -2.53(9)  \\ \cline{3-5}
& & 0 & 3.90(3) & -2.56(9)  \\ \cline{3-5}
& & 1 & 4.13(4) & -2.72(9)  \\ \cline{3-5}
& & 2 & 4.35(4) & -2.90(10) \\ \cline{3-5}
& & 3 & 4.19(4) & -2.92(10) \\ \hline
5.9044  & $16^{4}$ &  N. C. 
              & 2.833(18) & -2.40(16) \\ \hline
 5.9256 & $14^{3}\times28$ & N. C. 
              & 2.77(3) & -2.58(9)   \\ \cline{3-5}
        & & 0 & 2.69(2) & -2.58(9)    \\ \cline{3-5}
        & & 1 & 2.92(2) & -2.53(9)    \\ \cline{3-5}
        & & 2 & 3.15(3) & -2.70(9)    \\ \cline{3-5}
        & & 3 & 3.24(3) & -2.94(10)   \\ \cline{3-5}
        & & 4 & 3.21(3) & -3.04(11)   \\ \cline{3-5}
        & & 5 & 3.20(3) & -3.06(11)   \\ \hline
6.0000 & $14^{3}\times28$ &  N. C. 
              & 1.820(11) & -2.21(7)  \\ \cline{3-5}
        & & 0 & 1.794(9)  & -2.16(7)    \\ \cline{3-5}
        & & 1 & 1.928(11) & -2.41(8)    \\ \cline{3-5}
        & & 2 & 2.213(13) & -2.68(9)    \\ \cline{3-5}
        & & 3 & 2.350(15) & -2.81(10)   \\ \cline{3-5}
        & & 4 & 2.411(14) & -2.98(10)   \\ \cline{3-5}
        & & 5 & 2.431(14) & -2.97(10) \\  \cline{2-5}
 & $16^{3}\times32$ &  N. C. 
              & 2.023(18) & -2.58(9)  \\ \cline{3-5}
        & & 0 & 1.982(17) & -2.57(9)   \\ \cline{3-5}
        & & 1 & 2.026(17) & -2.62(9)   \\ \cline{3-5}
        & & 2 & 2.273(19) & -2.89(10)  \\ \cline{3-5}
        & & 3 & 2.41(2)  & -3.04(10)  \\ \cline{3-5}
        & & 4 & 2.414(19) & -3.07(10)  \\ \cline{3-5}
        & & 5 & 2.41(2)  & -3.09(11)  \\ \hline
6.0522 & $18^{3}\times32$ &  N. C. 
              &  1.535(13) & -2.45(8)    \\ \cline{3-5}
        & & 0 &  1.529(13) & -2.44(8)      \\ \cline{3-5}
        & & 1 &  1.629(14) & -2.62(9)      \\ \cline{3-5}
        & & 2 &  1.781(15) & -2.84(10)     \\ \cline{3-5}
        & & 3 &  1.921(18) & -3.03(10)     \\ \cline{3-5}
        & & 4 &  2.018(18) & -3.20(11)     \\ \cline{3-5}
        & & 5 &  2.03(2)   & -3.18(11)   \\  \hline
6.1366 & $20^{3}\times40$ &   N. C. 
              & 1.064(14) & -2.48(9) \\ \cline{3-5}
        & & 4 & 1.381(17)   & -3.11(11) \\ \cline{3-5}
        & & 5 & 1.50(2)   & -3.39(12) \\ \hline
      \end{tabular}
  \end{center}
\end{table}
\clearpage

\section{Numerical results of PCAC relation and renormalization constant $\hat{Z}_{S}$}\label{sec:pcac_zs}

\begin{table*}[htbp]
  \begin{center}
    \caption{The fitting results of the slopes $aA_{\text{PCAC}}^{(1)}$, $aA_{\text{PCAC}}^{(2)}$ and intercept $a^{2}B_{\text{PCAC}}$ by the functions $(am_{\pi})^{2} = a^{2}A_{\text{PCAC}}^{(1)}\bar{m}_{ud} + a^{2}B_{\text{PCAC}}$ and $(am_{\pi})^{2} = a^{2}A_{\text{PCAC}}^{(2)}\bar{m}_{ud}$. The calculated results of the renormalization constant $\hat{Z}_{S}$ for the scalar density.}\label{tb:prop_v12xx3x24_b5p84572_fit1_2}
     \begin{tabular}{|c|c|c|c|c|c|c|c|c|c|} \hline
      $\beta $ & $V$ & $m_{c}$ & $aA_{\text{PCAC}}^{(1)}$ & $a^{2}B_{\text{PCAC}}$ & $aA_{\text{PCAC}}^{(2)}$ & FR: $a\bar{m}_{ud}$ & $\frac{\chi^{2}}{\text{dof}}$ & $\frac{\chi^{2}}{\text{dof}}$ & $\hat{Z}_{S}$ \\
      & & & &  $\times10^{-3}$ & & $\times10^{-2}$ & $(aA^{(1)}, a^{2}B)$ & $(aA^{(2)})$ & \\ \hline
      5.8457 & $12^{3}\times24$ & Normal conf      
      & 1.955(18) & -6.0(1.0) & 1.854(4) & 3.7-8.9 & 11.7/12.0 & 46.2/12.0 & 1.16(4)  \\ \cline{3-10}
      & & 0 & 1.99(2)  & -8.6(1.3) & 1.833(5) &  3.7-7.0 & 8.4/9.0   & 52.8/10.0  & 1.18(4) \\ \cline{3-10}
      & & 1 & 1.987(15) & -7.7(7) & 1.825(4) & 2.5-7.0  & 13.0/13.0 & 139.2/14.0 & 1.18(4) \\ \cline{3-10}
      & & 2 & 1.944(16) & -7.1(8)   & 1.802(4) &  2.8-7.0 & 12.0/12.0 & 94.1/13.0  & 1.20(4) \\ \cline{3-10}
      & & 3 & 1.89(2) & -7.5(9) & 1.728(5) & 2.8-6.3 & 9.6/10.0 & 75.8/11.0 & 1.25(4)   \\\cline{3-10}
      & & 4 & 1.87(2) & -7.4(1.0) & 1.704(5) & 2.8-6.3 & 9.9/10.0 & 70.3/11.0 & 1.27(4) \\ \cline{3-10}
      & & 5 & 1.83(4) & -7.3(1.4) & 1.641(7) & 2.8-5.1 & 6.0/6.0 & 31.8/7.0 & 1.32(4) \\ \hline
       5.9256 & $14^{3}\times28$ &   Normal conf                                              
            &  1.839(18) & -4.0(8)   & 1.756(4) &  3.2-6.5 & 10.1/10.0 & 32.4/11.0  & 1.05(4) \\ \cline{3-10}
      & & 0 &  1.81(2)   & -3.2(1.0) & 1.743(4) &  3.2-5.7 & 7.6/8.0   & 17.8/9.0   & 1.06(4) \\ \cline{3-10}    
      & & 1 &  1.832(16) & -3.0(7)   & 1.766(3) &  2.9-6.5 & 11.2/11.0 & 29.7/12.0  & 1.05(3) \\ \cline{3-10}
      & & 2 &  1.82(2)   & -3.1(1.0) & 1.750(4) &  3.2-5.7 & 7.9/8.0   & 17.6/9.0   & 1.06(4) \\ \cline{3-10}
      & & 3 &  1.828(19) & -3.5(8)   & 1.744(4) &  2.6-5.4 & 9.0/9.0   & 29.9/10.0  & 1.06(4) \\ \cline{3-10}
      & & 4 &  1.838(16) & -5.2(6)   & 1.708(4) &  2.4-5.4 & 9.8/10.0  & 81.3/11.0  & 1.08(4) \\ \cline{3-10}
      & & 5 &  1.835(14) & -4.5(5)   & 1.710(4) &  2.1-5.4 & 11.1/11.0 & 94.9/12.0  & 1.08(4) \\ \hline
      6.0000 & $14^{3}\times28$ &  Normal conf                                                
            &  1.862(15) & -4.3(6)   & 1.761(3) &  2.8-6.2 & 10.7/11.0 & 59.2/12.0  & 0.92(3) \\ \cline{3-10}
      & & 0 &  1.857(16) & -3.6(7)   & 1.775(3) &  3.0-6.6 & 9.6/10.0  & 36.6/11.0  & 0.91(3) \\ \cline{3-10}
      & & 1 &  1.88(2)   & -4.9(6)   & 1.711(5) &  1.8-3.8 & 7.2/7.0   & 65.7/8.0   & 0.95(3) \\ \cline{3-10}
      & & 2 &  1.853(14) & -3.9(6)   & 1.764(3) &  2.8-6.2 & 11.4/11.0 & 52.8/12.0  & 0.92(3) \\ \cline{3-10}
      & & 3 &  1.852(19) & -2.9(8)   & 1.784(3) &  3.0-5.7 & 9.0/9.0   & 22.1/10.0  & 0.91(3) \\ \cline{3-10}
      & & 4 &  1.85(2)   & -4.0(6)   & 1.726(4) &  2.1-4.1 & 7.2/7.0   & 47.4/8.0   & 0.94(3) \\ \cline{3-10}
      & & 5 &  1.880(19) & -5.1(7)   & 1.750(3) &  2.8-5.0 & 8.0/8.0   & 55.9/9.0   & 0.93(3) \\ \cline{2-10}
      & $16^{3}\times32$ & Normal conf                                                        
            &  1.711(14) & -1.4(5)   & 1.676(3) &  2.5-5.2 & 10.2/10.0 & 17.1/11.0  & 0.97(3) \\ \cline{3-10}
      & & 0 &  1.725(11) & -3.0(4)   & 1.644(3) &  2.1-5.2 & 11.9/12.0 & 70.0/13.0  & 0.98(3) \\ \cline{3-10} 
      & & 1 &  1.711(9)  & -2.5(3)   & 1.644(3) &  1.6-5.2 & 14.6/14.0 & 69.9/15.0  & 0.99(3) \\ \cline{3-10} 
      & & 2 &  1.725(17) & -1.8(7)   & 1.679(3) &  2.6-5.2 & 9.5/9.0   & 17.5/10.0  & 0.96(3) \\ \cline{3-10}
      & & 3 &  1.718(17) & -1.2(7)   & 1.689(3) &  2.8-5.2 & 9.4/9.0   & 12.4/10.0  & 0.96(3) \\ \cline{3-10}
      & & 4 &  1.713(17) & -1.3(6)   & 1.679(3) &  2.5-4.8 & 8.0/8.0   & 12.0/9.0   & 0.96(3) \\ \cline{3-10}
      & & 5 &  1.710(18) & -1.6(7)   & 1.670(3) &  2.8-5.0 & 7.9/8.0   & 13.2/9.0   & 0.97(3) \\ \hline
      6.1366 & $20^{3}\times40$ & Normal conf
      & 1.505(15) & -2.0(4) & 1.434(3) & 1.7-3.6 & 8.1/8.0 & 33.1/9.0 & 0.90(3) \\ \cline{3-10}
      & & 4 & 1.518(15) & -1.1(4) & 1.482(3) & 2.0-3.8 & 8.0/8.0 & 13.8/9.0 & 0.87(3) \\ \cline{3-10}
      & & 5 & 1.514(12) & -9(3) & 1.482(3) & 1.8-3.8 & 9.3/9.0 & 16.8/10.0 & 0.87(3)  \\ \hline
    \end{tabular}
  \end{center}
\end{table*}
\clearpage

\section{Fitting results of $aF_{PS}$ and $(am_{PS})^{2}$ and the computed results of intersections $aF_{PS}^{\pi}$ and $am_{PS}^{\pi}$}\label{sec:fitting_res_fps_intersec_fpi_mpi}

\begin{table*}[htbp]
  \caption{The fitting results of the slope $a^{-1}A_{PS}$ and intercept $aB_{PS}$ by the function $aF_{PS} = a^{-1}A_{PS}x + aB_{PS}$, [$x = (am_{PS})^{2}$] together with the computed results of the intersections $aF_{PS}^{\pi}$ and $am_{PS}^{\pi}$.}\label{tb:fitresults_aps_ampi2_inter}
  \begin{center}
    \begin{tabular}{|c|c|c|c|c|c|c|c|c|} \hline
      $\beta $ & $V$ &  $m_{c}$ & $a^{-1}A_{PS}$ & $aB_{PS}$ & $FR: (am_{PS})^{2}$ & $\frac{\chi^{2}}{\text{dof}}$ & $aF_{PS}^{\pi}$ & $am_{PS}^{\pi}$ \\
 &  &  &  & $\times10^{-2}$   & $\times10^{-2}$ & & $\times10^{-2}$ & $\times10^{-2}$ \\ \hline
      5.8457 & $12^{3}\times24$ &  N. C.
               & 0.174(9) & 4.60(9)  & 4.2-17.6 & 3.9/17.0 &  4.69(9) &  7.09(14) \\ \cline{3-9}
     & &  0 & 0.171(8) & 4.66(9)  & 4.7-17.6 & 5.2/16.0 &   4.75(9) &  7.19(14) \\ \cline{3-9}
     & &  1 & 0.174(9) & 4.69(10)  & 4.8-17.6 & 3.1/16.0 &  4.79(10) &  7.24(15) \\ \cline{3-9}
     & &  2 & 0.170(10)  & 4.85(10) & 4.2-16.2 & 1.9/16.0 & 4.94(11) &  7.47(16) \\ \cline{3-9}
     & &  3 & 0.174(10)  & 4.87(10) & 4.0-15.8 & 1.3/16.0 & 4.96(10) &  7.51(15) \\ \hline
      5.9256 & $14^{3}\times28$ &  N. C.                     
          & 0.205(7)  & 3.89(6)  & 2.9-13.9 & 11.0/18.0  & 3.96(7) & 6.00(10)  \\ \cline{3-9}
      & & 0 & 0.206(7)  & 3.90(6)  & 2.9-13.8 & 9.6/18.0   & 3.97(6) & 6.01(10) \\ \cline{3-9}
    & & 1 & 0.197(8) & 3.99(6) & 2.9-13.2 & 8.0/17.0 & 4.07(7) & 6.15(10)   \\ \cline{3-9}
    & & 2 & 0.191(8)  & 4.17(7)  & 2.9-13.1 & 6.5/17.0   & 4.24(7) & 6.42(11)  \\ \cline{3-9}
    & & 3 & 0.192(10) & 4.20(8)  & 3.0-12.3 & 2.2/16.0   & 4.28(8) & 6.47(12)  \\ \cline{3-9}
    & & 4 & 0.188(8)  & 4.28(7)  & 2.9-13.0 & 5.1/17.0   & 4.36(7) & 6.60(11)  \\ \cline{3-9}
    & & 5 & 0.191(8)  & 4.22(7)  & 2.9-13.0 & 4.6/17.0   & 4.30(7) & 6.50(11)  \\  \hline
      6.0000 & $14^{3}\times28$ &  N. C.                                                                   
          & 0.218(7)  & 3.26(5)  & 2.9-12.4 & 12.1/17.0  & 3.31(5) & 5.01(8)   \\ \cline{3-9}
    & & 0 & 0.214(7)  & 3.24(5)  & 2.9-9.8 & 6.8/16.0   & 3.29(6) & 4.98(8)   \\ \cline{3-9}
    & & 1 & 0.211(8)  & 3.34(6)  & 3.0-9.8 & 4.8/16.0   & 3.39(6) & 5.13(9)   \\ \cline{3-9}
    & & 2 & 0.209(7)  & 3.50(5)  & 2.9-12.4 & 9.9/17.0   & 3.57(5) & 5.39(8)   \\ \cline{3-9}
    & & 3 & 0.197(8)  & 3.60(6)  & 2.6-9.1 & 3.6/16.0   & 3.66(6) & 5.54(9)   \\ \cline{3-9}
    & & 4 & 0.192(8)  & 3.70(5)  & 2.6-9.0 & 3.0/16.0   & 3.76(6) & 5.69(9)   \\ \cline{3-9}
    & & 5 & 0.193(7)  & 3.71(5)  & 2.6-9.8 & 5.7/17.0   & 3.77(5) & 5.71(7)   \\ \cline{2-9}  
    &   $16^{3}\times32$ &  N. C.                                                            
          & 0.240(7)  & 3.30(5)  & 2.1-9.4 & 15.6/19.0  & 3.37(5) & 5.09(7)  \\ \cline{3-9}
    & & 0 & 0.242(8)  & 3.31(5)  & 2.5-9.3 & 10.4/17.0  & 3.37(6) & 5.10(9)   \\ \cline{3-9}
    & & 1 & 0.240(8)  & 3.37(6)  & 2.9-9.4 & 8.8/17.0   & 3.44(6) & 5.20(10)  \\ \cline{3-9}
    & & 2 & 0.230(8)  & 3.53(5)  & 2.5-9.5 & 12.2/18.0  & 3.60(6) & 5.45(9)   \\ \cline{3-9}
    & & 3 & 0.229(8)  & 3.60(6)  & 2.5-9.5 & 10.5/18.0  & 3.67(6) & 5.55(9)   \\ \cline{3-9}
    & & 4 & 0.211(8)  & 3.70(5)  & 2.2-10.3 & 5.6/17.0   & 3.77(6) & 5.70(8)   \\ \cline{3-9}    
    & & 5 & 0.219(8)  & 3.69(5)  & 2.5-10.8 & 7.9/17.0   & 3.76(5) & 5.69(8)   \\  \hline
      6.1366 & $20^{3}\times40$ & N. C.
      & 0.292(8) & 2.69(4) & 1.4-7.3 & 18.9/18.0  & 2.74(4) & 4.15(6)  \\ \cline{3-9}
      & & 4 & 0.265(9) & 2.92(4) & 1.5-7.1 & 12.2/17.0  & 2.98(4) & 4.50(7)  \\ \cline{3-9}
      & & 5 & 0.260(9) & 3.00(4) & 1.5-7.1 & 13.7/17.0  & 3.05(4) & 4.62(6)  \\  \hline
   \end{tabular}
  \end{center}
\end{table*}
\clearpage

\section{The numerical results of the observables}\label{sec:comp_results_all}

\begin{table*}[htbp]
  \caption{The outcomes of the decay constant $F_{0}$, renormalized chiral condensate $\langle\bar{\psi}\psi\rangle^{\overline{\text{MS}}}$ and average mass of light quarks $\hat{\bar{m}}_{ud}^{\overline{\text{MS}}}$ in the $\overline{\text{MS}}$-scheme at 2 [GeV], pion mass $m_{\pi}$ and decay constant $F_{\pi}$, and ratios of the decay constants $\frac{F_{\pi}}{F_{0}}$. The predictions of these observables are indicated as Pre1 for $V_{\text{phys}}$ = 5.7845 [fm$^{4}$] and Pre2 for $V_{\text{phys}}$ = 9.8582 [fm$^{4}$] in the column of $\beta$.}\label{tb:computed_ops}
  \begin{center}
      \begin{tabular}{|c|c|c|c|c|c|c|c|c|} \hline
         &  & & $F_{0}$  & $\langle\bar{\psi}\psi\rangle^{\overline{\text{MS}}}$  & $\hat{\bar{m}}_{ud}^{\overline{\text{MS}}}$ & $m_{\pi}$ & $F_{\pi}$ & \\
        $\beta$ & $V$ & $m_{c}$  & [MeV] &  [GeV$^{3}$]  &   [MeV]  &  [MeV]  & [MeV] & $\frac{F_{\pi}}{F_{0}}$ \\
        &  &  & & $\times10^{-2}$ & & $\times10^{2}$  &   &  \\\hline
        Pre1 & - & N. C.
                  &  85.366 & -2.0280 & 3.5$_{+0.7}^{-0.3}$ &  1.395706(2) &  92.28(8)     & 1.0810(9)      \\\cline{3-9}
        &  & 1 &   88.666 & -2.1878 & 3.8$_{+0.8}^{-0.3}$ &  1.449656(2) &  95.84(9)     &  1.0809(10)    \\\cline{3-9}
        &  & 2 &   91.634 & -2.3367 & 4.0$_{+0.8}^{-0.3}$ &  1.498178(2) &  99.05(9)     &  1.0809(10)   \\ \cline{3-9}
        &  & 3 &   94.338 & -2.4766 & 4.3$_{+0.9}^{-0.4}$ &  1.542397(2) &  101.98(9)    &  1.0810(10)   \\ \cline{3-9}
        &  & 4 &   96.828 & -2.6091 & 4.5$_{+0.9}^{-0.4}$ &  1.583109(2) &  104.67(10)   &  1.0810(10)  \\ \cline{3-9}
        &  & 5 &   99.140 & -2.7352 & 4.7$_{+0.9}^{-0.4}$ &  1.620903(2) &  107.17(10)   &  1.0810(10)\\ \hline
        Pre2 & - & N. C.
               & 85.366 & -2.0280 & 3.5$^{+0.7}_{-0.3}$ & 1.395706(2) & 92.28(9)     &   1.0810(11)     \\\cline{3-9} 
        &  & 1 & 87.345 & -2.1231 & 3.7$^{+0.7}_{-0.3}$ & 1.428069(2) & 94.42(9)  &   1.0810(10)  \\\cline{3-9} 
        &  & 2 & 89.199 & -2.2142 & 3.8$^{+0.8}_{-0.3}$ & 1.458370(2) & 96.42(9)  &   1.0810(10)  \\\cline{3-9} 
        &  & 3 & 90.943 & -2.3016 & 4.0$^{+0.8}_{-0.3}$ & 1.486892(2) & 98.31(9)  &   1.0810(10)  \\\cline{3-9} 
        &  & 4 & 92.593 & -2.3859 & 4.1$^{+0.8}_{-0.4}$ & 1.513862(2) & 100.09(9) &   1.0810(10)  \\\cline{3-9} 
        &  & 5 & 94.159 & -2.4672 & 4.3$^{+0.9}_{-0.4}$ & 1.539462(2) & 101.78(9) &   1.0809(10)  \\\hline       
         5.8457 & $12^{3}\times24$ & N. C.
              & 91(3) &  -1.95(13) & 4.1(3) &  1.40(4) & 92(3) & 1.02(3)  \\\cline{3-9}
        && 0  & 92(3) &  -2.01(13) & 4.2(3) &  1.42(4) & 94(3) & 1.02(3)  \\\cline{3-9}
        && 1  & 93(3) &  -2.03(14) & 4.3(3) &  1.43(4) & 94(3) & 1.02(3)  \\\cline{3-9}
        && 2  & 95(3) &  -2.17(14) & 4.6(3) &  1.47(4) & 97(3) & 1.02(3)  \\\cline{3-9}
        && 3  & 96(3) &  -2.19(14) & 4.6(3) &  1.48(4) & 98(3) & 1.02(3)  \\\hline
        5.9256 & $14^{3}\times28$ &  N. C.
              & 91(2)  & -1.95(11) & 4.1(2) & 1.40(3)  & 92(2)  & 1.02(2)   \\\cline{3-9}
        & & 0 & 91(2)  & -1.96(11) & 4.1(2) & 1.40(3)  & 92(2)  & 1.02(2)  \\ \cline{3-9}
        & & 1 & 93(2)  & -2.05(12) & 4.3(3) & 1.43(3)  & 95(2)  & 1.02(2)  \\ \cline{3-9}
        & & 2 & 97(2)  & -2.24(13) & 4.7(3) & 1.49(4)  & 99(2)  & 1.02(2)  \\ \cline{3-9}
        & & 3 & 98(2)  & -2.27(14) & 4.8(3) & 1.51(4)  & 100(3) & 1.02(3)  \\ \cline{3-9}
        & & 4 & 100(2) & -2.36(13) & 5.0(3) & 1.54(4)  & 102(2) & 1.02(2)  \\ \cline{3-9}
        & & 5 & 98(2)  & -2.30(13) & 4.8(3) & 1.51(4)  & 100(2) & 1.02(2)  \\ \hline
       6.0000 & $14^{3}\times28$ &  N. C. 
              & 91(2)  & -1.96(11) & 4.1(2) &  1.40(3)  & 92(2)  & 1.02(2)      \\\cline{3-9}
        & & 0 & 90(2)  & -1.94(11) & 4.0(2) &  1.39(3)  & 92(2)  & 1.02(2)      \\ \cline{3-9}
        & & 1 & 93(2)  & -2.06(12) & 4.3(3) &  1.43(3)  & 95(2)  & 1.02(3)      \\ \cline{3-9}
        & & 2 & 98(2)  & -2.27(12) & 4.7(3) &  1.50(3)  & 99(2)  & 1.02(2)      \\ \cline{3-9}
        & & 3 & 100(2) & -2.40(13) & 5.0(3) &  1.54(4)  & 102(2) & 1.02(2)      \\ \cline{3-9}
        & & 4 & 103(2) & -2.52(14) & 5.3(3) &  1.58(3)  & 105(2) & 1.02(2)      \\ \cline{3-9}
        & & 5 & 103(2) & -2.54(13) & 5.3(3) &  1.59(3)  & 105(2) & 1.017(18)    \\ \cline{2-9}
        & $16^{3}\times32$  & N. C.
              & 90.6(1.8) & -1.95(10) &   4.1(2)  & 1.40(3)  & 92(2)  & 1.02(2)   \\\cline{3-9}
        & & 0 & 91(2)     & -1.96(11) &   4.1(2)  & 1.40(3)  & 92(2)  & 1.02(2)   \\ \cline{3-9}
        & & 1 & 92(2)     & -2.03(12) &   4.3(2)  & 1.42(3)  & 94(2)  & 1.02(3)   \\ \cline{3-9}
        & & 2 & 97(2)     & -2.23(12) &   4.7(3)  & 1.49(3)  & 99(2)  & 1.02(2)   \\ \cline{3-9}
        & & 3 & 99(2)     & -2.31(12) &   4.9(3)  & 1.52(3)  & 101(2) & 1.02(2)   \\ \cline{3-9}
        & & 4 & 101(2)    & -2.47(13) &   5.1(3)  & 1.56(3)  & 103(2) & 1.02(2)   \\ \cline{3-9}
        & & 5 & 101(2)    & -2.44(13) &   5.1(3)  & 1.56(3)  & 103(2) & 1.02(2)   \\ \hline
        6.1366 & $20^{3}\times40$ &  N. C. 
               & 90.6(1.8) & -1.95(10) & 4.1(2)  & 1.40(3) & 92.3(1.9) & 1.02(2)  \\\cline{3-9}
        &&  4  & 98(2)     & -2.30(12) & 4.8(3)  & 1.51(3) & 100(2)    & 1.02(2)  \\\cline{3-9}
        &&  5  & 101(2)    & -2.42(12) & 5.1(3)  & 1.55(3) & 103(2)    & 1.02(2)  \\\hline
      \end{tabular}
  \end{center}
\end{table*}
\clearpage
\twocolumn
\bibliographystyle{unsrt}
\bibliography{Chiral_symmetry_breaking_and_catalytic_effect_M.Hasegawa_05Sep2022_final_ver}

\end{document}